\documentclass[11pt,a4paper]{article}

\usepackage{amsfonts}
\usepackage{graphicx}

\title{\textbf{Generalized MSTB Models: \\ Structure and kink varieties.}}
\author{A. Alonso Izquierdo$^{(a)}$, J. Mateos Guilarte$^{(b)}$
\\ {\normalsize {\it $^{(a)}$ Departamento de Matematica
Aplicada and IUFFyM}, {\it Universidad de Salamanca,
SPAIN}}\\{\normalsize {\it $^{(b)}$ Departamento de Fisica
Fundamental and IUFFyM}, {\it Universidad de Salamanca, SPAIN}}}
\date{}

\begin{document}

\maketitle
\begin{abstract}
In this paper we describe the structure of a class of two-component
scalar field models in a (1+1) Minkowskian space-time which
generalize the well-known Montonen-Sarker-Trullinger-Bishop -hence
MSTB- model. This class includes all the field models whose static
field equations are equivalent to the Newton equations of
two-dimensional type I Liouville mechanical systems with a discrete
set of instability points. We offer a systematic procedure to
characterize these models and to identify the solitary wave or kink
solutions as homoclinic or heteroclinic trajectories in the
analogous mechanical system. This procedure is applied to a
one-parametric family of generalized MSTB models with a degree-eight
polynomial as potential energy density.
\end{abstract}

\section{Introduction}

Over the last decades topological defects or solitary waves have
played an essential r$\hat{\rm o}$le in the explanation of new
phenomena in diverse branches of Science, e.g., Cosmology and
Condensed Matter Physics. These non-linear waves are non-dispersive
spatially localized solutions of non-linear field equations
describing a given physical system. Focusing on one-dimensional wave
phenomena, we shall deal with variations of the relativistic
non-linear Klein-Gordon PDE \cite{Tasaki}
\begin{equation}
\frac{\partial^2 \phi_a}{\partial t^2}-\frac{\partial^2
\phi_a}{\partial x^2}+ \frac{\partial U}{\partial \phi_a}=0
\hspace{2cm} a=1,\dots, N \label{eq:KleGor}
\end{equation}
for $N$ scalar fields $\phi_a(x,t)$. The non-linear nature of these
equations is the key point to finding unexpected a priori
non-dispersive wave solutions. The simplest examples of relativistic
topological defects of this kind are the soliton of sine-Gordon
theory and the kink of the $(\phi)_2^4$ model. These systems
encompass only one scalar field and provide theoretical support to
explain, for instance, the appearance of superconductivity in type
II materials \cite{Eschen,Jona}, electric charge fractionization in
trans-polyacetylene (CH)$_x$ \cite{Harris}, the Josephson effect,
\cite{Dav}, etcetera. In several branches of Physics ranging from
condensed matter physics to cosmology, from fluid dynamics to
relativistic quantum field theory, more complex systems involving
several scalar fields arise. The search for topological defects,
solitary waves or kinks in models which involve several scalar
fields, as Rajaraman notices \cite{Rj}, is a very difficult
endeavor: \textit{This already brings us to the stage where no
general methods are available for obtaining all localized static
solutions (kinks), given the field equations. However, some
solutions, but by no means all, can be obtained for a class of such
Lagrangians using a little trial and error}. The move from
one-component scalar field models to two-component scalar field
models is thus a very important qualitative step, passing from the
integration of a single partial differential equation to the
integration of a system of partial differential equations. Although
there are no general methods to tackle the problem of solving
systems of non-linear partial differential equations, some
strategies for finding soliton solutions have been developed in both
the physical and mathematical literature over the last thirty years.

We now point out two of these strategies that lie at the core of the
conceptual framework to be developed in this paper:

\begin{itemize}
\item Since the models we are interested in
exhibit Lorentz invariance, see (\ref{eq:KleGor}), the search for
kinks or traveling waves is tantamount to the solution of an
analogous mechanical problem where the unit-mass point particle
motion in a certain potential $V(\phi_1,\phi_2)$ must be identified:
this analogy is established by thinking of the variable $x$ as \lq
\lq time" whereas the field components $(\phi_1,\phi_2)$ are
reinterpreted as the coordinates of the particle \cite{Rj} moving in
the Euclidean plane under the influence of the potential
$V(\phi_1,\phi_2)=-U(\phi_1,\phi_2)$. Moreover, the field
theoretical energy becomes the mechanical action. This
identification means that finite energy field configurations
correspond to finite action (in infinite time) paths in the
analogous mechanical system.

By this token one must solve only a system of ordinary
differential equations, instead of partial differential equations.
Traveling wave solutions to the PDE system are thus obtained by
applying Lorentz transformations (or Galilean, in non-relativistic
systems) to the static solutions of the ODE system.

\item In some models the potential density energy can be written as half the
square of the norm of the gradient of a \lq\lq superpotential"
$W(\phi_1,\phi_2)$:
\begin{equation}
U(\phi_1,\phi_2)={1\over 2}\left(\frac{\partial
W}{\partial\phi_1}\cdot\,\frac{\partial
W}{\partial\phi_1}+\frac{\partial
W}{\partial\phi_2}\cdot\,\frac{\partial
W}{\partial\phi_2}\right)\qquad . \label{eq:hjw}
\end{equation}

Knowledge of such a function, $W(\phi_1,\phi_2)$, allows the
Bogomolny'i-Prasad-Sommerfield arrangement \cite{BPS} for the energy
functional:
\begin{equation}
 E[\phi_1,\phi_2]=\frac{1}{2} \int dx  \sum_{a=1}^2 \left[ \frac{d
\phi_a}{dx}-\frac{\partial W}{\partial\phi_a} \right]^2 + \int dx
\sum_{a=1}^2 \frac{\partial W}{\partial \phi_a} \frac{d
\phi_a}{dx}\qquad . \label{eq:enersu}
\end{equation}
Because it is an exact differential if $W$ is well behaved, the
second term
\[
T=\int_{\mathbb R} dx \sum_{a=1}^2 \frac{\partial W}{\partial
\phi_a} \frac{d \phi_a}{dx}=\int_{\mathbb K} \, dW  \qquad ,
\qquad {\mathbb K}: {\mathbb R}\longrightarrow {\mathbb R}^2
\qquad ,
\]
where by ${\mathbb K}(x)\in Maps({\mathbb R},{\mathbb R}^2)$ we
denote a non-singular curve in ${\mathbb R}^2$, in (\ref{eq:enersu})
only depends on the values of $W$ at the end points of the curve:
$W[\phi_1(\pm\infty,t),\phi_2(\pm\infty,t)]$. If finite energy is
required, the following conditions at spatial infinity,
$\partial{\mathbb R}$, must be satisfied:
\[
(\phi_1(\pm\infty,t),\phi_2(\pm\infty,t))\in{\cal M}  \qquad ,
\qquad
\frac{d\phi_1}{dx}(\pm\infty,t)=0=\frac{d\phi_2}{dx}(\pm\infty,t)
\qquad ,
\]
where ${\cal M}$ is the set of zeroes of $U$, assuming a
non-negative $U$. Because the temporal evolution is a homotopy
transformation, when ${\cal M}$ is a discrete set
$W[\phi_1(\pm\infty,t),\phi_2(\pm\infty,t)]$ is time-independent,
and $T$ is a \lq\lq topological charge".

The second term in (\ref{eq:enersu}) is semi-definite positive and
the absolute minima of the energy are solutions of the first-order
system of ordinary differential equations:
\begin{equation}
\frac{d \phi_a}{d x}=\frac{\partial W}{\partial \phi_a}
\hspace{1cm} , \hspace{2cm} a=1,2 \qquad . \label{eq:fode}
\end{equation}
Non-dispersive extended solutions of (\ref{eq:fode}) are also stable
solutions of the second-order ODE  system (\ref{eq:sode}) obeying
the static field equations:
\begin{equation}
\frac{d^2 \phi_a}{d x^2}=\frac{\partial U}{\partial \phi_a}
\hspace{3cm} , \label{eq:sode}
\end{equation}
usually referred to as solitary waves, kinks, or one-dimensional
topological defects.

The link between these two strategies is hidden in equation
(\ref{eq:hjw}). This equation is no more than the reduced
Hamilton-Jacobi equation for zero mechanical energy trajectories of
the analogous mechanical system. As a consequence $W$ is the
Hamilton characteristic function, whereas the term superpotential
comes from the fact that only models with potential energy of the
form written in (\ref{eq:hjw}) are susceptible to supersymmetric
extensions.

In this way, the search for solitary waves in $N$ scalar field
theory is tantamount to the solving of a $N$-dimensional mechanical
system. Thus, the most favorable situation is to deal with
integrable mechanical systems. If $N=2$, there is a lot of
information about classes of integrable mechanical systems, see e.g.
\cite{Pere}. In particular, we shall choose field theoretical models
such that their analogous mechanical system is of Liouville Type I:
those two-dimensional mechanical systems such that their
Hamilton-Jacobi equation is separable using elliptic coordinates.
The simplest and best studied (1+1) dimensional two-component scalar
field theory model of this type is the MSTB model (after Montonen,
Sarker, Trullinger and Bishop, who first addressed this theory in
\cite{Monto1,Trulli1}).

Our goal in this paper is to analyze the structure of this broad
class of systems of the MSTB (the tip of the iceberg) type as well
as to study their manifold of solitary waves. Applications of
multi-component kinks to describe interesting physical systems
abound in the literature, see [7-15].
\end{itemize}

\subsection{The MSTB Model}

The MSTB model is a physical system with a proud history. In 1976
Montonen \cite{Monto1}, searching for charged solitons in a model
with one complex and one real scalar field,  discovered, by fixing
the time-dependent phase for the complex field, a (1+1)-dimensional
two real scalar field theory that provided the basic neutral
solitons. When such basic neutral solitons are embedded  in the
bigger system charged solitons arises. Thus, the MSTB model is a
(1+1)-dimensional relativistic theory of two scalar fields with
potential energy density{\footnote{We shall use non-dimensional
coordinates, fields, and parameters throughout the paper.}}:
\begin{equation}
U_{\rm MSTB}[\phi_1(x,t),\phi_2(x,t)]=\frac{1}{2}
(\phi_1^2(x,t)+\phi_2^2(x,t)-1)^2 +
\frac{\Omega^2}{2}\phi_2^2(x,t) \qquad .\label{eq:mstbpot}
\end{equation}
Considered as a function of the two scalar fields $\phi_1$ and
$\phi_2$, $U_{\rm MSTB}$ is a fourth-order polynomial isotropic in
quartic but anisotropic in quadratic terms. There is only one
non-dimensional parameter $\Omega$, which determines the intensity
of the anisotropy (as well as the time dependence of the phase of
the complex scalar field).

The kink solutions asymptotically connect elements of the set of
zeroes of $U_{\rm MSTB}(\phi_1,\phi_2)$, which in this case are two
: ${\cal M}=\{(\phi_1^+=1,\phi_2^+=0);\,
(\phi_1^-=-1,\phi_2^-=0)\}$. In his seminal paper \cite{Monto1}, the
author identified two solitary waves or kinks, joining the points
$\phi_+$ and $\phi_-$ asymptotically in the range of the coupling
constant: $\Omega^2 \in (0,1)$. The first kink was the old kink of
the $\phi^4_2$ theory with only one real scalar field:
\begin{equation}
\phi_1^{\rm TK1}(x)=\tanh \bar x \hspace{1cm} ,
\hspace{1cm}\phi^{\rm TK1}_2(x) = 0 \qquad ; \qquad
\bar{x}=\frac{(-1)^\alpha
(x-\gamma_2)-vt}{\sqrt{1-v^2}}\label{eq:mstbtk1}
\end{equation}
where $\alpha=0,1$, $v$ is a velocity slower than the speed of light
$c=1$, and $\gamma_2\in \mathbb{R}$ sets the kink center, see Figure
1. The dependence on $\bar{x}$ is due to the breaking of the
translational, Lorentz, and reflection symmetries of the model by
the kink solution. Because the second field is zero for this
solution, it is referred to with the abbreviated name of TK1 kink
(one-component topological kink): the topological charge is non-zero
and only one field component is non-null, whereas the solution lives
in the abscissa axis of the ${\mathbb R}^2$ field space, tracing the
orbit: $(\phi^{\rm TK1}_1\in (-1,1),\phi^{\rm TK1}_2=0)$, which
starts and ends at different points of ${\cal M}$.

The second and novel kink connects the points $\phi^+$ and
$\phi^-$ through the semi-elliptic orbits
\begin{equation}
\phi_1^2+\frac{\phi_2^2}{{\bar\Omega}^2}=1 \qquad \qquad , \qquad
\bar\Omega^2=1-\Omega^2 \label{eq:eor}
\end{equation}
in the ${\mathbb R}^2$ field space. Solitary wave solutions of
this type read:
\begin{equation}
\phi_1^{\rm TK2}(\bar{x})= \tanh \Omega \bar x \hspace{1cm} ,
\hspace{1cm} \phi_2^{\rm TK2}(\bar{x}) =(-1)^\beta \bar\Omega
\,{\rm sech}\,\Omega \bar x  \qquad ; \qquad \beta=0,1 \quad
,\label{eq:mstbtk2}
\end{equation}
see Figure 1. The abbreviated name for solitary waves of this type
is ${\rm TK2}$ kink (two-component topological kinks) because the
topological charge is not zero and the two field components are
non-null. It is clear that (\ref{eq:eor}) becomes a hyperbola when
$\Omega^2>1$ and ${\rm TK2}$ kinks do not exist in this range. ${\rm
TK1}$ kinks, however, are still traveling wave solutions for
$\Omega>1$.

To be precise, in a previous paper \cite{Raja1} Rajaraman and
Weinberg had identified the first class of these solutions and had
described the qualitative behavior of the second type of kinks in a
more general model. Sarker, Trullinger, and Bishop entered the game
by establishing the stability of the two kinds of solution from a
energetic point of view \cite{Trulli1}: if $\Omega^2 \in (0,1)$ the
TK2 kinks, being less energetic, are the stable traveling wave
solutions and the TK1 kinks should decay to them. If $\Omega^2>1$,
obviously the TK1 kinks are stable. In 1979, further analysis of
kink stability in this model were performed in References
\cite{Sarker79} and \cite{Trullinger79}. Study of the spectrum of
the second-order small fluctuation operator around ${\rm TK1}$ and
${\rm TK2}$ kinks confirmed that the latter are stable and the
former unstable. The lowest eigenvalue is positive (respectively,
negative) for ${\rm TK2}$ (respectively,${\rm TK1}$) kinks if
$\Omega^2<1$, and positive for ${\rm TK1}$ kinks if $\Omega^2>1$.

In the same year, Rajaraman \cite{Raja79} discovered a new kind of
kink solution for the particular value of the coupling constant
$\Omega=\frac{1}{2}$ whose orbits are the circle trajectories:
\[
(\phi_1\mp\frac{1}{4})^2+\phi_2^2=\frac{9}{16} \qquad \qquad .
\]
These ${\rm NTK2}$ kink solutions read
\[
\phi_1^{\rm NTK2}(x)=\pm\frac{3}{2} \tanh^2 \frac{\bar
x}{2}\mp\frac{1}{2} \hspace{2cm} \phi_2^{\rm NTK2}(x) =\frac{3}{2}\,
{\rm sech}\,\frac{\bar x}{2} \tanh \frac{\bar x}{2}
\]
and asymptotically reach $\phi^+$ , or, $\phi^-$ both at
$x=\pm\infty$, see Figure 1. Hence, these solitary waves are
referred to as two-component non-topological kinks, or, ${\rm NTK2}$
kinks.
\begin{figure}[htb]
\centerline{
\includegraphics[height=2.6cm]{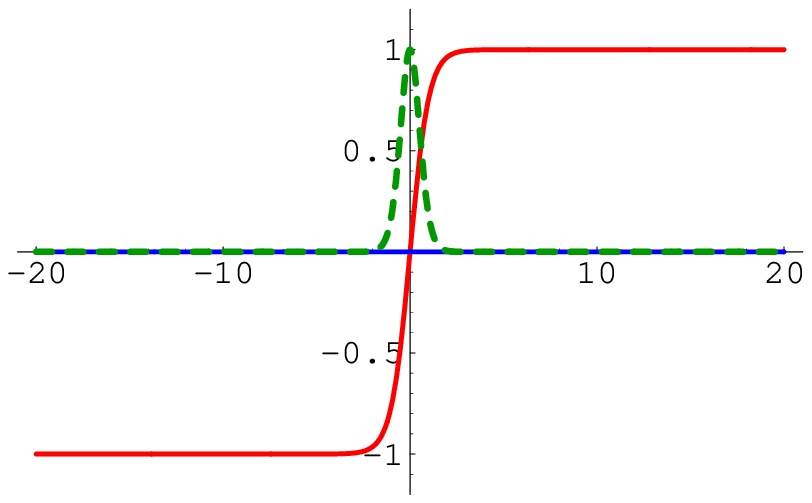}\hspace{1cm}
\includegraphics[height=2.6cm]{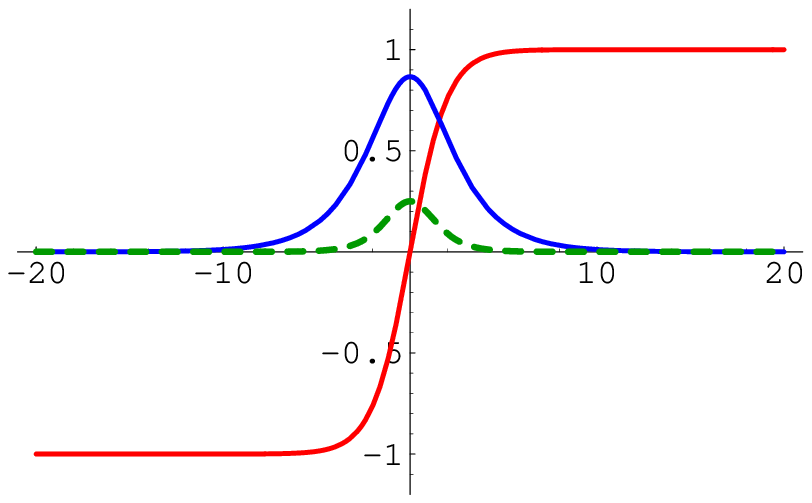}\hspace{1cm}
\includegraphics[height=2.6cm]{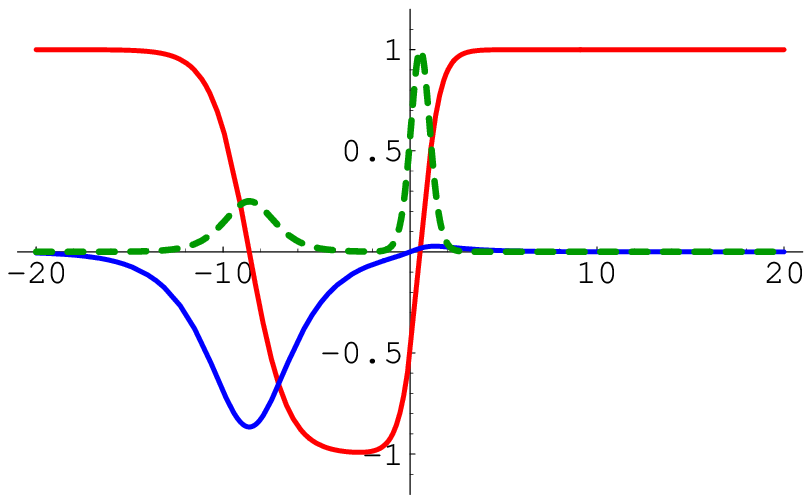}}
\caption{\small \textit{$\phi_1^K$ profile (red solid lines),
$\phi_2^K$ profile (blue solid lines), and energy density (green
dashed lines) of the TK1, TK2 and NTK2 kinks in the MSTB model.}}
\end{figure}

The discovery of this new type of solitary wave prompted several
numerical investigations by Subbaswamy and Trullinger; their
research focused on numerical explorations of the existence, energy,
and stability of non-topological kink solutions in a wide range of
$\Omega$ \cite{Trulli3,Trulli2}. These authors numerically found
that there exists a continuous family of non-topological kinks of
NTK2 type that describe closed orbits, all of them enclosed by the
ellipse ({\ref{eq:eor}) in the field space ${\mathbb R}^2$ if
$\Omega^2 \in (0,1)$. Moreover, they numerically checked an
astonishing but simple relationship between kink energies: the
energy of any of the non-topological kinks is the sum of the
energies of the two topological kinks:
\begin{equation}
E[{\rm NTK2}]=E[{\rm TK1}]+E[{\rm TK2}] \qquad \qquad
.\label{eq:kmsr}
\end{equation}

This identity, termed \textit{the kink mass sum rule}, was regarded
as a mysterious property of the model (consider the kink energy
densities in Figure 1). In 1984, Magyari and Thomas \cite{Magy1}
noticed a very important feature of the model: by finding a second
constant of motion, in involution with the mechanical energy, these
authors showed that the two-dimensional analogous mechanical system
is completely integrable. In fact, the analogous mechanical model is
a particular case of the Garnier integrable system, see \cite{Gar}.
Having two isolated unstable points -the minima of $U$ are the
maxima of $V$- of maximum potential energy, the heteroclinic
trajectories of finite mechanical action joining those points are
precisely the topological kinks. Homoclinic trajectories, however,
starting from and ending at the same maximum correspond to the
non-topological kinks. The intelligence of this fact plus the
existence of two invariants in involution provided an strong hint
about the origin of the kink mass sum rule as being due to
continuous degeneracy in the mechanical action of all the separatrix
trajectories sharing the critical values of the two constants of
motion.

In 1985, Ito \cite{Ito1} showed that the analogous mechanical system
is not only completely integrable but Hamilton-Jacobi separable by
using elliptic coordinates. The immediate application by Ito of the
Hamilton-Jacobi principle allowed him to describe the whole kink
orbit and kink profile varieties analytically by means of
complicated expressions between elliptic variables. In particular,
he verified the sum rule by analytically checking that all the
closed trajectories have the same mechanical action.

Although to the best of our knowledge nobody has been able to invert
Ito's implicit expressions, we recently succeed in doing this by
performing a shrewd reshuffling of the kink orbit equations, to
find:
\begin{eqnarray}
\phi_1^{\rm
NTK2}(\bar{x};\gamma_1)&=&\frac{(\Omega-1)(1+e^{2(1+\Omega)(\bar{x}+\Omega
\gamma_1)})+
(\Omega+1)(e^{2\Omega(\bar{x}+\gamma_1)}+e^{2(\bar{x}+\gamma_1
\Omega^2)})}{(\Omega-1)(1+e^{2(1+\Omega)(\bar{x}+\Omega\gamma_1)})-
(\Omega+1)(e^{2\Omega(\bar{x}+\gamma_1)}+e^{2(\bar{x}+\gamma_1
\Omega^2)})} \nonumber \\
\phi_2^{\rm NTK2}(\bar{x};\gamma_1)&=&\frac{2(\Omega^2-1)
e^{\Omega(\bar{x}+\gamma_1)}(e^{2(\bar{x}+\gamma_1
\Omega^2)}-1)}{(\Omega-1)(1+e^{2(1+\Omega)(\bar{x}+\Omega\gamma_1)})-
(\Omega+1)(e^{2\Omega(\bar{x}+\gamma_1)}+e^{2(\bar{x}+\gamma_1
\Omega^2)})} \qquad.
\label{eq:mstbntk}
\end{eqnarray}
This one-parametric family of solitary waves, depending on the
real integration constant $\gamma_1 \in \mathbb{R}$, is the whole
manifold of two-component non-topological kinks of the
model{\footnote{In fact, there are other two parameters, $\gamma_2
\in \mathbb{R}$ and $v\in (0,1)$, which are fixed by setting the
kink center of mass system\, .}}.

Next, Ito and Tasaki \cite{Tasaki} addressed the issue of the kink
stability of the kink solutions from a very interesting point of
view: applying the Morse index theorem to the kink orbits they
showed that the unique stable kink in the $\Omega^2<1$ range is the
TK2 kink. These authors realized that all the other kink orbits
starting from one of the vacuum points pass through one of the two
foci $\phi^{f_\pm}=(\phi_1^{f_\pm}=\pm\Omega,\phi_2^{f_\pm}=0)$ of
the ellipse (\ref{eq:eor}); these points are thus respectively
conjugate to one of the two unstable points
$\phi_\pm=(\phi^1_\pm=\pm 1,\phi^2_\pm=0)$ because each kink orbit
starting from $\phi_-$ (respectively $\phi_+$) passes through
$\phi^{f_+}$ (respectively $\phi^{f_-}$). In Reference \cite{J0}
Guilarte extended this idea to embedding the kink stability problem
in the degenerate Morse theory of the configuration space, bearing
in mind that, in this framework, the topology of an infinite
dimensional manifold -the configuration space of the MSTB model- is
determined from the critical points -the kinks- of a functional -the
energy of the MSTB model-. A full analysis of this approach was
achieved in Reference \cite{J1}, where the Morse's Lemma, governing
the small fluctuations around the kink solutions, suggests how to
deal with quantum corrections. This goal was fully attained from
this point of view in Reference \cite{ AMWJ}. To finish this
subsection about the history of the MSTB model, let us mention that
in \cite{J2} the decay amplitude from the TK1 to the TK2 kink was
computed in the quantum MSTB model by using the Witten-Smale version
of Morse theory.

Two properties of the kink variety of the MSTB model induced its
early perception as singular and unique: a) The existence of several
kinds of kink: the TK1 kink, the TK2 kink and the NTK2$(\gamma_1)$
kink family. b) The relationship between its energies (the kink mass
sum rule).

\subsection{The discovery of new models of the MSTB type}

The critical feature of the MSTB model allowing this perception to
be overcome is the fact that its analogous mechanical system is an
integrable system of Liouville type. In \cite{Aai2} Alonso,
Gonzalez, and Guilarte proposed and studied two models of the MSTB
kind with different potential energy densities, chosen in such a way
that their analogous mechanical system would be of Type I in the
first model and Type III Liouville integrable system in the second
case. Together with a selection of parameters ensuring the existence
of a discrete set of maxima for the mechanical potential -five in
the first, four in the second model- this work revealed for the
first time that the MSTB is by no means unique but, rather, a member
of a broad family of models. The configuration space is formed by
$25$ (in the first case) and $16$ (in the second model)
topologically disconnected
sectors{\footnote{In general, there are $N+2\cdot\,\left(\begin{array}{c} N \\
2 \end{array}\right)=N^2$ topologically disconnected sectors in
models of this kind with $N$ zeroes of the potential energy density
\quad . }} and it is expected that the kink varieties in these cases
will be richer than in the MSTB system. This is indeed the case, and
several families of stable and unstable topological kinks were
discovered with various new kink mass sum rules arising between the
different kink types.

Model B, the second model in \cite{Aai2}, is Liouville Type III,
i.e., its analogous mechanical system is a two-dimensional
Hamilton-Jacobi separable system in parabolic coordinates.
Remarkably, this system arises for a particular value of the
coupling constant of a model that is important in supersymmetric
field theory. The dimensionally reduced Wess-Zumino model discussed
by Bazeia et al. in \cite{B3,B5} is not in general of Liouville
Type. Only for two special values of the coupling constant does the
analogous mechanical system fall into this class: one of the values
leads to a mechanical system of Type III already mentioned, and the
other one gives a Liouville system of Type IV, Hamilton-Jacobi
separable in Cartesian coordinates. The important point is that,
given its supersymmetric origin, a solution of the Hamilton-Jacobi
equation (not complete) is always known and trajectories associated
to this solution (the superpotential) can be found. This strategy
was used in \cite{Mar1} to find a whole family of topological kinks
in the BNRT (acronym from \cite{B5}, in analogy with the MSTB
acronym from \cite{Monto1} and \cite{Trulli1}) model. The difference
with model B, a particular case, is the HJ-separability that allows
one to know a {\underline{complete}} solution of the Hamilton-Jacobi
equation and, therefore, knowledge of {\underline{all}} the
trajectories. The point is that for models with analogous mechanical
system of Liouville Type, complete integrability permits an
exhaustive knowledge of all the kink orbits, which, in turn,
provides a full understanding of the solitary wave manifold and kink
mass sum rules between them. Recently, this analysis has been
performed in \cite{AJ} for some Type II Liouville systems; the field
theoretical models analyzed have analogous mechanical systems HJ
separable in polar coordinates.

A general pattern emerges from the study of the manifold of kink
solutions in these models. There are two kinds of solitary waves:
basic kinks with energy density localized at one point in ${\mathbb
R}$ and families of composite kinks, non-linear superpositions of
several basic kinks, such that their energy densities are localized
at several points in ${\mathbb R}$. The distance between the basic
lumps of energy is determined by the parameter fixing the
corresponding kink orbit. The situation is perfectly illustrated in
Figure 1. By using a solid line in the three graphics we represent
the profile of the three kinds of solution in the MSTB model, i.e,
the TK1, the TK2 and a member of the NTK2 family of kinks
respectively. We also depict the energy density of these solutions
by using a dashed line. We notice that the energy density of the TK1
and TK2 kinks are localized around one point while the energy
density for the NTK2 kink is localized around two points. The NTK2
is composed of the other two kinks. The parameter $\gamma_1$ in
(\ref{eq:mstbntk}) fixes the distance between the TK1 and TK2 kinks.
In References \cite{Mar2,Modeloa} a description of the evolution of
this distance in time has been worked out for slow speeds. Although
we are dealing with non-linear systems, the adiabatic evolution of
composite kinks can be dealt with as geodesic motion in the kink
parameter space with respect to the metric induced by the field
kinetic energy density, thus providing a good description of kink
scattering at low energies. From this point of view, kink mass sum
rules become natural: these rules are forced by energy conservation
when radiative processes are absent, i.e., the energy of a composite
kink is the sum of the energies of the component basic kinks.

One might wonder at this point if similar structures arise in
systems with more scalar fields. In References \cite{Aai3,Aai5} a
positive answer addressing field theoretical systems with three
scalar fields was found. The analysis is even more difficult,
because three-dimensional Hamilton-Jacobi separable systems not
abound. Now, the composite kinks are formed by three basic kinks.
Stability issues are also more difficult to deal with, although in
\cite{Aai4} a thorough study of kink stability was achieved in the
generalization of the MSTB model to three fields. The adiabatic
dynamics of three slow moving kinks has also been recently unveiled
in \cite{AJ1}.

\subsection{Generalized MSTB Models}

The work in the papers referred  to above shows that the MSTB model
is not a unique or special model. Several models are known to share
essential characteristics with the MSTB model, in such a way that
the kink variety of each model and its properties can be identified
by a parallel analysis. Our goal in this paper is to study the
general structure of this type of models, which we shall call
generalized MSTB models. All these systems share the same critical
properties: their analogous mechanical system is a Type I Liouville
system having a discrete set of maxima in the mechanical potential
energy; these MSTB-type models are distinguished by the number of
zeroes -the maxima- of the non-positive mechanical potential energy.
The kink variety can always be identified as the manifold of
separatrix trajectories between the bounded and unbounded motion of
the integrable mechanical system of Liouville Type I. Remarkably,
the structure of the kink variety can be unveiled in all of these
models without solving any OD equation. Moreover, the kink mass sum
rules can also be guessed simply from the analytical form of the
potential density energy.

This paper is organized as follows: In Section \S. 2 we shall define
generalized MSTB models as field theoretical two scalar field models
with analogous mechanical system of Liouville Type I. These
two-dimensional mechanical systems are Hamilton-Jacobi separable in
elliptic coordinates. Thus, in this Section, we shall translate the
generalized MSTB models into elliptic coordinates, mapping the
internal field space ${\mathbb R}^2$ to the infinite strip ${\mathbb
E}^2\equiv{\mathbb R}^+\times {\mathbb
I}\equiv\,\,[\Omega,+\infty)\times[-\Omega,\Omega]$, where
$\Omega\in (0,1)$ is a real number between $0$ and $1$. In Section
\S. 3 we shall study the kink manifold that arises in this kind of
model. In order to do this, the distribution of the zeroes of the
potential energy density is crucial. In generalized MSTB models
these zeroes are located at the nodes of a grid or reticulum in the
\lq\lq elliptic strip" ${\mathbb E}^2$. The distribution of zeroes
of the potential determines the classification of generalized MSTB
models in four types. Generically, the reticulum is formed by gluing
several cells where the kinks are confined. The zeroes of the
potential energy density live at the vertices of the cells and/or at
the foci of the elliptic and hyperbolic curves that set the
curvilinear coordinate system. There are three types of cells
distinguished by the nature of the vertices. The type of cell
determines the behavior of the kink family confined in the cell. In
general, there are basic and isolated kinks living in the bars of
the grid surrounding the cell and families of kinks living inside.
Each member of one of these families is a non-linear superposition
of basic kinks: these are composite kinks and have their energy
density localized around several points, each of these component
lumps having the same energy density as a basic kink. In this
Section, we shall also explore new and generalized kink mass sum
rules and we shall investigate the relationship between the
integrability of the analogous mechanical systems and the existence
of superpotentials. In Section \S. 4 we shall describe the kink
variety of a specific model. The MSTB model \cite{Monto1} involves a
quartic algebraic potential energy density, whereas the potential
energy density of model A in \cite{Modeloa} is a polynomial in the
fields of sixth degree. In order to deal with completely new
generalized MSTB models, we shall analyze a model characterized by a
eighth degree polynomial in the two fields as the potential energy
density. The polynomial depends on two coupling constants in such a
way that a process of bifurcation arises: the number of elements in
${\cal M}$, static homogeneous solutions of the field theoretical
model or the fixed points of the Newton equations of the analogous
mechanical system, depends on the values of the coupling constants.
There are three regimes where we shall study the kinks  and also the
kink mass sum rules. Finally, the summary of the main results and
the outlook on future research in similar models are offered in
Section \S . 5.

\section{The structure of generalized MSTB models}

\subsection{The configuration space}

We shall study field theoretical models of two real scalar fields
defined in a (1+1) Minkowskian space-time, whose dynamics is
governed by the action functional:
\begin{equation}
S[\vec{\phi}]=\int dx dt \left[ \frac{1}{2} \sum_{a=1}^2
\partial^\mu \phi_a \partial_\mu \phi_a -U(\phi_1,\phi_2)\right]
\qquad . \label{eq:action}
\end{equation}
We use Einstein sum convention for Greek scripts and choose the
metric tensor to be:  $g_{00}=-g_{11}=1$, $g_{01}=g_{10}=0$. By
$\vec{\phi}=\sum_{a=1}^2\phi_a \vec{e}_a$, $a=1,2$, we denote the
scalar fields:
\[
\vec{\phi}(x_0,x_1)=\phi_1(x_0,x_1)\vec{e}_1+\phi_2(x_0,x_1)\vec{e}_2:
\mathbb{R}^{1,1} \rightarrow \mathbb{R}^2
\]
The vectors $\vec{e}$ form an orthonormal basis in ${\mathbb R}^2$
and $U(\phi_1,\phi_2)$ is the (non-negative) potential energy
density, which vanishes in a discrete set of points, ${\cal M}$. For
the sake of simplicity we assume that the magnitudes involved in
(\ref{eq:action}) are dimensionless.

A \textit{point} in the configuration space of the system is a
configuration of the field of finite energy; i.e., a picture of the
field at a fixed time such that the energy $E$, the integral over
the line of the energy density,
\begin{equation}
E [\vec{\phi}] =\int_{-\infty}^\infty dx \, {\cal E}[\vec{\phi}(x)]
\hspace{1cm} ; \hspace{1cm} {\cal E}[\vec{\phi}(x)] = \frac{1}{2}
\left( \frac{d\phi_1}{dx} \right)^2 + \frac{1}{2} \left(
\frac{d\phi_2}{dx} \right)^2 + U(\phi_1,\phi_2) \label{eq:denener}
\end{equation}
is finite. Therefore, the configuration space is the set of
continuous maps from $\mathbb{R}$ to $\mathbb{R}^2$ of finite
energy:
\[
{\cal C}=\{ \vec{\phi}(x) \in {\rm Maps}(\mathbb{R},\mathbb{R}^2) \,
/ \, E[\phi]< \infty \}
\]
In order to belong to ${\cal C}$, each configuration must comply
with the asymptotic conditions
\begin{equation}
\lim_{x\rightarrow \pm \infty} \vec{\phi}(x) \in {\cal M}
\hspace{2cm}, \hspace{2cm} \lim_{x\rightarrow \pm \infty}
\frac{d\vec{\phi}}{dx}=0 \label{eq:asymptotic}
\end{equation}
where ${\cal M}$ is the set of zeroes of the potential energy
density $U(\phi)$,
\[
{\cal M}=\{ \vec{\phi}^{(I)} \in \mathbb{R}^2 \subset {\rm
Maps}(\mathbb{R},\mathbb{R}^2) \, / \, U[\vec{\phi}^{(I)}]=0 \, \, ,
\, \, I=1,2, \cdots , N \} \qquad .
\]
We shall also choose $U(\vec{\phi})$ non-negative and such that
${\cal M}$ is also the set of absolute minima of $U$, i.e., the set
of the static homogeneous solutions of the model because the
Euler-Lagrange field equations of (\ref{eq:action})
\begin{equation}
\frac{\partial^2 \phi_a}{\partial t^2}-\frac{\partial^2
\phi_a}{\partial x^2}+\frac{\partial U}{\partial \phi_a}=0
\hspace{1cm} , \qquad a=1,2 \label{eq:eulerlagrange}
\end{equation}
are satisfied by the elements of ${\cal M}$. Moreover, bearing in
mind that time evolution is continuous -a homotopy transformation-,
if ${\cal M}$ is a discrete set the configuration space is the union
of disconnected topologically sectors ${\cal C}^{IJ}$:
\[
{\cal C}= \bigcup_{I,J \in {\cal M}} {\cal C}^{IJ} \qquad .
\]
The configurations in ${\cal C}^{IJ}$ are characterized as those
which respectively tend to the elements $\vec{\phi}^{(I)}$ and
$\vec{\phi}^{(J)}$ of ${\cal M}$ at minus and plus spatial infinity.

According to Rajaraman \cite{Rj}: \textit{A kink is a non-singular
solution of non-linear field equations (\ref{eq:eulerlagrange})
whose energy density, as well as being localized, has space-time
dependence of the form: $\varepsilon(t,x)=\varepsilon(x-\rm{v}
t)$, where {\rm v} is some velocity vector}. Because we are
dealing with a Lorentz invariant system, solutions with a temporal
dependence like this are obtained from static solutions by means
of a Lorentz velocity transformation: $L({\rm v})\phi(x)=
\phi^L(t,x)=\phi\left( \frac{x-{\rm v} t}{\sqrt{1-{\rm v}^2}}
\right)$. Bearing this in mind the PDE (\ref{eq:eulerlagrange})
can be restricted to the ODE
\begin{equation}
\frac{d^2 \phi_a}{dx^2}= \frac{\partial U}{\partial \phi_a}
\hspace{0.5cm} , \hspace{0.5cm} a=1,2 \hspace{0.8cm} \equiv
\hspace{0.8cm} \frac{d^2\vec{\phi}}{dx^2}=\vec{\nabla}U
\label{eq:ordinary} \qquad .
\end{equation}
in such a way that the search for some kinks or solitary waves is
tantamount to studying an analogous mechanical system, where we
interpret the static scalar field
$\vec{\phi}(x)=\phi_1(x)\vec{e}_1+\phi_2(x)\vec{e}_2$ as the
\textit{particle position vector}; $x$ as the \textit{particle
time}, and $U=-V$ as the \textit{particle potential}. Moreover, the
action and the Lagrangian of the mechanical system respectively play
the r$\hat{\rm o}$les of the energy functional and the energy
density in the field theory, see (\ref{eq:denener}). In sum,
solitary waves or kinks are derived from stationary points of the
functional energy (\ref{eq:denener}) that comply with the asymptotic
conditions (\ref{eq:asymptotic}). Therefore, kinks asymptotically
connect two elements of ${\cal M}$ or, homogeneous solutions of
(\ref{eq:ordinary}).

\subsection{Algebraic Type I Liouville potentials}

The potential energy density of the MSTB model is a positive quartic
algebraic expression in the fields, see (\ref{eq:mstbpot}). In this
sub-Section we shall characterize all the algebraic (polynomial)
potential energy densities that have Type I Liouville systems as the
analogous mechanical model. Because the polynomials will in general
be of higher order than four, the cardinal of ${\cal M}$ may be
greater than in the MSTB model, promising richer kink varieties.

{\bf Proposition 1:} {\it All the members in the ($q+2$)-parametric
family of models with algebraic energy density of the form:
\begin{equation}
U_q(\phi_1,\phi_2)=\sum_{n=0}^q b_n \prod_{j=1}^n \left(
\phi_1^2+\phi_2^2-2 \Omega \phi_1 \cos \frac{j \pi}{n+1}+\Omega^2
\right) \label{eq:genmstb}\qquad ,
\end{equation}
a polynomial in the $\phi_1$ and $\phi_2$ fields of degree $2 q$,
have analogous mechanical systems that are of Liouville Type I,
\cite{Pere}, and, therefore, these mechanical systems are
Hamilton-Jacobi separable in a system of elliptic coordinates.}

The proposition will be proved later in this sub-Section. In formula
(\ref{eq:genmstb}) $b_n$, $n=1,\dots,q$ and $\Omega$ are the
parameters or coupling constants distinguishing between the members
of the family. In order to support kinks/solitary waves, or
non-trivial separatrix trajectories in the analogous mechanical
system, the choice of $b_n$ parameters must be restricted in such a
way that $U(\phi_1,\phi_2)$ must be bounded below by a discrete
degenerate set of constant field configurations that are the
absolute minima of $U(\phi_1,\phi_2)$. Also, some transformations,
e.g. scaling of the space-time $\tilde{x}^\mu=\lambda x^\mu$,
scaling and rotations in internal space,
$\tilde{\phi}_a=\gamma\phi_a$, $\tilde{\phi}_1={\rm
cos}\alpha\phi_1-{\rm sin}\alpha\phi_2$, $\tilde{\phi}_2={\rm
sin}\alpha\phi_1+{\rm cos}\alpha\phi_2$ lead to equivalent models,
allowing us to restrict the values of three parameters.

Arguing as Magyari and Thomas for the MSTB model in \cite{Magy1}, we
notice that the analogous mechanical system for the potential energy
density (\ref{eq:genmstb}) is completely integrable: the static
field equations (\ref{eq:ordinary}) admit two independent invariants
in involution:
\begin{eqnarray}
I_1&=&
 \frac{1}{2} \left( \frac{d \phi_1}{d x} \right)^2 +
 \frac{1}{2} \left( \frac{d \phi_2}{d x} \right)^2 -
\sum_{n=0}^{q} b_n \prod_{j=1}^{n} \left( \phi_1^2+\phi_2^2-2
\Omega \phi_1 \cos \frac{j \pi}{n+1}+\Omega^2 \right) \qquad
,\label{eq:i1} \\ I_2 & = & \frac{1}{2} \left[ \left( \phi_1
\frac{d \phi_2}{d x} - \phi_2 \frac{d \phi_1}{d x} \right)^2 -
\Omega^2 \left( \frac{d \phi_2}{d x} \right)^2 \right]+
\sum_{n=0}^{q} b_n \left[- \phi_1^2 \prod_{j=1}^{n-1} \left(
\phi_1^2+\phi_2^2- \right. \right. \nonumber\\ & & \left. \left. -
2 \Omega \phi_1 \cos \frac{i \pi}{n}+\Omega^2 \right)+\Omega^2
\prod_{j=1}^{n} \left( \phi_1^2+\phi_2^2-2 \Omega \phi_1 \cos
\frac{j \pi}{n+1}+\Omega^2 \right) \right] \label{eq:i2}
\end{eqnarray}
The first invariant is the energy of the mechanical system and the
second one is also quadratic in the velocities, a certain
deformation of a first-integral of Runge-Lenz type. Therefore, the
field theoretical model is associated to an analogous mechanical
system of St\"ackel type. By choosing an appropriate system of
coordinates dynamical systems of this kind are always
Hamilton-Jacobi separable. The form of the second invariant dictates
that elliptic coordinates will produce a separable Hamilton-Jacobi
equation (Type I Liouville systems) and all the solutions can be
found by quadratures. In the MSTB model, this property was used by
Ito \cite{Ito1}) to find all the solitary wave solutions.

To study algebraic Type I Liouville systems we define Euler elliptic
coordinates by means of the map:
\[
\begin{array}{cccc}
\xi_\Omega^{\pm}: &  {\mathbb E}^2 \equiv  [-\Omega,\Omega]\times
[\Omega,\infty) &\longrightarrow & {\mathbb R}^2 / {\mathbb Z}_2
 \\
 & (v,u) &\longrightarrow &(\phi_1, \phi_2) \end{array}
\]
{\small\[
 \xi_\Omega^\pm (u)=\frac{\sqrt{(\phi_1-\Omega)^2+\phi_2^2}+
 \sqrt{(\phi_1+\Omega)^2+\phi_2^2}}{2} \qquad , \qquad  \xi_\Omega^\pm
 (v)=\frac{\sqrt{(\phi_1+\Omega)^2+\phi_2^2}-
 \sqrt{(\phi_1-\Omega)^2+\phi_2^2}}{2}\qquad ,
\]}which is a diffeomorphism between the elliptic open strip
$(-\Omega,\Omega)\times (\Omega,+\infty)$ and the Cartesian upper
half-plane $\phi_2>0$. Explicitly, the inverse map from ${\mathbb
R}^2 / {\mathbb Z}_2$ to ${\mathbb E}^2$, is:
\begin{equation}
\begin{array}{lcr}
{\xi_\Omega^{\pm}}^{*}(\phi_1)=\frac{1}{\Omega} u v & \hspace{0.5cm}
, \hspace{0.5cm} & {\xi_\Omega^{\pm}}^{*}(\phi_2)= \pm
\frac{1}{\Omega} \sqrt{(u^2-\Omega^2)(\Omega^2-v^2)}
\end{array}
\label{eq:defelip}
\end{equation}
such that the coordinating curves are the hyperbolae and ellipses
\[
\frac{\phi_1^2}{v^2}-\frac{\phi_2^2}{\Omega^2-v^2}=1 \qquad \quad ,
\qquad \quad \frac{\phi_1^2}{u^2}+\frac{\phi_2^2}{u^2-\Omega^2}=1
\]
whose foci are the points: $F_\pm=(\pm \Omega, 0)$. It is necessary
to use two copies of the elliptic strip ${\mathbb E}^2$ to cover the
whole Cartesian internal plane: $\xi_{\Omega}^+$ maps ${\mathbb
E}^2$ into the $\phi_2\geq 0$ half-plane whereas $\xi_\Omega^-$ maps
the other copy of ${\mathbb E}^2$ into the $\phi_2\leq 0$
half-plane. Translation of information from ${\mathbb E}^2$ to the
Cartesian ${\mathbb R}^2$ internal plane must take this fact into
account by smoothly gluing data across the border between the two
copies. In particular, the straight lines $u=\Omega$ and $v=\pm
\Omega$ in ${\mathbb E}^2$ determine the axis $\phi_1$ in ${\mathbb
R}^2$; $u=\Omega$ is the $\phi_1\in [-\Omega,\Omega],\phi_2=0$
interval and $v=\pm \Omega$ are respectively the $\phi_1\in
[-\infty,-\Omega]$ (- sign) and $[\Omega,\infty],\phi_2=0$ (+ sign)
half-lines, all together forming the abscissa axis in ${\mathbb
R}^2$. Note that the parameter $\Omega$ setting the foci of the
coordinating curves is chosen to be the coupling constant $\Omega$
in (\ref{eq:genmstb}).

Applying the map (\ref{eq:defelip}) to the potential energy density
(\ref{eq:genmstb}) we obtain:
\begin{eqnarray}
\xi_\Omega^{*} U(\phi_1,\phi_2)&=&\frac{1}{u^2-v^2} \sum_{n=0}^{q}
 b_n \left( u^{2n+2}-v^{2n+2} \right) =\frac{1}{u^2-v^2}
\left( f(u)+g(v) \right) \label{eq:tres} \\
& & f :  [\Omega, \infty)  \rightarrow  \mathbb{R} \hspace{0.4cm} ;
\hspace{0.4cm} u \mapsto f(u)=\displaystyle\sum_{n=0}^q b_n u^{2n+2}
\\ & & g :  [-\Omega, \Omega]  \rightarrow  \mathbb{R} \hspace{0.4cm}
; \hspace{0.4cm} v \mapsto g(v)=-\displaystyle\sum_{n=0}^q b_n
v^{2n+2} \label{eq:fyg} \qquad .
\end{eqnarray}
In elliptic coordinates the potential energy density (\ref{eq:tres})
is the product of the $\frac{1}{u^2-v^2}$ factor times the sum of
two polynomials, which depend only on one of the elliptic variables.
$\frac{1}{u^2-v^2}$ is in all places positive in the strip ${\mathbb
E}^2$ except at the foci $F_\pm \equiv (u,v)=(\pm \Omega, \Omega)$
(the points $(\phi_1,\phi_2)=(\pm \Omega, 0)$ in ${\mathbb R}^2$),
where the denominator vanishes. Despite this, the potential energy
density is not singular at the foci $F$ because the tendency to zero
is at least as strong in the numerator of (\ref{eq:tres}) as in the
denominator. This can be seen in the factorization:
\begin{equation}
u^{2n+2}-v^{2n+2}=(u^2-v^2)(u^2-2 u v \cos
\frac{\pi}{n+1}+v^2)(u^2-2 u v \cos \frac{2 \pi}{n+1}+v^2)\cdot
\dots \cdot (u^2-2 u v \cos \frac{n \pi}{n+1}+v^2) \,\, \,
,\label{eq:epol}
\end{equation}
which also explains the symmetric form of the functions $f$ and $g$,
i.e., $g(z)=-f(z)$. The inverse image in Cartesian coordinates of
(\ref{eq:tres}) provides the strange form of the potential energy
density (\ref{eq:genmstb}) $U_q(\phi_1,\phi_2)$.

The main point is that the static energy functional
(\ref{eq:denener}) in elliptic variables reads:
\begin{equation}
\xi_\Omega^{*} E[\phi_1,\phi_2]=\int dx \left\{\frac{1}{2}
\frac{u^2-v^2}{u^2-\Omega^2} \left( \frac{du}{dx}
\right)^2+\frac{1}{2} \frac{u^2-v^2}{\Omega^2-v^2} \left(
\frac{dv}{dx} \right)^2  + \frac{f(u)+g(v)}{u^2-v^2}   \right\}
\label{eq:cuatro} \qquad .
\end{equation}
The functional $\xi_\Omega^{*} E[\phi_1,\phi_2]:{\rm Maps}({\mathbb
R},{\mathbb E}^2)\longrightarrow {\mathbb R}$ can be interpreted as
the functional action for a particle moving in ${\mathbb E}^2$ under
a mechanical potential energy $V[u(x),v(x)]=-\xi_\Omega^{*}
U[\phi_1(x),\phi_2(x)]$ if the particle time is $x$. In this
conceptual framework, the change to elliptic coordinates induces a
flat, but non-Euclidean, metric:
\[
g^{uu}(u,v)=\frac{u^2-v^2}{u^2-\Omega^2} \qquad , \qquad
g^{vv}(u,v)=\frac{u^2-v^2}{\Omega^2-v^2} \qquad , \qquad
g^{uv}(u,v)=0=g^{vu}(u,v) \qquad .
\]
In any case, it is well known that mechanical systems with dynamics
ruled by an action such as (\ref{eq:cuatro}) are Type I Liouville
systems, see \cite{Pere}. Therefore, there are two first integrals:
\begin{equation}
\xi_\Omega^{*}I_1=\frac{1}{2}  \frac{u^2-v^2}{u^2-\Omega^2} \left(
\frac{du}{dx} \right)^2+\frac{1}{2} \frac{u^2-v^2}{\Omega^2-v^2}
\left( \frac{dv}{dx} \right)^2
 - \frac{f(u)+g(v)}{u^2-v^2}
\label{eq:cinco}
\end{equation}
\begin{equation}
{\displaystyle \xi_\Omega^{*}I_2=\frac{1}{2} (v^2-u^2) \left[
\frac{\Omega^2-v^2}{u^2-\Omega^2} \left( \frac{du}{dx} \right)^2-
\frac{u^2-\Omega^2} {\Omega^2-v^2} \left( \frac{dv}{dx} \right)^2
\right]}  {\displaystyle - \frac{(u^2-\Omega^2) g(v)}{u^2-v^2}
+\frac{(\Omega^2-v^2) f(u)}{u^2-v^2} }, \label{eq:seis}
\end{equation}
which allow us to solve the motion equations. Back in Cartesian
coordinates, (\ref{eq:cinco}) and (\ref{eq:seis}) become
respectively (\ref{eq:i1}) and (\ref{eq:i2}), fully proving the
Proposition.

\subsection{Generalized MSTB models}

Generalized MSTB models are a subset of those models having
algebraic Type I Liouville systems as analogous mechanical system.
In order to characterize with precision this kind of model we recall
how the MSTB model arises between the $q=2$ algebraic Type I
Liouville systems:
\begin{eqnarray*}
U_{q=2}(\phi_1,\phi_2)&=&b_0+b_1\left( \phi_1^2+\phi_2^2-2 \Omega
\phi_1 \cos \frac{\pi}{2}+\Omega^2 \right)\\&+&b_2\left(
\phi_1^2+\phi_2^2-2 \Omega \phi_1 \cos \frac{ \pi}{3}+\Omega^2
\right)\left( \phi_1^2+\phi_2^2-2 \Omega \phi_1 \cos \frac{2
\pi}{3}+\Omega^2 \right) \\&=&a_1 (\phi_1^2+\phi_2^2+a_2)^2+a_3
\phi_1^2+a_4 \phi_2^2 \qquad .
\end{eqnarray*}
The criterion is that, up to equivalences, $U_{q=2}$ must be
non-negative, having a discrete set of zeroes with more than one
element; these properties give a hint of the existence of solitary
waves. Demanding $U_{q=2}\geq 0$ with a discrete set of minima
requires $a_1>0$, $a_3\geq 0$ and $a_4\geq 0$. Also $a_1$ can be
fixed by re-scaling the space-time, $a_2$ by re-scalings of the
fields and $a_3$ by rotations of the fields. In the MSTB model, we
choose $a_1={1\over 2}$, $a_2=-1$, $a_3=0$ such that only a positive
parameter, $a_4$, is left and there are two zeroes of $U_{q=2}$.

Another model belonging to this kinship has been dealt with in the
Literature.  The $q=3$ family of potential energy densities is:
\begin{eqnarray*}
&&U_{q=3}(\phi_1,\phi_2)=b_0+b_1\left( \phi_1^2+\phi_2^2-2 \Omega
\phi_1 \cos \frac{\pi}{2}+\Omega^2 \right)\\&+&b_2\left(
\phi_1^2+\phi_2^2-2 \Omega \phi_1 \cos \frac{ \pi}{3}+\Omega^2
\right)\left( \phi_1^2+\phi_2^2-2 \Omega \phi_1 \cos \frac{2
\pi}{3}+\Omega^2 \right) \\&+&b_3\left( \phi_1^2+\phi_2^2-2 \Omega
\phi_1 \cos \frac{ \pi}{4}+\Omega^2 \right)\left(
\phi_1^2+\phi_2^2-2 \Omega \phi_1 \cos \frac{2 \pi}{4}+\Omega^2
\right)\left( \phi_1^2+\phi_2^2-2 \Omega \phi_1 \cos \frac{3
\pi}{4}+\Omega^2 \right)\\&=& a_1
(\phi_1^2+\phi_2^2)^2(\phi_1^2+\phi_2^2+a_2)+(a_3
\phi_2^2+a_4)(\phi_1^2+\phi_2^2)+ \frac{a_3 (2 a_1 a_2+a_3)}{4 a_1}
\phi_2^2+a_5 \qquad .
\end{eqnarray*}
$U_{q=3}$ is a five-parametric algebraic polynomial in $\phi_1$ and
$\phi_2$ of sixth order that can be understood as a deformation of
the Chern-Simons-Higgs potential energy density:
\[
U_{\rm
CSH}(\phi_1,\phi_2)=(\phi_1^2+\phi_2^2)(\phi_1^2+\phi_2^2-1)^2
\qquad \quad .
\]
The A model investigated in Reference \cite{Modeloa} is a special
member of this family of $q=3$ with five zeroes of $U_{q=3}$.
Therefore, ${\cal M}$ has five elements in the A model and the
configuration space is the union of 25 disconnected topological
sectors.

The potential energy densities in the MSTB (\ref{eq:mstbpot}) and A
model \cite{Modeloa} in elliptic coordinates take the form of
formula (\ref{eq:tres}) by choosing $f(u)$ and $g(v)$ respectively
to be:
\begin{eqnarray*}
f(u)=\frac{1}{2} (u^2-1)^2 (u^2-\Omega^2) \qquad \quad &,& \qquad
\quad g(v)=\frac{1}{2} (v^2-1)^2 (\Omega^2-v^2) \\ f(u)=\frac{1}{2}
u^2 (u^2-1)^2 (u^2-\Omega^2) \qquad \quad &,& \qquad \quad
g(v)=\frac{1}{2} v^2 (v^2-1)^2 (\Omega^2-v^2) \qquad .
\end{eqnarray*}

{\bf Definition 1:} {\it Generalized MSTB models are
(1+1)-dimensional field theoretical models of two scalar fields such
that their potential energy densities have the following form in
elliptic coordinates:}
\begin{equation}
f(u)=(u^2-\Omega^2)\, u^{2\alpha_0} \prod_{j=1}^{n+m}
(u^2-\sigma_j^2)^{2\alpha_j} \hspace{1cm} , \hspace{1cm}
g(v)=(\Omega^2-v^2)\, v^{2\alpha_0} \prod_{j=1}^{n+m}
(v^2-\sigma_j^2)^{2\alpha_j} \qquad . \label{eq:deta}
\end{equation}
Here $\alpha_0,\alpha_j, n, m \in \mathbb{N}$ are natural numbers
and there are $n+m+1$ coupling constants: $\sigma_j,j=1,2, \cdots ,
n+m$, and $\Omega$. With no loss of generality, we order the
coupling constants in such a way that:
\[
0<\sigma_1<\sigma_2<\cdots<\sigma_{n-1}<\sigma_n=\Omega<\sigma_{n+1}<\cdots<\sigma_{n+m}
\qquad \quad .
\]
This choice is guided by the following criteria:
\begin{itemize}
\item The potential energy density is positive for all the values of
the coupling constants because all the terms involve even-power
factors. Notice that the factors $(u^2-\Omega^2)$ and
$(\Omega^2-v^2)$ in the functions $f(u)$ and $g(v)$ are
semi-definite positive because of the range of the elliptic
coordinates $u\in [\Omega,\infty]$ and $v\in [-\Omega,\Omega]$. The
presence of these factors allows us to comply with the condition
$f(z)=-g(z)$, preserving the positiveness of the potential energy
density.

\item The functions $f(u)$ and $g(v)$ have been also chosen in order to
maximize the cardinal of the set of minima or zeroes ${\cal M}$. For
this reason these functions involve only real roots. In this way the
models will have a richer kink variety .
\end{itemize}
Because solitary waves come from separatrix trajectories complying
with the asymptotic conditions (\ref{eq:asymptotic}), kinks will
arise when $\xi_\Omega^{*}I_1=0=\xi_\Omega^{*}I_2$. Indeed the
condition $\xi_\Omega^{*}I_1=0$ (\ref{eq:cinco}) is tantamount to
the virial theorem and the kink energy density reads:
\begin{equation}
{\cal E}[u,v]=\frac{2}{u^2-v^2} \left[f(u)+g(v) \right]=
\frac{u^2-v^2}{u^2-\Omega^2} \left( \frac{du}{dx} \right)^2+
\frac{u^2-v^2}{\Omega^2-v^2} \left( \frac{dv}{dx} \right)^2
\label{eq:vir} \qquad .
\end{equation}

\subsection{The structure of the set ${\cal M}$ of zeroes of $U(\phi_1,\phi_2)$ }

Bearing in mind that solitary wave solutions asymptotically connect
two homogeneous minima of $U(\phi_1,\phi_2)$, in this sub-Section we
shall study the distribution of elements of ${\cal M}$ in the
internal plane ${\mathbb R}^2$. We shall show that the minima are
located at the nodes of a rectangular mesh or reticulum both on the
elliptic strip and on the Cartesian plane. This grid comprises a
number of cells that depends on each specific model. Kinks are
confined inside the cells and their qualitative behavior is
determined by knowing only the type of cells where they are
constrained. Moreover, on the edges of the cells single kinks
emerge, while inside the cells families of composite kinks
(non-linear combinations of single kinks) arise.

In order to systematize the analysis of generalized MSTB models we
shall start by describing the set $\bar{{\cal M}}$ of zeroes of
$\xi_\Omega^{*} U(\phi_1,\phi_2)$ on the elliptic strip ${\mathbb
E}^2$. From (\ref{eq:tres}), one notices that the zeroes of the
potential energy density are roots of the functions $f(u)$ and
$g(v)$. We shall denote respectively by
\[
\bar{\cal M}_g=\{v \in [-\Omega,\Omega] \, / \, g(v)=0\}\hspace{1cm}
\mbox{and} \hspace{1cm} \bar{\cal M}_f=\{u \in [\Omega,\infty) \, /
\, f(u)=0\}
\]
the sets of real roots of $g(v)$ and $f(u)$. We stress the following
points about the elements of $\bar{\cal M}_g$ and $\bar{\cal M}_f$:
\begin{itemize}
\item If $\sigma_j$ is
a root of $g(v)$, $-\sigma_j$ is another root, see formula
(\ref{eq:deta}). We shall simply denote: $\sigma_{-j}=-\sigma_j$.

\item Although $f(u)$ and $g(v)$ are almost identical functions their domains are  different:
the domain of $g$ is a finite interval, ${\rm Dom}\,
g=[-\Omega,\Omega]$, whereas the domain of $f$ is the infinite
half-line, ${\rm Dom}\, f=[\Omega,\infty)$. For this reason, in
general the numbers of roots of these functions are also different.
We shall denote by $n$ and $m+1$, respectively the number of
positive roots of $g(v)$ and $f(u)$. Therefore, ${\rm card}\,
\bar{\cal M}_f=m+1$ and, either ${\rm card}\, \bar{\cal M}_g=2 n+1$
if $v=0$ is a root of $g(v)$, or, ${\rm card}\, \bar{\cal M}_g=2 n$
if $0$ is not a root. We shall show later that the pair $(n,m)$
characterizes the qualitative behavior of the kink variety in the
generalized MSTB models; models with the same pair $(n,m)$ have kink
varieties of the same type.

\item It is convenient to order together the roots $\sigma_j$ of the
two functions $g(v)$ and $f(u)$ from the lowest to the highest. Let
$J$ be the set of labels: ${\rm J}=\{-n,-(n-1),...,n-1,n\}$. The set
of roots of $g(v)$ is: $\bar{\cal M}_g=\{\sigma_j \in
[-\Omega,\Omega] \, / \, j\in {\rm J} \}$. If $g(v)$ has no zero
root $j=0 \notin {\rm J}$. Analogously, let $K$ be another set of
labels: ${\rm K}=\{n,n+1,...,n+m\}$. The set of roots of $f(u)$ is:
$\bar{\cal M}_f=\{\sigma_k \in [\Omega,\infty) \, / \, k \in {\rm
K}\}$. It is convenient to use the global set of labels ${\rm J}
\cup {\rm K}$ because  $j \in {\rm J} \cup {\rm K}=\{-n,-(n-1),...,
n,n+1,...,n+m\}$ jointly label the roots of the two functions $g(v)$
and $f(u)$ (see Figure 2).

\item It is clear that $v=\pm \Omega$ are the lowest and highest roots of $g(v)$ whereas
$u=\Omega$ is the lowest root of $f(u)$. Therefore, the (increasing)
ordering of the positive or zero roots of $\bar{\cal M}_g$ and
$\bar{\cal M}_f$ is as follows:

\centerline{$0<\sigma_1 <
\sigma_2<...<\sigma_{n-1}<\sigma_n=\Omega<
\sigma_{n+1}<...<\sigma_{n+m} \qquad \quad .$}
\noindent\begin{figure}[htb] \centerline{
\includegraphics[height=2.cm]{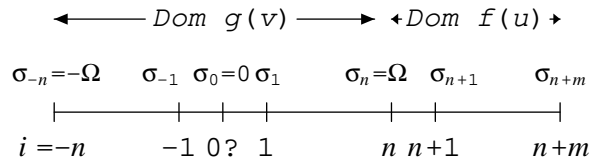}}
\caption{\small \textit{Ordering of the roots of the functions
$g(v)$ and $f(u)$.}}
\end{figure}
\end{itemize}
\noindent The set ${\cal M}$ of zeroes of $U(\phi_1,\phi_2)$ has a
straightforward connection with the zeroes of the sum of $f(u)$ and
$g(v)$, (the numerator in (\ref{eq:tres})), in ${\mathbb E}^2$:
\[
\bar{\cal M}_{f+g}=\bar{\cal M}_g \times \bar{\cal M}_f =
\displaystyle\bigcup_{(j,k) \in {\rm J} \times {\rm K}}
(v=\sigma_j,u=\sigma_k) \qquad ,
\]
or, back in ${\mathbb R}^2$:
\[
{\cal M}_{f+g}= \xi_\Omega^{\pm*} \bar{\cal M}_{f+g} =
\displaystyle\bigcup_{(j,k) \in {\rm J} \times {\rm K}}
\left(\phi_1=\frac{\sigma_j \sigma_k}{\Omega} ,\phi_2=\pm
\frac{1}{\Omega} \sqrt{(\sigma_k^2-\Omega^2)(\Omega^2-\sigma_j^2)}
\right) \qquad .
\]
The elements of ${\cal M}$ are obviously zeroes of $f(u)+g(v)$ but
the converse statement is false: ${\cal M} \subset {\cal M}_{f+g}$.
The reason for this is that the foci $F_\pm$ of the coordinating
curves belong to ${\cal M}_{f+g}$ but the potential energy density
(\ref{eq:tres}) might not vanish at these points because of the
${1\over u^2-v^2}$ factor in (\ref{eq:tres}).

\subsection{Types of generalized MSTB models}

There are two pairs of excluding  possibilities, (1) or (2), (a) or
(b), prompting a classification of the generalized MSTB models:
\begin{itemize}
\item[-] (1) $v=\pm \Omega$ are simple roots of the function
$g(v)$, or, (2) $v=\pm \Omega$ are multiple roots of the function
$g(v)$.

\item[-] (a) The function $g(v)$ has no zero roots, or, (b)
the function $g(v)$ has zero roots.
\end{itemize}

Therefore, there are four kinds of generalized MSTB models:
\begin{itemize}
\item {\it TYPE I-1a}: All the roots of $g(v)$ are non zero and $v=\sigma_{\pm
n}=\pm \Omega$ are simple roots. In this case the foci $F_\pm$ do
not belong to the set $\xi_\Omega^{\pm*}{\cal M}=\bar{\cal M}$ of
minima of the potential energy density in ${\mathbb E}^2$ and:
\[
\bar{\cal M}= \bar{\cal M}_{f+g}-\{(\pm \Omega,\Omega)\}=
\bigcup_{(j,k)\in {\rm J} \times {\rm K}}
\{(\sigma_j,\sigma_k)\}-\{(\pm
\Omega,\Omega)\}=\displaystyle\bigcup_{(j,k) \in {\rm J} \times {\rm
K}} \bar{v}_{j,k} \qquad .
\]
Note that ${\rm J}=\{-n,...,-1,1,...,n\}$, i.e., the element $j=0$
is not accounted for in the set of labels ${\rm J}$ because there is
no zero root of $g(v)$. The number of elements of the minima of the
potential energy density is straightforward in ${\mathbb E}^2$:
${\rm card}(\bar{\cal M})=2n(m+1)-2$. To give the corresponding
number in ${\mathbb R}^2$ is a tricky business. In general, to each
minima in ${\mathbb E}^2$ correspond two minima in ${\mathbb R}^2$:
\[
\xi^\pm_\Omega(\sigma_{\pm j},\sigma_k)=(\frac{\sigma_{\pm
j}\sigma_k}{\Omega},\pm{1\over\Omega}\sqrt{(\sigma_k^2-\Omega^2)(\Omega^2-\sigma_j^2)}\,)={v}_{j,k}^\pm
\quad , \quad k=n+1,n+2, \cdots , n+m
\]
This is only true in the interior of ${\mathbb E}^2$. Each minima
living at the boundary maps only in one minima in ${\mathbb R}^2$
sitting on the abscissa axis: $\xi^\pm_\Omega(\sigma_{\pm
j},\Omega)=(\sigma_{\pm j},0)$, $j=1,2,\cdots , n-1$,
$\xi^\pm_\Omega(\pm\Omega , \sigma_k)=(\pm\sigma_k,0)$,
$k=n+1,n+2,\cdots , n+m$. Therefore, the number of minima of the
potential energy density in ${\mathbb R}^2$ is: ${\rm card}( {\cal
M})=4nm+2(n-m)-2$. In Figure 3 the distribution of all these points
is shown, both in ${\mathbb E}^2$ and ${\mathbb R}^2$. The straight
lines joining the minima in the elliptic strip (hyperbolae and
ellipses in the Cartesian plane) form the reticulum associated with
this type of model.

\begin{figure}[htb]
\centerline{
\includegraphics[height=3.5cm]{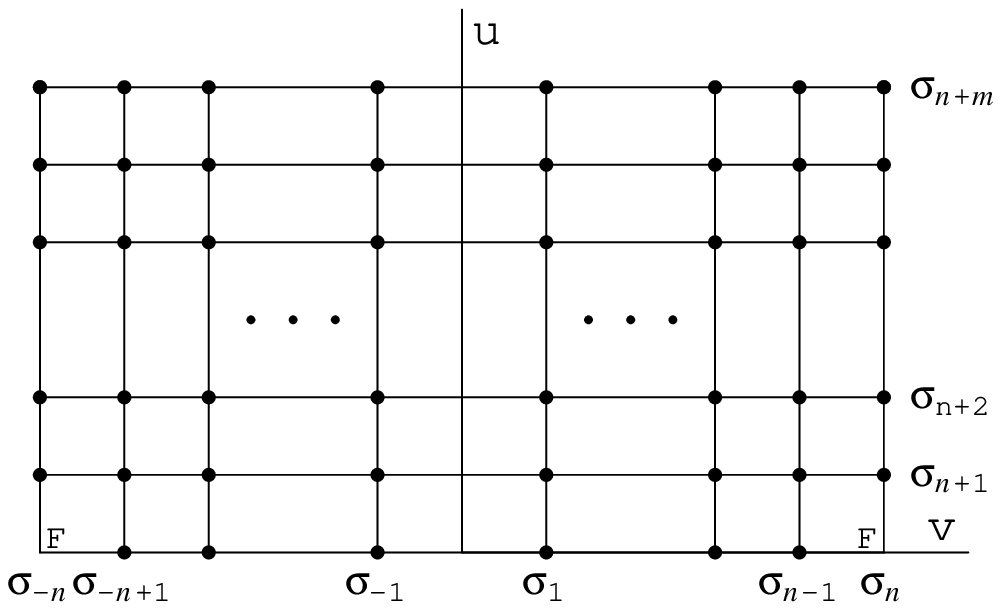}\hspace{1cm}
\includegraphics[height=3.5cm]{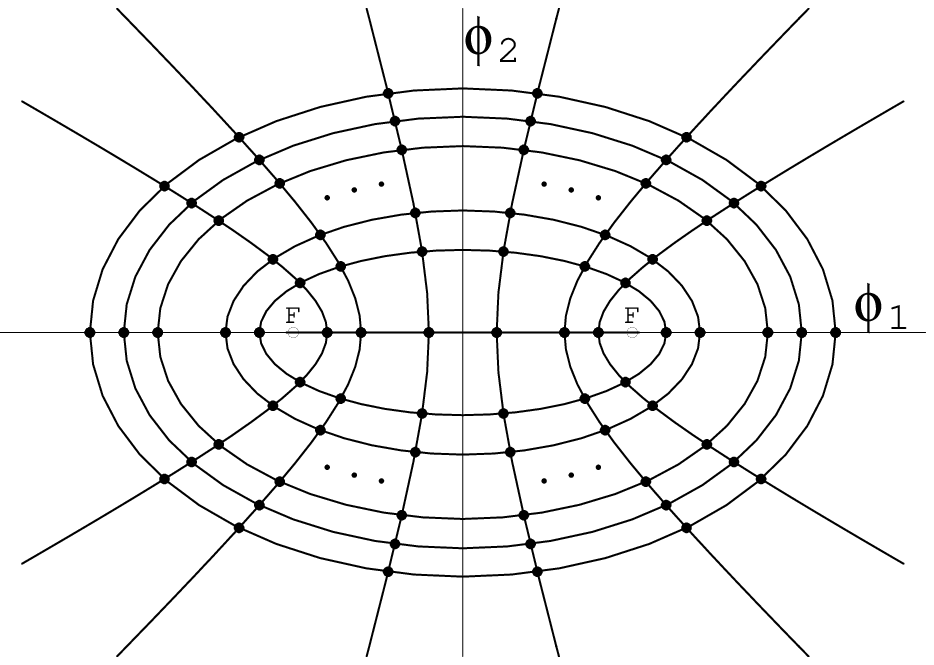}}
\caption{\small \textit{$\bar{\cal M}$ (points) and the reticulum
$\bar{\cal R}$ (solid lines) in ${\mathbb E}^2$ (left), ${\cal M}$
(points) and the reticulum ${\cal R}$ (solid lines) in ${\mathbb
R}^2$ (right) for the Type I-1a models.}}
\end{figure}

\item {\it TYPE I-1b}:  In this case
$v=\sigma_0=0$ is a root and $\sigma_{\pm n} = \pm \Omega$ are
simple roots of $g(v)$. The foci $F_\pm$ do not belong to $\bar{\cal
M}$ in this type of model either and:
\[
\bar{\cal M}= \bar{\cal M}_{f+g}-\{(\pm \Omega,\Omega)\}=
\bigcup_{(j,k)\in {\rm J} \times {\rm K}}
\{(\sigma_j,\sigma_k)\}-\{(\pm
\Omega,\Omega)\}=\displaystyle\bigcup_{(j,k) \in {\rm J} \times {\rm
K}} \bar{v}_{j,k} \qquad ,
\]
where now ${\rm J}=\{-n,...,-1,0,1,...,n\}$ because $v=0$ is a root
of $g(v)$. The number of elements of $\bar{\cal M}$ is: ${\rm
card}(\bar{\cal M})=2n(n+1)+m-1$. In the count of minima in
${\mathbb R}^2$ we must add to the minima of Type I-1a models the
points: $\xi_\Omega^\pm
(0,\sigma_k)=(0,\pm\sqrt{\sigma_k^2-\Omega^2}\,\,)$, $k=n, n+1,
\cdots , n+m$, all of them living on the ordinate $\phi_1=0$ axis.
Therefore, ${\rm card}( {\cal M})=4nm+2n-1$ because all the minima
in the $v=0$ line of ${\mathbb E}^2$ go to two minima in ${\mathbb
R}^2$, except the $(k=n,j=0)$ minimum which is mapped at the origin
of ${\mathbb R}^2$. In Figure 4 the minima of the potential energy
density in this type of model is  depicted both in the elliptic
strip and in the Cartesian plane. The reticulum of straight lines
joining these points is also plotted. Different cells with respect
to the Type I-1a cells arise close to the ordinate axis due to the
new Type I-1b minima. .

\begin{figure}[htb]
\centerline{
\includegraphics[height=3.5cm]{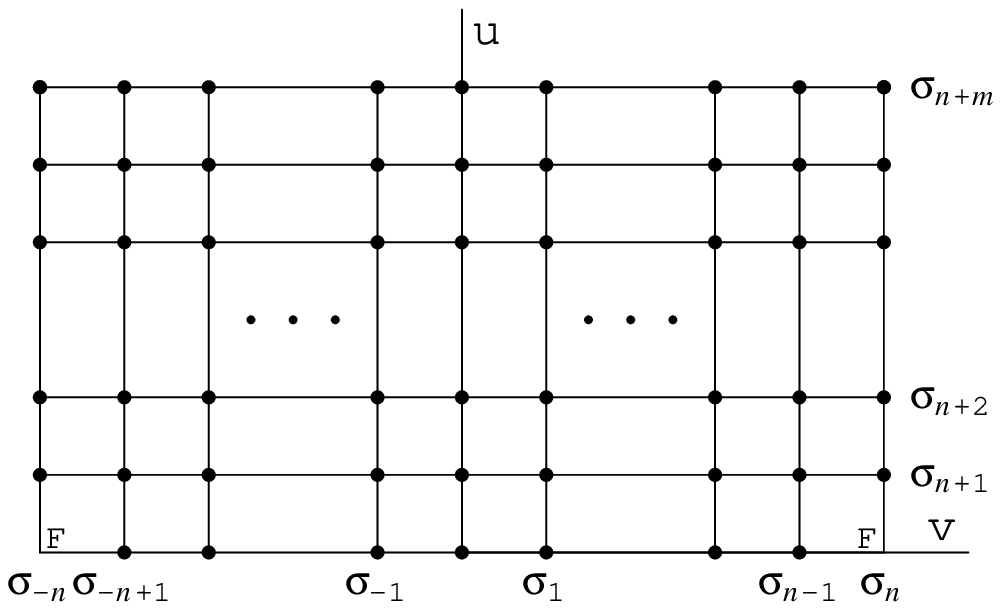}\hspace{1cm}
\includegraphics[height=3.5cm]{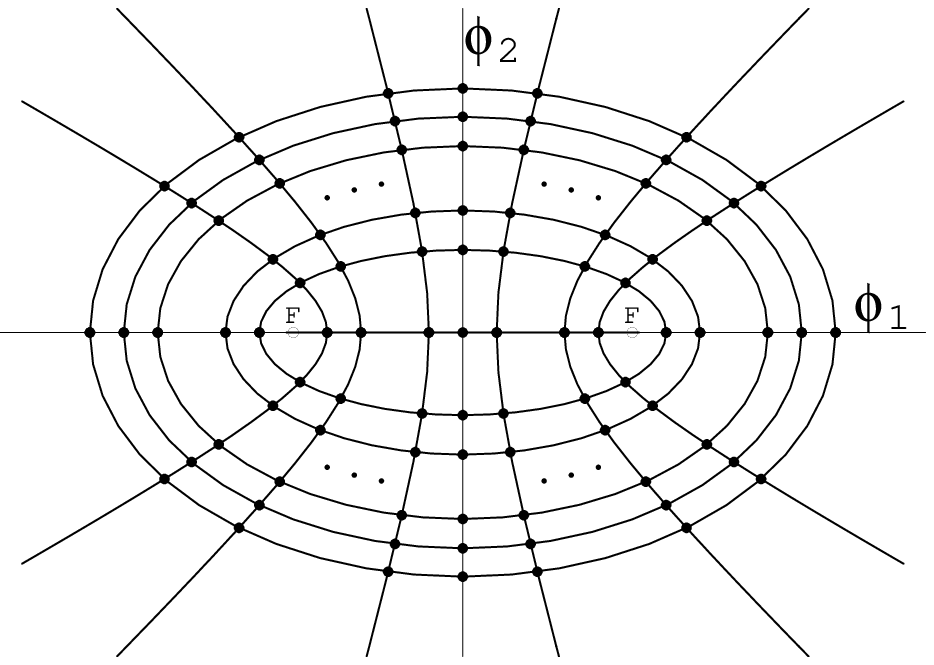}}
\caption{\small \textit{$\bar{\cal M}$ (points) and the reticulum
$\bar{\cal R}$ (solid lines)(left), ${\cal M}$ (points) and the
reticulum ${\cal R}$ (solid lines) (right) for the Type I-1b
models.}}
\end{figure}

\item {\it TYPE I-2a}: In this type of model $v=\sigma_{\pm n}=\pm \Omega$ are
multiple roots of the function $g(v)$, such that the foci $F_\pm$
belong to the set $\bar{\cal M}$ whereas $v=0$ is not a root.
Therefore,
\[
\bar{\cal M}= \bar{\cal M}_{f+g}= \bigcup_{(j,k)\in {\rm J} \times
{\rm K}} \{(\sigma_j,\sigma_k)\}=\displaystyle\bigcup_{(j,k) \in
{\rm J} \times {\rm K}} \bar{v}_{j,k} \qquad ,
\]
where ${\rm J}=\{-n,...,-1,1,...,n\}$. The number of elements in
$\bar{\cal M}$ is: ${\rm card}(\bar{\cal M})=2n(m+1)$. In ${\mathbb
R}^2$, however, this number is: ${\rm card}( {\cal M}) =
4nm+2(n-m)$, because the two new minima at the foci in ${\mathbb
E}^2$ are one-to-one mapped to the foci in ${\mathbb R}^2$, see
Figure 5.

\begin{figure}[htb]
\centerline{
\includegraphics[height=3.5cm]{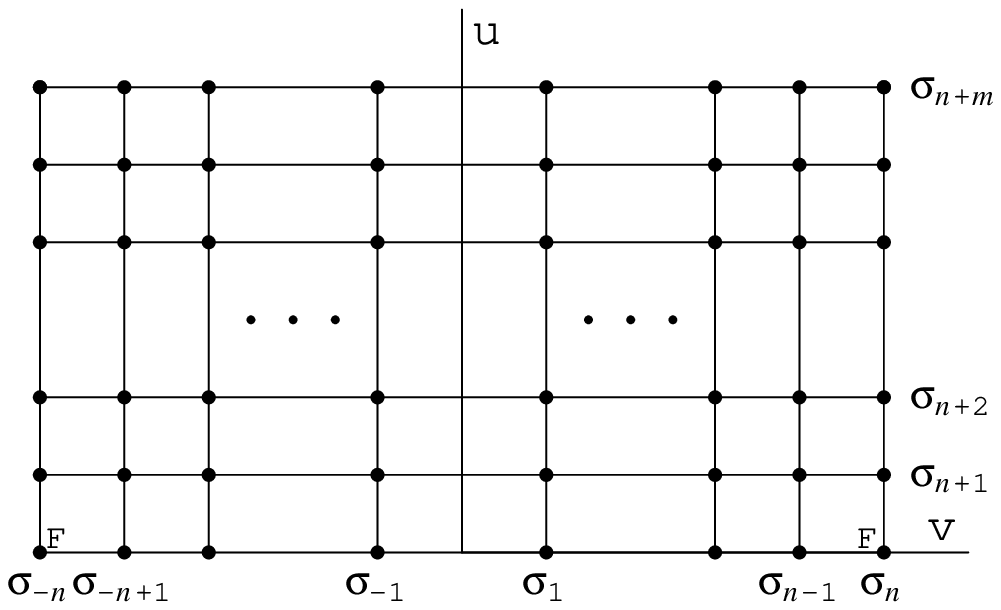}\hspace{1cm}
\includegraphics[height=3.5cm]{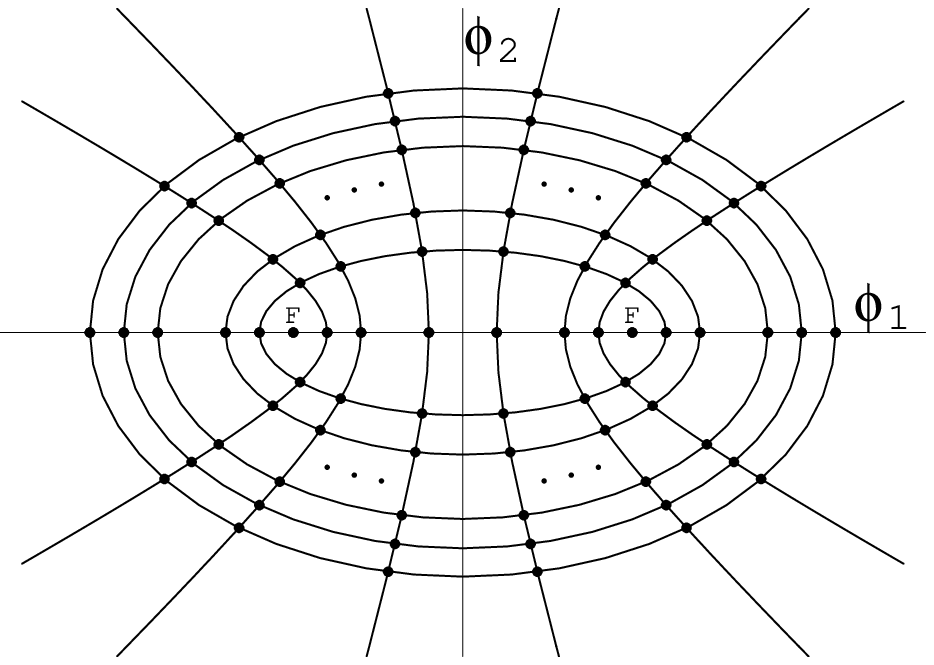}}
\caption{\small \textit{$\bar{\cal M}$ (points) and the reticulum
$\bar{\cal R}$ (solid lines)(left) ${\cal M}$ (points) and the
reticulum ${\cal R}$ (solid lines)(right) for the Type I-2a
models.}}
\end{figure}

\item {\it TYPE I-2b}: In the last type, $v=\sigma_0=0$ is a root and $v=\sigma_{\pm n}=\pm \Omega$ are
multiple roots of the function $g(v)$. There are minima of the
potential energy density in ${\mathbb E}^2$ at the foci $F_\pm$ and
on the $v=0$ line. Therefore,
\[
\bar{\cal M}= \bar{\cal M}_{f+g}= \bigcup_{(j,k)\in {\rm J} \times
{\rm K}} \{(\sigma_j,\sigma_k)\}=\displaystyle\bigcup_{(j,k) \in
{\rm J} \times {\rm K}} \bar{v}_{j,k} \qquad ,
\]
where ${\rm J}=\{-n,...,-1,0,1,...,n\}$ and: ${\rm card}(\bar{\cal
M})=(2n+1)(m+1)$. Passing to Cartesian coordinates, each minimum at
the two foci and the $(v=0,u=\Omega)$ point go to one minimum in the
$\phi_2=0$ abscissa axis, but the $m$ remaining new minima
$(v=0,u=\sigma_k)$, $k=n+1,n+2, \cdots , n+m$, have two image
minimum in ${\mathbb R}^2$. Accordingly, the count of minima in the
Cartesian plane is: ${\rm card}( {\cal M})= 4nm+2n+1$, see Figure 6.
\noindent\begin{figure}[htb] \centerline{
\includegraphics[height=3.5cm]{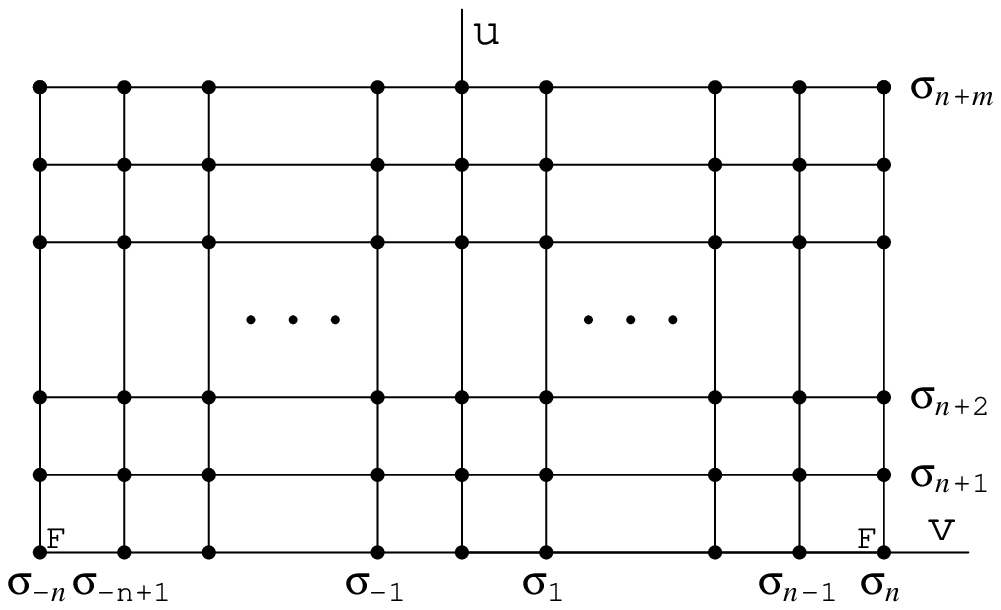}\hspace{1cm}
\includegraphics[height=3.5cm]{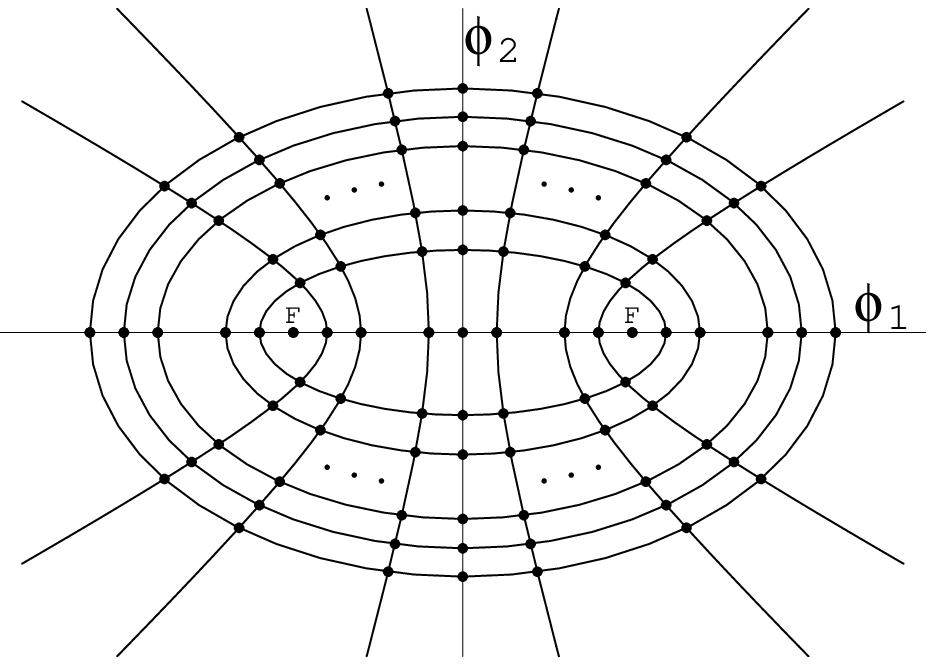}}
\caption{\small \textit{$\bar{\cal M}$ (points) and the reticulum
$\bar{\cal R}$ (solid lines)(left), ${\cal M}$ (points) and the
reticulum ${\cal R}$ (solid lines)(right) for the Type I-2b
models.}}
\end{figure}
\end{itemize}
In all four cases, the network in ${\mathbb E}^2$ is formed by the
horizontal segments
\[
r^{k,k}_{j,j+1}=\{(v,u) \in E / \sigma_{j}\leq v \leq
\sigma_{j+1},u=\sigma_k\} \qquad ,
\]
joining the points $(\sigma_j,\sigma_k)$, and
$(\sigma_{j+1},\sigma_k)$ and the vertical segments
\[
r^{k,k+1}_{j,j}=\{(v,u) \in E / v=\sigma_j, \sigma_{k}\leq u \leq
\sigma_{k+1}\} \qquad ,
\]
which join the points $(\sigma_j,\sigma_k)$ and
$(\sigma_j,\sigma_{k+1})$. Recall that $j \in {\rm J}$, $k \in {\rm
K}$ and $j+1$ means the following element in the set J. The reticula
${\cal R}$ in ${\mathbb R}^2$ become the sets of orthogonal
hyperbolae and ellipses:
\[
\frac{\phi_1^2}{\sigma^2_i}+\frac{\phi_2^2}{\sigma^2_i-\Omega^2}=1
\qquad , \qquad i\in {\rm J}\cup {\rm K} \qquad .
\]

\subsection{Types of cells}

The reticulum $\bar{\cal R}$ on the elliptic plane is formed by
rectangular cells, see Figures 3-6. We shall denote by
$\bar{C}_{j,j+1}^{k,k+1}$ the cell whose vertices are the elements
of $\bar{\cal M}_{f+g}$ $(\sigma_j,\sigma_k)$,
$(\sigma_j,\sigma_{k+1})$, $(\sigma_{j+1},\sigma_k)$ and
$(\sigma_{j+1},\sigma_{k+1})$. Therefore, the edges of this cell are
the segments $r^{k,k}_{j,j+1}$, $r^{k,k+1}_{j,j}$,
$r^{k,k+1}_{j+1,j+1}$ and $r^{k+1,k+1}_{j,j+1}$. We distinguish the
following types of cells:

\vspace{0.2cm}

$\bullet$ Cells of Type 1: The four vertices of the cell are
elements of $\bar{\cal M}$, see Figure 7a. This type of cell arises
in all the kinds of generalized MSTB models and it is the usual
pattern for sufficiently high values of $m$ and $n$, see Figures 3,
4, 5 and 6.

\vspace{0.2cm}

$\bullet$ Cells of Type 2: Three of the vertices of the cell are
elements of $\bar{\cal M}$ whereas the fourth vertex is one of the
foci of the model, see Figure 7b. The cells $C^{n,n+1}_{\pm(n-1),\pm
n}$ that arise in the type I-1a and I-1b models are of this type. In
this case the foci $F_\pm$ are not zeroes of the potential energy
density.

\vspace{0.2cm}

$\bullet$ Cells of Type 3: Only two of the vertices of the cell
belong to $\bar{\cal M}$ whereas the other two are the foci of the
coordinate system, see Figure 7c. This type of $C_{-1,1}^{1,2}$ cell
arises in the special case $(n=1,m)$ of Type I-1a models. Note that
$(n=1,m=1)$ is precisely the MSTB model.

\noindent\begin{figure}[htb] \centerline{
\includegraphics[height=3.cm]{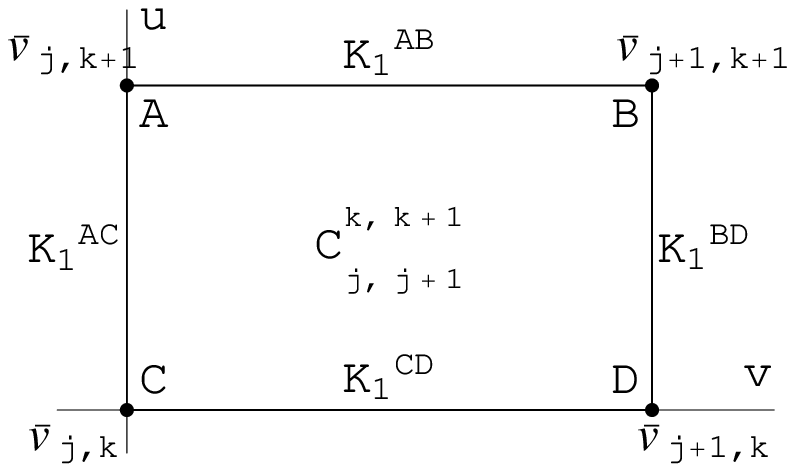}\hspace{0.5cm}
\includegraphics[height=3.cm]{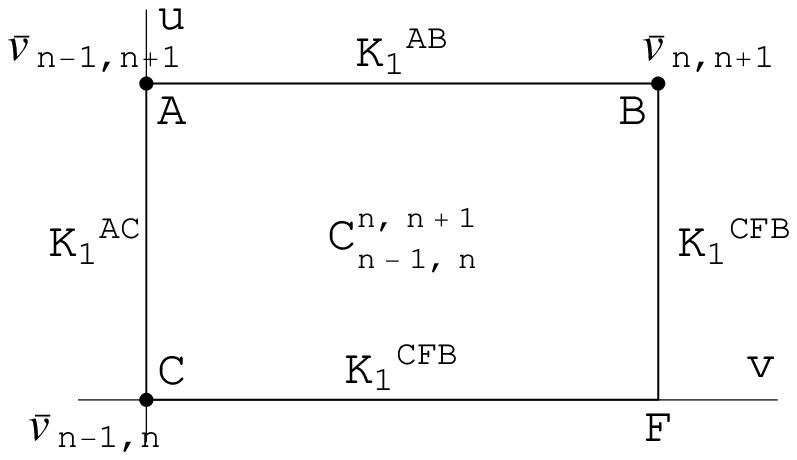}\hspace{0.5cm}
\includegraphics[height=3.cm]{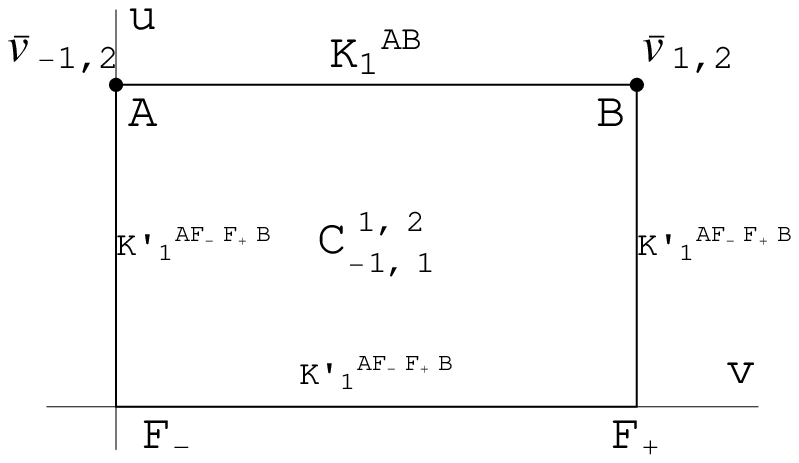}}
\caption{\small \textit{Types of cells: (a) Type 1 cell (the four
vertices are zeroes of $\xi_\Omega^\pm U$), (b) Type 2 cell (three
of the four vertices are zeroes of $\xi_\Omega^\pm U$) and (c) Type
3 cell (two of the four vertices are zeroes of $\xi_\Omega^\pm U$).
The edges of the cells have also been plotted as well as the
singular Kinks living on them.}}
\end{figure}

In sum, Type I-2a and I-2b models exhibit only cells of Type 1, Type
I-1a and I-1b models have cells of both Type 1 and Type 2. Cells of
Type 3 arise only in $(n=1,m)$ Type I-1a models.

\section{The structure of the kink varieties}

This Section is devoted to analyzing the general structure of kink
varieties in the generalized MSTB models. In order to alleviate the
notation we shall denote the elements of $\bar{\cal M}$ in a generic
cell as $A=v_{j,k+1}$, $B=v_{j+1,k+1}$, $C=v_{j,k}$ and
$D=v_{j+1,k}$, see Figure 7. The notation for solitary waves will be
$K_{\cal N}^{vv'}$ where the superscripts designate the elements of
$\bar{\cal M}$ (or ${\cal M}$) that are asymptotically connected by
the kink solutions, and ${\cal N}$ stands for the number of energy
lumps supported by the traveling wave. Bearing in mind the
invariance of generalized MSTB models under spatial reflections, $x
\rightarrow -x$, one may identify $K_{\cal N}^{vv'}$ and $K_{\cal
N}^{v'v}$ as a single type of solution. Also, we shall use the term
\textit{single kinks} to refer to those solutions whose energy
density is localized around a single point (${\cal N}=1$) whereas we
use \textit{composite kinks} to refer to those solutions whose
energy density can be localized at several points (${\cal N}>1$).

There are two main categories or kinds of solitary wave solutions:
(1) Singular (isolated) kinks, which are the single or basic kinks
of generalized MSTB models. (2) One-parametric families of composite
kinks, which are non-linear combinations of the single kinks.
Composite kinks are formed by several lumps whose separation is
determined by the value of the parameter of the family.

\subsection{Singular kinks}

\hspace{0.5cm} {\bf Proposition 2}. {\it There exist kinks whose
orbits are located on the border of the cells $C^{k,k+1}_{j,j+1}$,
$j\in {\rm J}$, $k\in {\rm K}$, i.e., on the segments
$r^{k,k}_{j,j+1}$ or $r^{k,k+1}_{jj}$  of the elliptic strip
${\mathbb E}^2$ (pieces of hyperbolae and ellipses in the
${\mathbb R}^2$ Cartesian plane). For reasons to become clear
later we shall refer to these solutions as singular kinks.}

\vspace{0.1cm}

\textit{Proof:} The critical value
$\xi_\Omega^{*}I_1=0=\xi_\Omega^{*}I_2$, where the two invariants
cease to be independent, is reached by all the trajectories or
orbits satisfying the system of first-order ODE's:
\begin{equation}
\frac{du}{dx}=(-1)^\alpha \frac{\sqrt{2(u^2-\Omega^2)
f(u)}}{u^2-v^2} \hspace{1cm} , \hspace{1cm} \frac{dv}{dx}=(-1)^\beta
\frac{\sqrt{2(\Omega^2-v^2) g(v)}}{u^2-v^2} \qquad  ; \quad
\alpha,\beta=0,1\label{eq:fis}
\end{equation}
see (\ref{eq:cinco}) and (\ref{eq:seis}). We try the orbits
$u=\sigma_k$, $k \in {\rm K}$ (the segments $r^{k,k}_{jj+1}$) or
$v=\sigma_j$, $j\in {\rm J}$ (the segments $r^{k,k+1}_{jj}$) into
the (\ref{eq:fis}) equations. Alternatively, one of the two
equations is satisfied if either $\frac{du}{dx}=0$ or
$\frac{dv}{dx}=0$, and the other may be immediately integrated by
means of a quadrature. In sum, there are

$-$ Kink orbits living on pieces of hyperbolae $r_{j,j}^{k,k+1}$:
\begin{equation}
v(x)=\sigma_j \hspace{0.5cm} , \hspace{0.5cm} u(x)=h^{-1}(\bar{x})
\hspace{0.5cm} , \hspace{0.5cm} h(u)=\int \frac{(u^2-\sigma_j^2) du
}{\sqrt{2 f(u)(u^2-\Omega^2)}} \label{eq:tkv}
\end{equation}
$-$ Kink orbits living on pieces of ellipses $r_{j,j+1}^{k,k}$:
\begin{equation}
v(x)=h^{-1}(\bar{x}) \hspace{0.5cm} , \hspace{0.5cm} u(x)=\sigma_k
\hspace{0.5cm} , \hspace{0.5cm} h(v)=\int
\frac{(\sigma_k^2-v^2)dv}{\sqrt{2 g(v)(\Omega^2-v^2)}} \qquad ,
\label{eq:tku}
\end{equation}
and the Proposition is proved.

\vspace{0.1cm}

The type of singular kink, however, depends on the type of cell
where singular kink orbits arise.

\begin{itemize}
\item In type 1 cells there are four kinds of kinks. The four vertices $A$, $B$, $C$ and $D$
in these cells are elements of $\bar{\cal M}$: $A=v_{j,k+1}$,
$B=v_{j+1,k+1}$, $C=v_{j,k}$ and $D=v_{j+1,k}$, see Figure 7a. Kink
orbits connect two adjacent corners through the edges of the cell.
Thus, singular kink orbits are the segments $r^{k+1,k+1}_{j,j+1}$
($u=\sigma_{k+1}$) and $r^{k,k}_{j,j+1}$ ($u=\sigma_{k}$)
respectively joining $A$ to $B$ and $C$ to $D$. In the ${\mathbb
R}^2$ Cartesian plane these kink orbits become elliptic arcs.

There are also singular kink orbits connecting $A$ to $C$ and $B$ to
$D$ along the segments $r^{k,k+1}_{j,j}$ ($v=\sigma_{j}$) and
$r^{k,k+1}_{j+1,j+1}$ ($v=\sigma_{j+1}$) (arcs of hyperbolae in the
${\mathbb R}^2$). We shall denote these four kinds of kinks as
$K_1^{AB}$, $K_1^{CD}$, $K_1^{AC}$ and $K_1^{BD}$. Their associated
profiles are respectively obtained from the quadratures
(\ref{eq:tkv}) and (\ref{eq:tku}).

The main feature of this type of kink is that they are {\it single
kinks}. To show this, note that:
\begin{itemize}
\item The kink solutions are monotonic (increasing or
decreasing) functions $u(x)$ and $v(x)$ of $\bar{x}$. In the
first-order equations (\ref{eq:fis}) one immediately realizes this
point because $\frac{du}{dx}\neq 0$ and $\frac{dv}{dx}\neq 0$ for
every point in ${\mathbb E}^2$ except at the zeroes of $f(u)$ and
$g(v)$, which are reached only asymptotically (at $x=\pm\infty$).

\item The kink energy density (\ref{eq:vir}) becomes a polynomial in
a single elliptic variable for this type of kink orbits. For
instance, for $v=\sigma_j$ orbits the energy density reads:
\begin{eqnarray*}
{\cal E}[u,\sigma_j]&=& 2 u^{2\alpha_0} (u^2-\sigma_1^2)^{2\alpha_1}
\cdot \dots \cdot (u^2-\sigma^2_{j-1})^{2\alpha_{j-1}}
(u^2-\sigma_j^2)^{2\alpha_j-1}
(u^2-\sigma_{j+1}^2)^{2\alpha_{j+1}}\\&& \cdot \dots \cdot
(u^2-\sigma_{n-1}^2)^{2\alpha_{n-1}}
(u^2-\Omega^2)(u^2-\sigma_{n+1}^2)^{2\alpha_{n+1}} \cdot \dots \cdot
(u^2-\sigma_{n+m}^2)^{2\alpha_{n+m}} \,\, \, .
\end{eqnarray*}
Because this polynomial $p(u)={\cal E}[u,\sigma_j]$ is a
semi-definite positive function in the $u\in[\sigma_k,\sigma_{k+1}]$
range that vanishes only at $u=\sigma_k$ and $u=\sigma_{k+1}$,
Rolle's theorem ensures that $p(u)$ reaches a maximum. To show that
this maximum is the only critical point in this interval requires a
little more work. In fact, it is better to consider the polynomial
$p(u)$ as a function of $z=u^2$:
\[
p(z)=\Pi_{i=0}^{n+m}\, (z-z_i)^{n_i} \qquad , \qquad z_0=0 \qquad  ,
\quad z_i=\sigma^2_i \quad , \quad i=1,2,  \cdots  , \, n+m \qquad .
\]
The multiplicities $n_i$ of the roots $z_i$ are{\footnote{Strictly,
$z_0=0$ is not a root because $z\in [1,\infty)$, but this does not
affect our argument.}}: $n_i=2\alpha_i$ except $n_0=\alpha_0$,
$n_j=2\alpha_j-1$, $n_n=1$. Let us consider the $\log p(z)$ function
and its derivative:
\[
\eta(z)=\log p(z)=\sum_{i=i}^{n+m}\, n_i \log (z-z_i) \qquad ,
\qquad \eta^\prime(z)=\frac{p^\prime(z)}{p(z)}=\sum_{i=0}^{n+m}\,
\frac{n_i}{z-z_i} \qquad .
\]
Excluding the points $z_1,z_2, \cdots , z_{n+m}$ of the half-line
$[1,+\infty)$ the zeroes of $\eta^\prime$ and $p^\prime$ are the
same. We consider a variation of $z$ asymptotically connecting
$\sigma_k^2$ with $\sigma^2_{k+1}$. Because $
\eta^{\prime\prime}(z)=-\sum_{i=0}^{n+m}\frac{n_i}{(z-z_i)^2}$ is
negative, $\eta^\prime(z)$ is a decreasing function of $z$. This,
together with the limiting values
\[
\lim_{z\rightarrow\sigma_k^2}\eta^\prime(z)=+\infty \qquad ,
\qquad \lim_{z\rightarrow\sigma_{k+1}^2}\eta^\prime(z)=-\infty
\]
means that $\eta^\prime$ intersects the abscissa axis at only one
point and, consequently, $p(z)$ has only one critical point.
Therefore, the kink energy density is localized around one point and
this single kink is a basic lump or particle.
\end{itemize}

\item In Type 2 cells, $C^{n,n+1}_{\pm(n-1),\pm n}$ three of the vertices
are elements of $\bar{\cal M}$ but the fourth vertex is one of the
foci $F_\pm$; $A=v_{\pm (n-1),n+1}$, $B=v_{\pm n,n+1}$, $C=v_{\pm
(n-1),n}$ and $D=F_\pm$. Kink orbits $K_1^{AB}$ and $K_1^{AC}$
respectively joining $A$ to $B$ and $A$ to $C$ through the segments
$u=\sigma_{n+1}$ and $v=\sigma_{n-1}$, see Figure 7b, are identical
to those in Type 1 cells, providing similar basic kinks.
Additionally, there is a new kind of kink orbit connecting $B$ and
$C$ along two edges of the cell: the orbit starts from $C$, follows
a segment where $u=\Omega$, crosses one of the foci $F_\pm$, runs
along a $v=\pm\Omega$ segment, and, eventually, arrives at $B$. The
path is the union of the trajectories $r_{n-1,n}^{nn} \cup
r_{n,n}^{n,n+1}$ or $r_{-n+1,-n}^{nn} \cup r^{n,n+1}_{-n,-n}$, an
interval on the $\phi_1$ axis in the ${\mathbb R}^2$ Cartesian
plane, such that the solution in ${\mathbb R}^2$ is continuous and
differentiable.

Because the orbits travel two edges of the cell, one guesses that
this new type of kink will be combinations of two basic kinks.
Therefore, we shall term them $K_2^{BF\bigvee FC}$. Let $\bar{x}^c$
be the point on the line where the kink profile crosses the focus.
Recall that in this cell the range of elliptic variables is $u \in
[\Omega, \sigma_{n+1}]$ and $v=[\sigma_{n-1},\Omega]$. Thus, the
energy density reads:
\[
{\cal E}[K_2^{BF\bigvee FC}](u,v)=\left\{ \begin{array}{l}
\displaystyle\prod_{j=1,j\neq n}^{n+m}
(v(x)^2-\sigma_j^2)^{2\alpha_j} \hspace{0.5cm} \mbox{if}
\hspace{0.5cm} \bar{x}\leq \bar{x}^c \\
\displaystyle\prod_{j=1,j\neq n}^{n+m}
(u(x)^2-\sigma_j^2)^{2\alpha_j} \hspace{0.5cm} \mbox{if}
\hspace{0.5cm} \bar{x}>\bar{x}^c
\end{array} \right. \qquad \qquad .
\]
Taking into account that the polynomial $p(z)=\prod_{i=1,i\neq
n}^{n+m} (z^2-\sigma_i^2)^{2\alpha_i}$ has roots at the values
$z=\sigma_{n-1}$ and $z=\sigma_{n+1}$, we expect a relative maximum
in the interval $[\sigma_{n-1},\sigma_{n+1}]$, which is achieved
either in the interval $[\sigma_{n-1},\Omega]$ or in
$[\Omega,\sigma_{n+1}]$. An identical argument to that applied to
the other type of kink shows that the maximum is the only critical
point of the energy density. Thus, this kind of solution exhibits
one lump, which may seem contradictory to the above guess about the
composite nature of these kinks. We shall confirm later that these
solutions are formed by two overlapping basic lumps by observing the
kink mass sum rules in the following Section.

\item In Type 3 cells of $(n=1,m)$ Type I-1a generalized MSTB
models only the vertices $A=v_{-1,2}$ and $B=v_{1,2}$ are elements
of $\bar{\cal M}$, whereas the other two vertices are the foci
$F_\pm$ of the model. There are two kinds of kinks: the $K_1^{AB}$
kinks akin to the kinks in Type 1 cells running on the $AB$ edge,
the segment $r^{2,2}_{-1,1}$, and the ${K}_1^{BF_+\bigvee
F_+F_-\bigvee F_-A}$ kinks. The orbit of this new kind of kink is
formed by the other three edges of the cell, see Figure 7c. The kink
orbit starts at the point $A$, following the straight line
$v=-\Omega$. After crossing the focus $F_-$ the orbit runs through
the $u=\Omega$ segment and reaches the focus $F_+$. Finally, the
kink orbit arrives at the point $B$, traveling  the segment
$v=\Omega$. In sum, this three-stage kink orbit is:
$r_{-1,-1}^{2,1}\cup r^{1,1}_{-1,1} \cup r_{1,1}^{1,2}$. In the
Cartesian ${\mathbb R}^2$ plane, this trajectory is the $[A,B]$
interval of the $\phi_1$ axis. Denoting by $\bar{x}_\pm$ the points
in the spatial line where the kink orbit in the elliptic strip
${\mathbb E}^2$ passes through the foci $F_{\pm}$ we write the
energy density of kinks of this type in the form:
\[
{\cal E}[{K}_1^{BF_+\bigvee F_+F_-\bigvee F_-A}](u,v)=\left\{
\begin{array}{l} \displaystyle\prod_{i=1,i\neq n}^{n+m}
(u(x)^2-\sigma_i^2)^{2\alpha_i} \hspace{0.5cm} \mbox{if}
\hspace{0.5cm} \bar{x}\leq \bar{x}_- \\
\displaystyle\prod_{i=1,i\neq n}^{n+m}
(v(x)^2-\sigma_i^2)^{2\alpha_i}
\hspace{0.5cm} \mbox{if}  \hspace{0.5cm} \bar{x}_- < \bar{x}<\bar{x}_+ \\
\displaystyle\prod_{i=1,i\neq n}^{n+m}
(u(x)^2-\sigma_i^2)^{2\alpha_i} \hspace{0.5cm} \mbox{if}
\hspace{0.5cm} \bar{x}_+\leq \bar{x}
\end{array} \right.
\]
One can argue that this energy density reaches a maximum value at
only one point inside the orbit whereas it tends to zero at both
endpoints along the lines shown for the other singular kink orbits.
In this bizarre case the kink is not a composite of another two
kinks because the end points are shared with the only other kink in
the cell. Thus, despite being built from three stage orbits this
kink is truly a basic kink.
\end{itemize}

In this Section we have shown that there are kink orbits living on
the border of each cell $C_{j,j+1}^{k,k+1}$. It remains to decide
whether or not there exist solitary wave solutions whose orbits are
confined inside the cells. The Picard Theorem about the existence
and uniqueness of solutions of ordinary differential equations
guarantees that there are no solutions of the analogous mechanical
system crossing the edges of the cells, except through their
vertices, where the theorem is not applicable, see (\ref{eq:fis}).
Therefore, the existence of singular kink orbits on the reticulum
rules that any other kink orbit must be confined inside the cells in
such a way that different orbits can meet only at the vertices of
the cell.

\subsection{Generic kink varieties}

Full knowledge of the kink manifold in this type of model is
achieved by finding all the separatrix trajectories (finite
mechanical action) of the Newton equations (\ref{eq:ordinary})
equivalent to the static field equations. The most effective method
of solving (\ref{eq:ordinary}) is the Hamilton-Jacobi procedure
because these models are chosen in such a way that elliptic
coordinates allow Hamilton-Jacobi separability. From the mechanical
Lagrangian
\begin{equation}
{\cal L}=\xi_\Omega^{*}{\cal E} [\phi_1,\phi_2]= \left\{\frac{1}{2}
\frac{u^2-v^2}{u^2-\Omega^2} \left( \frac{du}{dx}
\right)^2+\frac{1}{2} \frac{u^2-v^2}{\Omega^2-v^2} \left(
\frac{dv}{dx} \right)^2  + \frac{f(u)+g(v)}{u^2-v^2}   \right\}
\label{eq:cuatroc}
\end{equation}
the generalized momenta are determined:
\[
\begin{array}{lcr}
{\displaystyle p_u=\frac{\partial {\cal L}}{\partial \left( \frac{d
u}{d x} \right)}=\frac{u^2-v^2}{u^2-\Omega^2} \; \frac{d u}{d x}  }
&  \hspace{2cm} & ,\hspace{2cm} {\displaystyle p_v=\frac{\partial
{\cal L}}{\partial \left( \frac{d v}{d x}\right)
}=\frac{u^2-v^2}{\Omega^2-v^2} \; \frac{d v}{d x} } \qquad .
\end{array}
\]
The mechanical Hamiltonian ${\cal H}=\xi_\Omega^{*}I_1$ reads:
\[
{\cal H}= \frac{1}{u^2-v^2} \; (h_u+h_v) \,\,\, ; \quad
h_u=\frac{1}{2} (u^2-\Omega^2) p_u^2-f(u) \quad , \quad
h_v=\frac{1}{2} (\Omega^2-v^2) p_v^2-g(v)\qquad .
\]
The Hamilton-Jacobi PDE equation
\begin{equation}
\frac{\partial {\cal J}}{\partial x}+ {\cal H} \left( \frac{\partial {\cal J}}{\partial u},
 \frac{\partial {\cal J}}{\partial v},u,v \right)=0
\label{eq:siete}
\end{equation}
corresponding to this Hamiltonian is separable. Denoting
${u}'=\frac{d u}{d x}$ , ${v}'=\frac{d v}{d x}$, the separation
ansatz ${\cal J}={\cal J}_x(x)+{\cal J}_u(u)+{\cal J}_v(v)$ provides
solutions of the PDE (\ref{eq:siete}) in terms of the solutions of
the ODE's:
\begin{equation}
{\cal J}_x^\prime=-E \quad , \quad {\cal J}_u^\prime={\rm sign}\,
({u}')\sqrt{\frac{2(F+E u^2+f(u))}{u^2-\Omega^2}}  \quad , \quad
{\cal J}_v^\prime={\rm sign}\, ({v}')\sqrt{\frac{2(-F-E
v^2+g(v))}{\Omega^2-v^2}} \label{eq:shj}
\end{equation}
because (\ref{eq:siete}) becomes:
\[
{1\over 2}(u^2-\Omega^2)\left({\cal
J}_u^\prime\right)^2-Eu^2-f(u)=F=-{1\over
2}(\Omega^2-v^2)\left({\cal J}_v^\prime\right)^2-Ev^2+g(v) \qquad .
\]
$E$ is thus the mechanical energy; the value of the first invariant
$I_1=E$. The relation of the separation constant $F$ with the two
first integrals is a little bit more involved. From
\[
I_2=\xi^*_\Omega
I_2=\frac{1}{u^2-v^2}\left\{(u^2-\Omega^2)h_v-(\Omega^2-v^2)h_u\right\}
\qquad ; \qquad h_u=F+Eu^2  \quad  , \quad \quad h_v=-F-Ev^2
\]
one deduces that $F$ is a linear combination of the two invariants:
$F=-\Omega^2 I_1-I_2$.

The first equation in (\ref{eq:shj}) is integrated immediately to
find ${\displaystyle {\cal J}_x=-E x}$ whereas the solution of the
other two equations in (\ref{eq:shj}) are the quadratures:
\begin{equation}
{\displaystyle {\cal J}_u= {\rm sign}\, ({u}')  \int du
\sqrt{\frac{2(F+E u^2+f(u))}{u^2-\Omega^2}} } \quad , \quad
{\displaystyle {\cal J}_v= {\rm sign} \,({v}')  \int dv
\sqrt{\frac{2(-F-E v^2+g(v))}{\Omega^2-v^2}} } \label{eq:sshj}
\qquad .
\end{equation}
Hamilton-Jacobi theory  prescribes that "particle" orbits are
determined by (\ref{eq:nueve}) (left) and "time schedules" ruled by
(\ref{eq:nueve}) (right):
\begin{equation}
\frac{\partial {\cal J}}{\partial F}=\gamma_1 \hspace{2cm}
,\hspace{2cm} \frac{\partial {\cal J}}{\partial E}=\gamma_2
\label{eq:nueve} \qquad ,
\end{equation}
where ${\cal J}$ is the Hamilton principal function, the solution of
(\ref{eq:siete}) and (\ref{eq:shj}), and $\gamma_1$, $\gamma_2$, are
integration constants.

Finite action trajectories require that $E=F=0$ and give the
solitary wave solutions, finite energy, in the field theoretical
model. Therefore, the kink orbits, corresponding to finite action
trajectories, and the kink profiles, given by the time schedules of
these orbits, are the quadratures:
\begin{equation}
{\rm sign} ({u}') \int \frac{du}{\sqrt{(u^2-\Omega^2) f(u)}} - {\rm
sign} ({v}') \int \frac{dv}{\sqrt{(\Omega^2-v^2) g(v)}}=\sqrt{2}
\gamma_1 \label{eq:diez}
\end{equation}
\begin{equation}
{\rm sign} ({u}') \int \frac{u^2 du}{\sqrt{(u^2-\Omega^2) f(u)}} -
{\rm sign} ({v}') \int \frac{v^2 dv}{\sqrt{(\Omega^2-v^2)
g(v)}}=\sqrt{2} (x+\gamma_2) \label{eq:once} \qquad .
\end{equation}
Thus, $\gamma_2$ obeys time translation and is associated to the
first invariant. In the field theoretical context this parameter
sets the center of the kink and/or the lump of energy. The meaning
of $\gamma_1$ is more subtle: orbits for different $\gamma_1$ are
related to each other by non-linear transformations generated by the
second invariant $I_2$.

\subsubsection{Linear stability analysis}

The qualitative structure of the moduli space of kink solutions
(\ref{eq:diez})-(\ref{eq:once}) parametrized by the
$(\gamma_1,\gamma_2)$ constants is explained by a linear (or local)
stability analysis. We write the first-order ODE system
(\ref{eq:fis}) in the form:
\begin{equation}
\frac{du}{dx}=(-1)^\alpha F(u,v) \hspace{2cm} , \hspace{2cm}
\frac{dv}{dx}=(-1)^\beta G(u,v)  \qquad . \label{eq:fis1}
\end{equation}
Note that $(-1)^\alpha={\rm sign} ({u}')$ and $(-1)^\beta={\rm sign}
({v}')$. The fixed points in the trajectories are identified as the
zeroes of the functions $F$ and $G$: $F(u_0,v_0)=0=G(u_0,v_0)$.
Thus, the points in ${\cal M}$ are fixed points and one can study
the behavior of the trajectories near those points by linearization
of the ODE system (\ref{eq:fis}):
\begin{equation}
\left(\begin{array}{l} \frac{d\delta u}{dx} \\
\frac{d\delta
v}{dx}\end{array}\right)=M^{(\alpha,\beta)}(u_0,v_0)\left(\begin{array}{l}
\delta u(x) \\ \delta v(x) \end{array}\right) \hspace{0.2cm} ;
\hspace{0.2cm} M^{(\alpha,\beta)}(u_0,v_0)=\left(\begin{array}{cc}
(-1)^\alpha\frac{\partial F}{\partial u} & (-1)^\alpha\frac{\partial
F}{\partial v}\\ (-1)^\beta\frac{\partial G}{\partial u} &
(-1)^\beta\frac{\partial G}{\partial
v}\end{array}\right)(u_0,v_0)\,\, .
\end{equation}
It is easily checked that, depending on the choice of $\alpha$ and
$\beta$, the fixed points belonging to ${\cal M}$ fall into three
categories: 1) stable nodes (the two eigenvalues of
$M^{(\alpha,\beta)}$ are different and negative), 2) unstable
nodes (distinct eigenvalues, both positive) 3) saddle points (one
positive , one negative eigenvalue). The trajectories flow inwards
and end in the fixed point (case 1), start in the fixed point and
flow outwards (case 2), and flow first inwards, skip the fixed
point, and go outwards (or viceversa, case 3).

Moreover, taking the quotient of the two equations in
(\ref{eq:fis1}), one obtains the differential equation
\begin{equation}
\frac{du}{dv}=\frac{(-1)^\alpha}{(-1)^\beta}
\frac{F(u,v)}{G(u,v)}=\frac{(-1)^\alpha}{(-1)^\beta}\sqrt{\frac{(u^2-\Omega^2)
f(u)}{(\Omega^2-v^2) g(v)}} \qquad , \label{eq:flujo}
\end{equation}
showing that the flow is undefined,
$\frac{du}{dv}\left|_{(u_0,v_0)}\right.=\frac{0}{0}$, precisely at
the fixed points. Thus, there are pencils of trajectories
parametrized by $\gamma_1$ flowing in, flowing out, or skipping the
fixed points.

To distinguish which points of ${\cal M}$ are nodes or saddle points
one must focus on the values of $\alpha$ and $\beta$. If
$\alpha=\beta$, the flow induced by (\ref{eq:flujo}) runs along
monotonic increasing curves in the elliptic strip ${\mathbb E}^2$
because $\frac{du}{dv}$ is always positive. On applying the linear
stability analysis to a type 1 cell, Figure 7(a), one finds that $C$
and $B$ are alternatively stable or unstable nodes if
$\alpha=\beta=0$ or $\alpha=\beta=-1$ (red orbits in Figure 8(a))
whereas $D$ and $A$ are saddle points. The r$\hat{\rm o}$les of
$C,B$ and $A,D$ are exchanged (blue orbits) if $\alpha\neq\beta$.
Given the existence of solutions on the borders of the cell,
Picard's theorem confines inside the cell the pencils of kink orbits
connecting the fixed points either on the vertices $C$ and $B$ or on
$D$ and $A$, depending on the relative values of $\alpha$ and
$\beta$, see Figure 8(a). These two families of kink orbits are
parametrized by the integration constant $\gamma_1\in \mathbb{R}$.

\subsubsection{Global stability and the boundary of the moduli space of kinks}

The stability of the solitary waves is inherited from the global
-rather than local- stability of the kink orbits. By global
stability we mean the stability of the trajectory as a whole, not
merely the identification of the character of the fixed points along
the trajectory. According to Morse Theory, see \cite{Tasaki},
\cite{J0}, \cite{J1}, \cite{Aai4}, global stability (or instability)
is essentially accounted for by the number of conjugate or focal
points crossed by a pencil of trajectories between the starting and
ending points. Conjugate points are the points where trajectories
forming a congruence meet. Except in the Type I-2a and Type I-2b
models, where the foci belong to ${\cal M}$ and are fixed points,
the foci are not fixed points of (\ref{eq:fis1}). The flow
(\ref{eq:flujo}) at the foci is always undefined:
$\frac{du}{dv}\left|_{(u_0=1,v_0=\pm 1)}\right.=\frac{0}{0}$. In
sum, in Type I-1a and Type I-1b models the foci are conjugate points
to the stable and unstable nodes on the other vertices of the cell,
and families of globally unstable trajectories cross these points.
Therefore, the blue and red kink orbits in Figure 8(a) are stable,
whereas the blue orbits in Figure 8(b) and the blue and red orbits
in Figure 8(c) are unstable. \noindent\begin{figure}[htb]

\centerline{
\includegraphics[height=3.cm]{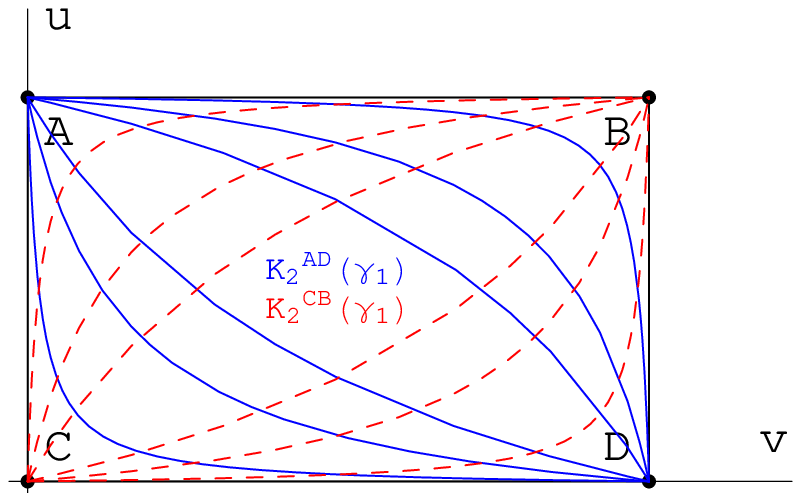}\hspace{0.5cm}
\includegraphics[height=3.cm]{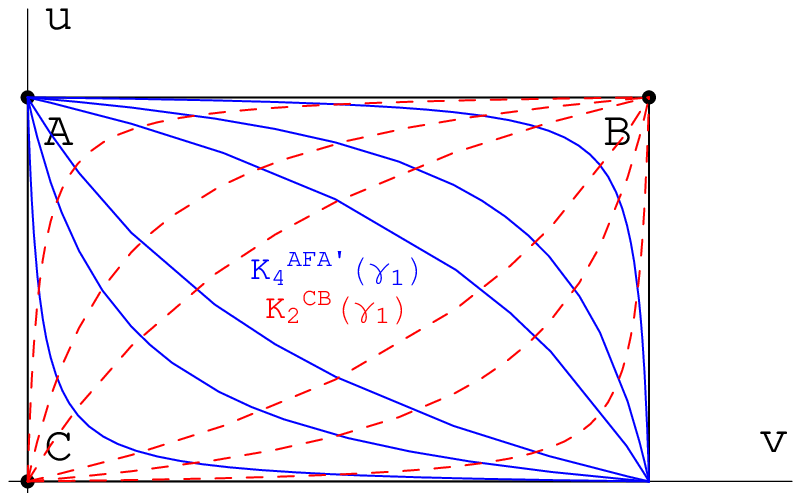}\hspace{0.5cm}
\includegraphics[height=3.cm]{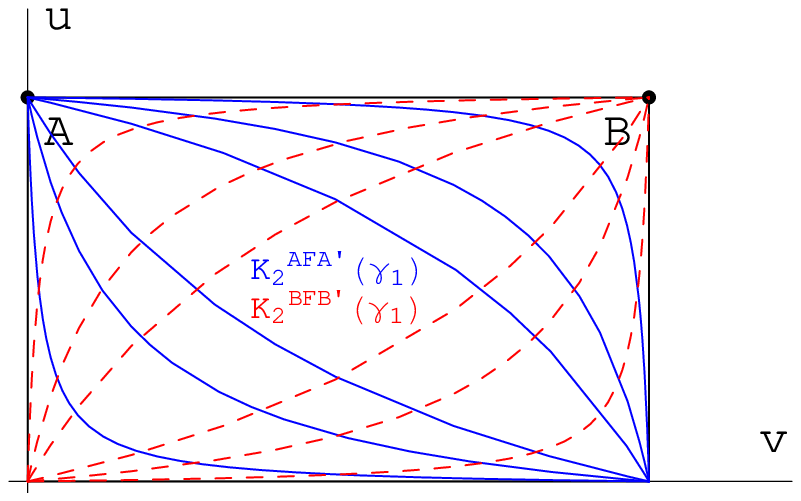}}
\caption{\small \textit{Families of kink orbits in each type of
cell: a) $K_2^{AD}(\gamma_1)$ and $K_2^{BC}(\gamma_1)$ in Type 1
cells, b) $K_2^{BC}(\gamma_1)$ and $K_4^{AA}(\gamma_1)$ in Type 2
cells, and c) $K_2^{AA}(\gamma_1)$ and $K_2^{BB}(\gamma_1)$ in
Type 3 cells.}}
\end{figure}

Bearing this in mind, we shall describe the families of kink
orbits inside the cells. In the cells where one or two of the foci
are vertices that do not belong to ${\cal M}$ , kink orbits
flowing according to (\ref{eq:flujo}) and connecting two elements
of ${\cal M}$ are formed by gluing two families of solutions that
correspond to two different choices of $\alpha$ relative to
$\beta$. The gluing must be performed in such a way that the kink
orbits back in the ${\mathbb R}^2$ Cartesian plane are continuous.
We now describe the different types of kink orbits arising in
different types of cells.

\begin{itemize}

\item In type 1 cells there are two families of kink orbits
connecting opposite vertices in the cells, distinguished by ${\rm
sign}((-1)^{\alpha+\beta})$ on (\ref{eq:flujo}). We shall refer to
these two types respectively as $K_2^{AD}(\gamma_1)$ when ${\rm
sign}((-1)^{\alpha+\beta}) =-1$, and $K_2^{CB}(\gamma_1)$, if
${\rm sign}((-1)^{\alpha+\beta})=+1$. When ${\rm
sign}((-1)^{\alpha+\beta})=-1$ $A$ is a stable node, and $D$ is an
unstable node, or viceversa.  In this case $B$ and $C$ are saddle
points. If ${\rm sign}((-1)^{\alpha+\beta})=+1$ the pair of points
$A$-$D$ exchanges its r$\hat{\rm o}$le with the $B$-$C$ couple,
see Figure 8(a). The parameter $\gamma_1$ labels one member in
each of the two families of kink orbits. The subscript suggests
that these kinks are formed by two basic kinks/lumps, which can be
confirmed by taking the asymptotic values: $\gamma_1 \rightarrow
\pm \infty$.

To prove this last statement, we consider a solution characterized
by the value $\gamma_1$. Therefore for generalized MSTB models
(\ref{eq:diez}) the orbit of this solution complies with:
\begin{equation}
(-1)^\alpha \int \frac{du}{(u^2-\Omega^2)u^{\alpha_0} \prod_{i=1}^N
|u^2-\sigma_i^2|^{\alpha_i}}- (-1)^\beta \int
\frac{dv}{(\Omega^2-v^2)|v|^{\alpha_0} \prod_{i=1}^N
|v^2-\sigma_i^2|^{\alpha_i}}=\sqrt{2} \gamma_1
\label{eq:k2orbits}\quad .
\end{equation}
Linearization of (\ref{eq:k2orbits}) respectively near the point
$C$ reads:
\begin{equation}
\sqrt{2} \gamma_1= (-1)^\alpha  \int \, du \, \frac{C_u(\sigma_1,
\dots , \sigma_{n+m})}{(u-\sigma_k)^{\alpha_k}}- (-1)^\beta \int
\, dv \, \frac{C_v(\sigma_1, \dots ,
\sigma_{n+m})}{(v-\sigma_j)^{\alpha_j}} \label{eq:k2linC}\quad ,
\end{equation}
where we have defined: {\footnotesize\begin{eqnarray*}
C_u(\sigma_1, \dots ,
\sigma_{n+m})&=&\frac{1}{(\sigma_k^2-\Omega^2)\sigma_k^{\alpha_0}\prod_{i=1}^{k-1}(\sigma_k^2-\sigma_i^2)^{\alpha_i}
(\sigma_{k+1}^2-\sigma_k^2)^{\alpha_{k+1}}\prod_{i=k+2}^{n+m}(\sigma_i^2-\sigma_k^2)^{\alpha_i}(2\sigma_k)^{\alpha_k}}
 \\
C_v(\sigma_1, \dots ,
\sigma_{n+m})&=&\frac{1}{(\Omega^2-\sigma_j^2)\sigma_j^{\alpha_0}\prod_{i=1}^{j-1}(\sigma_j^2-\sigma_i^2)^{\alpha_i}
(\sigma_{j+1}^2-\sigma_j^2)^{\alpha_{j+1}}\prod_{i=j+2}^{n+m}(\sigma_i^2-\sigma_j^2)^{\alpha_i}(2\sigma_j)^{\alpha_j}}
\end{eqnarray*}}
Likewise, linearization of (\ref{eq:k2orbits}) respectively near
the point $B$ reads:
\begin{eqnarray}
\sqrt{2}\gamma_1&=& (-1)^\alpha \int \, du \, \frac{B_u(\sigma_1,
\dots ,
\sigma_{n+m})}{(\sigma_{k+1}-u)^{\alpha_{k+1}}}-(-1)^\beta\int \,
dv \, \frac{B_v(\sigma_1, \dots ,
\sigma_{n+m})}{(\sigma_{j+1}-v)^{\alpha_{j+1}}}
\label{eq:k2linB}\,\, ,
\end{eqnarray}
where now we have {\footnotesize \begin{eqnarray*} B_u(\sigma_1,
\dots , \sigma_{n+m})&=&\frac{1}{(\sigma_{k+1}^2-\Omega^2)
\sigma_{k+1}^{\alpha_0}\prod_{i=1}^{k-1}
(\sigma_{k+1}^2-\sigma_i^2)^{\alpha_i}
(\sigma_{k+1}^2-\sigma_k^2)^{\alpha_{k}}\prod_{i=k+2}^{n+m}
(\sigma_i^2-\sigma_{k+1}^2)^{\alpha_i}(2\sigma_{k+1})^{\alpha_{k+1}}}
\\ B_v(\sigma_1, \dots ,
\sigma_{n+m})&=&\frac{1}{(\Omega^2-\sigma_{j+1}^2)\sigma_{j+1}^{\alpha_0}\prod_{i=1}^{j-1}
(\sigma_{j+1}^2-\sigma_i^2)^{\alpha_i}
(\sigma_{j+1}^2-\sigma_j^2)^{\alpha_{j}}\prod_{i=j+2}^{n+m}(\sigma_i^2-\sigma_{j+1}^2)^{\alpha_i}
(2\sigma_{j+1})^{\alpha_{j+1}}}\,\, .
\end{eqnarray*}}
In (\ref{eq:k2linC}) and (\ref{eq:k2linB}), $C_u>0$, $C_v>0$,
$B_u>0$,$B_v>0$ and the pertinent integrals are simple although we
must distinguish between the cases $\alpha_r=1$ or $\alpha_r>1$
with $r=j,j+1,k,k+1$. Whichever the case, the limit $\gamma_1
\rightarrow \infty$ in (\ref{eq:k2linC}) and (\ref{eq:k2linB})
provides us, for instance in the case $\alpha=\beta=1$, with the
orbit $v=\sigma_j$ in (\ref{eq:k2linC}), and $u=\sigma_{k+1}$ in
(\ref{eq:k2linB}), while the limit $\gamma_1 \rightarrow -\infty$
yields the orbit $v=\sigma_{j+1}$ in (\ref{eq:k2linB}) and
$u=\sigma_{k}$ in (\ref{eq:k2linC}). This means that the
combination ${K}_1^{BD}+{K}_1^{DC}$ of singular kinks arises as
$\gamma_1\rightarrow -\infty$ in the moduli space of $K_2^{BC}$
kinks while the combination ${\rm K}_1^{BA}+{ K}_1^{AC}$ lives at
the $\gamma_1=\infty$ boundary of the $K_2^{BC}$ kink moduli
space. The same analysis can be tediously repeated in the rest of
the cases. In sum, we find that:
\[
\lim_{|\gamma_1| \rightarrow \infty} K_2^{AD}(\gamma_1)=\left\{
\begin{array}{l} K_1^{AC}+K_1^{CD} \\ K_1^{AB}+K_1^{BD}
\end{array}\right. \hspace{1cm} , \hspace{1cm} \lim_{|\gamma_1| \rightarrow \infty}
K_2^{BC}(\gamma_1)=\left\{
\begin{array}{l} K_1^{BD}+K_1^{DC} \\ K_1^{BA}+K_1^{AC}
\end{array}\right. \qquad .
\]
The $\gamma_1$ parameter, giving the orbit in the mechanical
problem, measures the distance between the centers in the
combination of basic kinks. Finite values of the parameter
characterize the non-linear combinations of single kinks as the
finite separation between lump centers. The kink families
$K_2^{AD}(\gamma_1)$ and $K_2^{CB}(\gamma_1)$ are formed by
different compositions of two kinks at different distances that
become the basic or singular kinks when the inter-center distance
is infinite.

\item In type 2 cells there are also two families of kink orbits
but one of the families is of a new type. If ${\rm
sign}((-1)^{\alpha+\beta})=+1$, a $K_2^{BC}(\gamma_1)$ family
arises with identical properties to the $K_2^{BC}(\gamma_1)$
family in type 1 cells. $B$ and $C$ are respectively stable and
unstable nodes, or viceversa, and $A$ is a saddle point, see
Figure 8(b).

The novelty comes from the fact that in this type of cell, $D$ is
a focus, $F$, that does not belong to ${\cal M}$ because it is not
a fixed point of the trajectory flow. $F$ not being a fixed point,
neither the $K_2^{AF}(\gamma_1)$ nor the $K_2^{FA}(\gamma_1)$ kink
orbits -expected when ${\rm sign}((-1)^{\alpha+\beta})=-1$\, and
$A$ is either a stable or unstable node- can exist. In Figure 8(b)
the apparent homoclinic kink trajectories $K_4^{AA}(\gamma_1)$ are
plotted in blue. These trajectories correspond to one
$K_2^{AF}(\gamma_1)$ and one $K_2^{FA}(\gamma_1)$ kink orbit glued
at the focus. One might think that such orbits fail to be
derivable at the focus and consequently that their action is
infinite, yielding field theoretical solutions of infinite energy.
To examine this issue closely, it is convenient to go back to the
Cartesian plane.

Recall that the inverse map from ${\mathbb E}^2$ to ${\mathbb R}^2$
induced by the change to elliptic coordinates is one-to-two, see
(\ref{eq:defelip})): ${\mathbb E}^2$ is mapped into the upper
half-plane $\phi_2>0$ by $\xi^{+*}$ and also into the lower
half-plane $\phi_2<0$ by $\xi^{-*}$. Therefore, the inverse image of
the point $A\equiv (\sigma_{n-1},\sigma_{n+1})$ is the two-element
set: $A^\pm\equiv (\frac{\sigma_{n-1}\sigma_{n+1}}{\Omega},\pm
\frac{\sqrt{(\sigma_{n+1}^2-\Omega^2)(\Omega^2-\sigma_{n-1}^2)}}{\Omega})$.
The $K_2^{AF}(\gamma_1)$ can be chosen to be mapped in the upper
half-plane of the Cartesian plane ${\mathbb R}^2$ or viceversa.
These two choices give rise to a family of continuous and derivable
kink trajectories connecting the stable node $A^+$ with the unstable
node $A^-$ (or viceversa when $A^-$ is stable and $A^+$ is unstable)
{\footnote{ If the two $K_2^{AF}$, $K_2^{FA}$ families of the
connected sum $K_4^{AA}(\gamma_1)=K_2^{AF}(\gamma_1)\bigvee
K_2^{FA}(\gamma_1)$ are mapped into the same half-plane of ${\mathbb
R}^2$, the kink trajectories fail to be derivable at the focus and
develop infinite action.}}. Therefore, every kink trajectory of this
type is of finite energy, heteroclinic despite of appearing to be
homoclinic in ${\mathbb E}^2$, and crosses from the upper half-plane
to the lower half plane or viceversa through a single point: the
focus $F$. The last point confirms that $F$ is a conjugate point to
both $A^+$ and $A^-$ and, as anticipated, these trajectories are
unstable according to the Morse index theorem. A laborious analysis,
but conceptually identical to that shown for type 1 cells, shows
that at the $|\gamma_1|\rightarrow\infty$ limit the following kink
combination arises:
\[
\lim_{|\gamma_1| \rightarrow \infty} K_4^{AA}(\gamma_1)=
K_1^{AC}+K_2^{CF\bigvee FB}+K_1^{BA} \qquad \qquad ,
\]
where by $K_2^{CF\bigvee FB}$ we denote the two-step singular kink
orbit living respectively at the edges $u=\Omega$, and $v=\Omega$
in two different periods of time, the second period synchronized
with the previous one at the focus $F\equiv (\Omega, \Omega)$.
Note that
\[
K_2^{CF\bigvee FB}=\lim_{|\gamma_1|\rightarrow\infty}
K_2^{CB}(\gamma_1)
\]
and the two last limits above account for the subscript $4$ in the
nomenclature. In ${\mathbb R}^2$, the $K_2^{CF\bigvee FB}$ kink
orbit lives on the $\phi_2=0$ axis.

The translation from this analysis in the analogous mechanical
system to the field theory is as follows: 1) To this
one-parametric family of finite-action heteroclinic kink orbits
corresponds a one-parametric family of finite-energy topological
kinks. 2) Because the kink orbits are unstable, the topological
kinks are also unstable. 3) These kinks are composed of four basic
kinks.

\item In type 3 cells only two of the four vertices, $A\equiv
(-\Omega,\sigma_2)$, $B\equiv (\Omega,\sigma_2)$, are elements of
${\cal M}$ whereas the other two vertices are the foci,
$F_\pm\equiv (\pm \Omega,\Omega)$, of the elliptic coordinate
system, that do not belong to ${\cal M}$. If ${\rm
sign}((-1)^{\alpha+\beta})=+1$, $A$ is a stable or unstable node
whereas $B$ is a saddle point. When ${\rm
sign}((-1)^{\alpha+\beta})=-1$ the r$\hat{\rm o}$les of $A$ and
$B$ are exchanged. Therefore, except for singular kink orbits all
the generic kink trajectories cross one of the two foci, which are
not fixed points but display indefinite flow.

For these reasons, the features of the kink trajectories are
better understood in the Cartesian plane. In ${\mathbb R}^2$ the
vertices of the cell become: $A\equiv (-\sigma_2,0)$, $B\equiv
(\sigma_2,0)$, and $F_\pm\equiv (\pm\Omega,0)$. Kink trajectories
in the $K_2^{AA}(\gamma_1)$ blue family start from $A$, run along
the upper or the lower half-plane, meet at the $F_+$ focus, where
they pass to the other half-plane, and come back to $B$. This tour
is performed either clock- or counter-clockwise. The red kink
trajectories in Figure 8(c) are identical to the blue kink
trajectories of the same figure, but $B$ is the starting and
ending point, and the bridge between half-planes is the $F_-$
focus. The kink orbits in type 3 cells are thus unstable
homoclinic trajectories: each trajectory starts from either $A$ or
$B$ and ends at either $A$ or $B$. The foci $F_+$ and $F_-$ are
respectively conjugate points to $A$ and $B$ where all the members
of the family meet. Again, it is hard but merely routine work to
prove that the $|\gamma_1|=\infty$ limit is the following
combination of singular kinks:
\begin{eqnarray*}
&&\lim_{|\gamma_1| \rightarrow \infty} K_2^{AA}(\gamma_1)=
K_1^{AB}+{K}_1^{AF_-\bigvee F_-F_+\bigvee F_+B} \qquad  \qquad ,\\
&& \lim_{|\gamma_1| \rightarrow \infty} K_4^{BB}(\gamma_1)=
K_1^{BA}+{K}_1^{BF_+\bigvee F_+F_-\bigvee F_-A}  \qquad \qquad .
\end{eqnarray*}
${K}_1^{AF_-\bigvee F_-F_+\bigvee F_+B}$ denotes the three-step
singular kink orbit living respectively at the edges $v=-\Omega$,
$u=\Omega$, and $v=\Omega$ in three different periods of time, the
second period synchronized with the previous and later periods at
the foci. Until now we have met only kinks for which the number of
the steps coincides with the number of lumps. For instance, the kink
orbit $K_2^{CF\bigvee FB}$ runs over two edges, and even though the
energy density of these stable kinks is formed by a single lump, it
is exactly the superposition of one $K_1^{AC}$ and one $K_1^{CB}$
kinks with coinciding centers. ${K}_1^{AF_-\bigvee F_-F_+\bigvee
F_+B}$ kinks differ in two ways: 1) Combinations of three $K_1^{AB}$
do not form these kinks. 2) ${K}_1^{AF_-\bigvee F_-F_+\bigvee F_+B}$
kinks are unstable because upon crossing one of the two foci they
pass a conjugate point of the congruence of kink orbits they belong
to. In sum, there are no reasons for this configuration to be
considered as a composite kink and, moreover, ${K}_1^{AF_-\bigvee
F_-F_+\bigvee F_+B}$ kinks are unstable.

The translation from this analysis in the analogous mechanical
system to the field theory is as follows: 1) To this
one-parametric family of finite action homoclinic kink orbits
corresponds a one-parametric family of finite energy
non-topological kinks{\footnote{Amazingly, the non-topological
kinks of the MSTB model were discovered at the very beginning of
the search for solitary waves in field theories with two real
fields. Of all possible kinks that we have just described these
are by far the most bizarre.}}. 2) Because the kink orbits are
unstable the non-topological kinks are unstable. 3) These kinks
are composed of one basic kink and one unstable singular kink, see
\cite{J2}.

\end{itemize}

The previous results can be summarized as follows:

\vspace{0.1cm} {\bf Proposition 3.} {\it There are families of
kink orbits confined in each cell in the elliptic plane. The kink
trajectories connect elements of ${\cal M}$ at opposite cell
vertices. We shall denote these generic kink orbits by labelling
the vertices connected by them: $K_2^{AD}(\gamma_1)$,
$K_2^{BC}(\gamma_1)$. There are two special cases: either one or
two cell vertices are not zeroes of $V(\phi_1,\phi_2)$. In type
I-1a and type I-1b generalized MSTB models type 2 cells arise: in
these cells there is a family of topological kink orbits
connecting the elements $A^+=\xi^+ (\sigma_{n-1},\sigma_{n+1})$
and $A^-=\xi^- (\sigma_{n-1},\sigma_{n+1})$ of ${\cal M}$ in
${\mathbb R}^2$; every orbit in this family passes through one of
the two foci. In $n=1$ type I-1a generalized MSTB models there is
one cell of type 3; in this cell there is a family of
non-topological kink orbits joining the points
$A=(-\Omega,\sigma_2)$ or $B=(\Omega,\sigma_2)$ with themselves.
Each non-topological kink orbit crosses one of the two $F_\pm$
foci.}

\subsection{Sum rules of kink masses}

The action of the mechanical trajectory is the energy of the static
field profile. Therefore,
\begin{eqnarray*}
{E}[(\phi_1,\phi_2]&=& \int_{\xi^{*}\phi}  \hspace{-0.2cm} du \,\,
{\rm sign}({u}')
\sqrt{\frac{2(F+E u^2+f(u))}{u^2-\Omega^2}} + \\
&+& \int_{\xi^{*}\phi} \hspace{-0.2cm} dv \, \,{\rm sign} ({v}')
\sqrt{\frac{2(-F-E v^2+g(v))}{\Omega^2-v^2}}-\left. E
x\right|_{\xi^{*} \phi} \qquad ,
\end{eqnarray*}
which is the Hamilton principal function -the action- of the
mechanical trajectories, is also the energy of the static field
solutions. Solitary wave solutions have finite energy
$E[\phi_1,\phi_2]$ and correspond to trajectories with zero
mechanical energy and zero separation constant: $E=F=0$.

\vspace{0.1cm}

{\bf Proposition 4.} {\it The energy of the kinks in generalized
MSTB models is the mechanical action of the kink orbits:
\begin{equation}
{E}[\phi_1,\phi_2]= \left| \int_{\pi_u (\xi^{*} \vec{\phi}) }
\hspace{-0.6cm} du \hspace{0.2cm} \sqrt{\frac{2 f(u)}{u^2-\Omega^2}}
\right|+ \left| \int_{\pi_v (\xi^{*} \vec{\phi}) } \hspace{-0.6cm}
dv \hspace{0.2cm} \sqrt{\frac{2 g(v)}{\Omega^2-v^2}} \right| \qquad
,\label{eq:ener1}
\end{equation}
where $\pi_u(\xi^{*} \vec{\phi})$ and $\pi_v(\xi^{*} \vec{\phi})$
are respectively the projections of the point $\xi^{*} \vec{\phi}$
of the kink orbit into the axes $u$ and $v$.}

\vspace{0.1cm}

Therefore, all the members of the kink family are iso-energetic.
Formula (\ref{eq:ener1}) allows us to state the following:

\begin{enumerate}

\item The kink energy does not depend on the fine details of the solitary waves:
kinks with very different profiles share the same energy. The kink
energy is determined only by the projections of the kink
trajectories into the axes.

\item  The important implication of this is that the energy of any
member of a family of composite kinks is uniquely determined as
the sum of the energies of the basic or singular kinks. The total
energy of the singular kinks is:
\begin{equation}
{E}[K_1^{AB}]={E}[K_1^{CD}]= \left| \int_{\sigma_j}^{\sigma_{j+1}}
\hspace{-0.6cm} dv \hspace{0.1cm}
\sqrt{\frac{2g(v)}{\Omega^2-v^2}}\,\right|; \hspace{0.3cm} {
E}[K_1^{AC}]={E}[K_1^{BD}]= \left| \int_{\sigma_k}^{\sigma_{k+1}}
\hspace{-0.6cm} du \hspace{0.1cm} \sqrt{\frac{2f(u)}{u^2-\Omega^2}}
\,\right|
 \label{eq:ener}
\end{equation}
A little more care is needed when the foci are crossed by a
congruence of kink trajectories. Here, line integrals must be
computed piecewise: from the starting point to the focus and from
the focus to the endpoint. For instance, the energy of ${\rm
K}_2^{CF\bigvee FB}$ kinks in Type I-1a and I-1b generalized MSTB
models is:
\begin{equation}
{E}[K_2^{CF\bigvee FB}]= \left| \int_{\sigma_{n-1}}^{\Omega}
\hspace{-0.4cm} dv \hspace{0.1cm}
\sqrt{\frac{2g(v)}{\Omega^2-v^2}}\,\right| +\left|
\int_{\Omega}^{\sigma_{n+1}} \hspace{-0.6cm} du \hspace{0.1cm}
\sqrt{\frac{2f(u)}{u^2-\Omega^2}} \,\right| \qquad .
 \label{eq:ener5}
\end{equation}
Simili modo, the energy of unstable singular kinks
${K}_1^{AF_-\bigvee F_-F_+\bigvee F_+A}$ in  type 3 cells of $n=1$
Type I-1a generalized MSTB models is also computed piecewise:
\begin{equation}
 E[{K}_1^{AF_-\bigvee F_-F_+\bigvee F_+B}]= \left|
\int_{-\Omega}^{\Omega} \hspace{-0.3cm} dv \hspace{0.1cm}
\sqrt{\frac{2g(v)}{\Omega^2-v^2}}\,\right| +2 \left|
\int_{\Omega}^{\sigma_{2}} \hspace{-0.3cm} du \hspace{0.1cm}
\sqrt{\frac{2f(u)}{u^2-\Omega^2}} \,\right| \label{eq:ener6} \qquad
.
\end{equation}

\item Kinks or combinations of kinks that have the same
projections on the elliptic axes have the same energy. This is the
crux of the matter of the kink mass sum rules. We now write all
the kink mass sum rules observed in the generalized MSTB models.
These sum rules connect the energy of the kinks solutions confined
in a cell, and hence we must distinguish three possible cases:
\begin{eqnarray*}
 & \bullet& \mbox{Kinks confined in cells of Type 1:} \\
&& \hspace{0.7cm} E[K_1^{AB}]=E[K_1^{CD}] \hspace{0.5cm} ;
\hspace{0.5cm} E[K_1^{AC}]=E[K_1^{BD}]  \\ && \hspace{0.7cm}
E[K_2^{AD}(\gamma_1)]= E[K_2^{BC}(\gamma_1)]=E[K_1^{AB}]+E[K_1^{AC}]
\\  & \bullet& \mbox{Kinks confined in cells of Type 2:} \\
&& \hspace{0.7cm} E[K_2^{CF\bigvee FB}]=E[K_1^{AB}]+E[K_1^{AC}]=E[K_2^{CB}(\gamma_1)]  \\
&& \hspace{0.7cm} E[K_4^{AA}(\gamma_1)]=2 E[K_1^{AB}]+2 E[K_1^{AC}]
\\
& \bullet & \mbox{Kinks confined in cells of Type 3:} \\
&& \hspace{0.7cm} E[K_2^{AA}(\gamma_1)]=E[K_1^{AB}]
+E[{K}_1^{AF_-\bigvee F_-F_+\bigvee F_+B}]=E_2^{BB}(\gamma_1) \\
&&\hspace{0.7cm} E_2^{BB}(\gamma_1)=E[K_1^{BA}]+E[{K}_1^{BF_+\bigvee
F_+F_-\bigvee F_-A}]
\end{eqnarray*}
These kink mass sum rules are due to the Hamilton-Jacobi
separability in elliptic coordinates of the mechanical systems
ruling finite energy static solutions in the generalized MSTB
models.

\end{enumerate}

\subsection{Kink orbits as gradient flow lines}

It is interesting to address the results of the previous sub-Section
from the point of view sketched in the Introduction. The
Bogomolny'i-Prasad-Sommerfield understanding of several types of
topological defects in field theory, such as kinks, special classes
of vortices and magnetic monopoles, and instantons, can be applied
to some solitary waves of generalized MSTB models. We recall that
the BPS approach is possible if the potential energy density is
equal to half the norm of the square of the norm of a function
$W(\phi_1,\phi_2)$, see (\ref{eq:hjw}):
\begin{equation}
U(\phi_1,\phi_2)={1\over 2}\left(\frac{\partial
W}{\partial\phi_1}\cdot\,\frac{\partial
W}{\partial\phi_1}+\frac{\partial
W}{\partial\phi_2}\cdot\,\frac{\partial
W}{\partial\phi_2}\right)={1\over
2}\vec{\nabla}W\cdot\vec{\nabla}W\qquad . \label{eq:hjw1}
\end{equation}
Writing the static energy in the form (\ref{eq:enersu})
\begin{equation}
 E[\vec{\phi}]=\frac{1}{2} \int dx  \left[ \frac{d
\vec{\phi}}{dx}-\vec{\nabla} W \right]\cdot \left[ \frac{d
\vec{\phi}}{dx}-\vec{\nabla} W \right]+ \int dW \qquad .
\label{eq:enersuv}
\end{equation}
one sees that solutions of the first-order ODE system
\begin{equation}
\frac{\vec{d\phi}}{dx}=\vec{\nabla}W(\phi_1,\phi_2)=\frac{\partial
W}{\partial\phi_1}\vec{e}_1+\frac{\partial
W}{\partial\phi_2}\vec{e}_2 \label{eq:grf}
\end{equation}
are absolute minima of the energy if $W$ behaves well enough. Stable
kink orbits are thus the flow lines induced by the gradient of
$W${\footnote{The minima of $U$, the elements of ${\cal M}$, are
either minima, maxima, or saddle points of $W$. Accordingly, the
critical points of $W$ are either stable nodes, unstable nodes, or
saddle points of the gradient flow.}}, usually referred to as the
superpotential in physicists' literature, because of the central
r$\hat{\rm o}$le that this function plays in supersymmetric models.
A closer look to (\ref{eq:hjw1}) reveals that $W$ is no more than a
solution of the time-independent Hamilton-Jacobi equation of the
analogous mechanical system with zero mechanical energy: i.e., the
superpotential is the $E=0$ Hamilton characteristic function in
Hamiltonian dynamical systems.

For mechanical potential energies of Liouville Type I systems,
(\ref{eq:hjw1}) in elliptic coordinates reads:
\begin{equation}
\frac{1}{2(u^2-v^2)} \left\{ (u^2-\Omega^2) \left( \frac{\partial
W}{\partial u} \right)^2+(\Omega^2-v^2) \left( \frac{\partial
W}{\partial v} \right)^2 \right\}=\frac{1}{u^2-v^2}\left( f(u)+g(v)
\right) \qquad . \label{eq:hjwe}
\end{equation}
The separation ansatz $W(u,v)=W_1(u)+W_2(v)$ reduces the PDE
(\ref{eq:hjwe}) to two independent first-order differential
equations
\[
\frac{d W_1}{d u}=(-1)^\alpha \sqrt{\frac{2 f(u)}{u^2-\Omega^2}}
\hspace{2cm} \frac{d W_2}{d v}=(-1)^\beta \sqrt{\frac{2
g(v)}{\Omega^2-v^2}} \qquad ,
\]
which can be integrated by quadratures:
\begin{equation}
W^{(\alpha,\beta)}(u,v)= (-1)^\alpha \int du \sqrt{\frac{2
f(u)}{u^2-\Omega^2}}+(-1)^\beta \int dv \sqrt{\frac{2
g(v)}{\Omega^2-v^2}} \qquad , \qquad \alpha,\beta=0,1
\label{eq:lipsup} \qquad .
\end{equation}

\vspace{0.1cm}

{\bf Proposition 5:} {\it Generalized MSTB models are built from the
complete integral -(\ref{eq:lipsup})- of the time-independent
Hamilton-Jacobi equation with zero mechanical energy for the
analogous mechanical system. In modern language, we characterize the
generalized MSTB model as those admitting four different
superpotentials according to the freedom of choice of signs in
(\ref{eq:lipsup})}.

More precisely,

{\bf Proposition 6:} {\it  In ${\mathbb E}^2$ the four
superpotentials of generalized MSTB models have the form:
\begin{equation}
\xi_\Omega^{*} W=W_1(u)+W_2(v) \hspace{1cm} \equiv \hspace{1cm}
\frac{\partial^2 (\xi_\Omega^{*} W)}{\partial u \,\partial v}=0
\label{eq:essep} \qquad .
\end{equation}
In ${\mathbb R}^2$ the PDE (\ref{eq:essep}) becomes the cumbersome
PDE satisfied by generalized MSTB superpotentials:}
\begin{equation}
\phi_1 \phi_2 \left( \frac{\partial^2 W}{\partial \phi_1 \partial
\phi_1}-\frac{\partial^2 W}{\partial \phi_2 \partial \phi_2}\right)+
(\phi_2 \phi_2-\phi_1 \phi_1+\Omega^2) \frac{\partial^2 W}{\partial
\phi_1 \partial \phi_2}+\phi_2 \frac{\partial W}{\partial
\phi_1}-\phi_1 \frac{\partial W}{\partial \phi_2}=0 \qquad .
\label{eq:curl}
\end{equation}
The concept of superpotential allows us to write the invariants
(\ref{eq:ordinary}) of the mechanical system  in the surprisingly
simpler form:
\begin{eqnarray}
I_1&=&\frac{1}{2} \left(\frac{d \phi_1}{dx}\right)^2 + \frac{1}{2}
\left(\frac{d \phi_2}{dx}\right)^2 -  \frac{1}{2} \left(
\frac{\partial W}{\partial \phi_1} \right)^2 - \frac{1}{2}
\left(\frac{\partial W}{\partial \phi_2}\right)^2 \label{eq:i1w}
\\ I_2&=&\frac{1}{2} \left[ \left(\phi_2 \frac{d \phi_1}{dx} - \phi_1
\frac{d \phi_2}{dx} \right)^2- \Omega^2 \frac{d \phi_2}{dx} \frac{d
\phi_2}{dx}  - \left(\phi_2 \frac{\partial W}{\partial \phi_1} -
\phi_1 \frac{\partial W}{\partial \phi_2}\right)^2+ \Omega^2
\frac{\partial W}{\partial \phi_2}\frac{\partial W}{\partial \phi_2}
\right] \label{eq:i2w} \quad ,
\end{eqnarray}
explicitly showing that $I_1=0=I_2$ for the solutions of
(\ref{eq:grf}).

Obviously, because four superpotentials are available there are four
different ways of saturating the BPS bound \cite{BPS}. For instance,
in ${\mathbb E}^2$ these equations are:
\begin{equation}
\frac{du}{dx}=(-1)^\alpha \frac{\sqrt{2 (u^2-\Omega^2)
f(u)}}{u^2-v^2} \hspace{1cm} , \hspace{1cm} \frac{dv}{dx}=(-1)^\beta
\frac{\sqrt{2 (\Omega^2-v^2) g(v)}}{u^2-v^2} \label{eq:lipsup2}
\qquad .
\end{equation}
It is clear that for kink orbit solutions of (\ref{eq:lipsup}) the
kink energy
\[
E[\vec{\phi}_K]=\int_K \, dW \qquad \qquad ,
\]
the integral of the exterior derivative of the superpotential along
the kink trajectories, is precisely given by formulas
(\ref{eq:ener1}), (\ref{eq:ener}), (\ref{eq:ener5}), and
(\ref{eq:ener6}) for the different types of kinks. If the kink orbit
does not cross points where $W$ is not differentiable Stoke's
theorem rules that
\[
E[\vec{\phi}_K]=\left|W(\vec{\phi}_I)-W(\vec{\phi}_J)\right| \qquad
\quad ,
\]
and the energy of stable kinks is a topological charge. The
situation is more subtle for kink orbits that cross non-fixed
conjugate points. Stoke's theorem must be applied piece-wise, and we
find:
\[
E[\vec{\phi}_K]=\left|W(\vec{\phi}_I)-W(\mp
\Omega\vec{e}_1))\right|+\left|W(\mp\Omega\vec{e}_1)-W(\vec{\phi}_J)\right|
\qquad , \qquad
E[\vec{\phi}_K]=2\left|W(\vec{\phi}_I)-W(\mp\Omega\vec{e}_1)\right|
\, \, \, ,
\]
respectively for unstable kink orbits in Type 2 and Type 3 cells.
The important point is that these kink orbits are not solutions of
(\ref{eq:lipsup2}) for the same choice of $\alpha$ and $\beta$ over
the whole real line ${\mathbb R}$. The unstable trajectories solve
(\ref{eq:lipsup}) for a given $\beta$ relative to $\alpha$ between
$x=-\infty$ and $x=x_0$, the point where the focus is reached.
Between $x_0$ and $x=+\infty$, the first-order equations satisfied
correspond to the other relation between $\beta$ and $\alpha$. The
unstable kink orbits, however, satisfies the static second-order
equations (\ref{eq:ordinary}) but the unstable kinks do not saturate
the BPS bound.

We show explicitly  that the existence of four superpotentials is
connected to the Hamilton-Jacobi separability of the analogous
mechanical system to generalized MSTB models in elliptic
coordinates. Formulas (\ref{eq:cinco}) and (\ref{eq:seis}) for the
two invariants in elliptic coordinates due to HJ separability
provide the following identities for the square of $u$ and $v$
derivatives:
\begin{eqnarray*}
\left(\frac{du}{dx} \right)^2&=&\frac{2 (u^2-\Omega^2) \left[ -I_2+
I_1 (u^2-\Omega^2)+f(u) \right]}{(u^2-v^2)^2} \\ \left(\frac{dv}{dx}
\right)^2&=&\frac{2 (\Omega^2-v^2) \left[ I_2+I_1
(\Omega^2-v^2)+g(v) \right]}{(u^2-v^2)^2} \qquad .
\end{eqnarray*}
The separatrix trajectories between bound and unbound motion arise
for $I_1=0$, $I_2=0$, values for which the above equations reduce
to:
\begin{equation}
\left(\frac{du}{dx} \right)^2=\frac{2 f(u) (u^2-\Omega^2)
}{(u^2-v^2)^2} \hspace{1cm} , \hspace{1cm} \left(\frac{dv}{dx}
\right)^2=\frac{2 g(v) (\Omega^2-v^2) }{(u^2-v^2)^2} \qquad ,
\label{eq:ecu88}
\end{equation}
equivalent to (\ref{eq:lipsup2}) with the four sign combinations.
Complete Arnold-Liouville integrability is accompanied by HJ
separability, allowing us to obtain all the separatrix trajectories
- all the kink profiles- from the complete solution of the HJ
equation: the four superpotentials. The partial solvability of the
HJ equation would permit only a partial identification of the kink
manifold. The complete solution of the HJ equation for arbitrary
values (non-zero) of the invariants,
\[
W^{(\alpha,\beta)}=(-1)^\alpha \int du
\sqrt{\frac{2(-I_2+I_1(u^2-\Omega^2)+f(u))}{u^2-\Omega^2}} \,\,+
(-1)^\beta \int du
\sqrt{\frac{2(I_2+I_1(\Omega^2-v^2)+g(v))}{\Omega^2-v^2}}
\]
also reduces to algebraic equations between quadratures all the
other trajectories - either periodic or unbounded- of the analogous
mechanical system. Back in field theory, the periodic orbits are
interpreted as (unstable) kinks on a circle (the space-time being
${\mathbb R}\times {\mathbb S}^1$ with hyperbolic metric) whereas
unbound mechanical motion corresponds to static field solutions of
infinite energy.

To finish this Section and a thorough study of generalized MSTB
models we now address this issue in ${\mathbb R}^2$, i.e., we
analyze the $I_1=0=I_2$ condition in Cartesian coordinates. The
surprise is that not only it is satisfied by solutions of the
first-order ODE system (\ref{eq:grf}), but that solutions of the new
first-order ODE system {\large\begin{eqnarray} \frac{d
\phi_1}{dx}&=&\mp\frac{(\Omega^2-\phi_1^2+\phi_2^2) \frac{\partial
W}{\partial \phi_1} -2 \phi_1 \phi_2 \frac{\partial W}{\partial
\phi_2}}{\sqrt{ [ (\phi_1-\Omega)^2+\phi_1^2]
 [(\phi_1+\Omega)^2+\phi_2^2]}} \nonumber \\ \frac{d \phi_2}{dx}&=&\pm\frac{(\Omega^2-\phi_1^2+\phi_2^2)
\frac{\partial W}{\partial \phi_2} +2 \phi_1 \phi_2 \frac{\partial
W}{\partial \phi_1}}{\sqrt{ [ (\phi_1-\Omega)^2+\phi_1^2]
 [(\phi_1+\Omega)^2+\phi_2^2]}}
\label{eq:bogosegundo}
\end{eqnarray}}
also comply with $I_1=0=I_2$!. This implies the existence of a
second superpotential $\tilde{W}$. The vector field from the right
hand sides of (\ref{eq:bogosegundo}):
\[
\vec{A}(\phi_1,\phi_2)=\mp\left[\frac{(\Omega^2-\phi_1^2+\phi_2^2)
\frac{\partial W}{\partial \phi_1} -2 \phi_1 \phi_2 \frac{\partial
W}{\partial \phi_2}}{\sqrt{ [ (\phi_1-\Omega)^2+\phi_1^2]
 [(\phi_1+\Omega)^2+\phi_2^2]}}\vec{e}_1-\frac{(\Omega^2-\phi_1^2+\phi_2^2)
\frac{\partial W}{\partial \phi_2} +2 \phi_1 \phi_2 \frac{\partial
W}{\partial \phi_1}}{\sqrt{ [ (\phi_1-\Omega)^2+\phi_1^2]
 [(\phi_1+\Omega)^2+\phi_2^2]}}\vec{e}_2\right]\quad ,
\]
despite its horrible aspect, is curl-less because equation
(\ref{eq:curl}) is valid in Type I Liouville models. Therefore,
Green's theorem ensures that there exists a function $\tilde{W}$
such that (\ref{eq:bogosegundo}) is equivalent to the first-order
ODE system: $\frac{d\phi_1}{dx}=
\frac{\partial\tilde{W}}{\partial\phi_1}$, $
\frac{d\phi_2}{dx}=\frac{\partial\tilde{W}}{\partial\phi_2}$. Four
different superpotentials, $\pm W$, $\pm\tilde{W}$, are also
available in Cartesian coordinates.

\section{Eighth-order generalized MSTB models}

The general analysis of the previous Section affords us a practical
method of obtaining the kink variety of generalized MSTB models by
means of a few computations. The references [16-28] are devoted to
the original MSTB model, in which the potential energy density is a
quartic polynomial in the two scalar fields. In the paradigm of
generalized MSTB models, the functions $f$ and $g$ are respectively:
$f(u)=\frac{1}{2} (u^2-1)^2(u^2-\Omega^2)$ and $g(v)=\frac{1}{2}
(v^2-1)^2(\Omega^2-v^2)$. Therefore, $\bar{\cal M}_g=\{
\sigma_{-1}=-\Omega; \sigma_{1}=\Omega \}$ and $\bar{\cal M}_f=\{
\sigma_{1}=\Omega ; \sigma_{2}=1 \}$ such that $m=1$ and $n=1$, a
type I-1a generalized MSTB model that involves only the type 3 cell
$C_{-11}^{12}$ in the reticulum $\cal{R}$. There are only two
singular kinks: $K_1^{AB}$ and ${K}_1^{AF_-\bigvee F_-F_+\bigvee
F_+B}$ (respectively denoted as TK2 and TK1 kinks in the
literature). The remaining kinks form a family of non-topological
(unstable) kinks:  $K_2^{AA}$ (usually referred to as NTK2 kinks).
In other papers, [29-31], a generalized MSTB model where the
potential energy density is a degree-six polynomial in the fields is
discussed. The $f$ and $g$ functions in this model are:
$f(u)=\frac{1}{2} u^2 (u^2-1)^2(u^2-\sigma^2)$ and $g(v)=\frac{1}{2}
v^2 (v^2-1)^2(\sigma^2-v^2)$. A type I-1b generalized MSTB model
with $m=1$ and $n=1$ arises. The set of zeroes of $g$ and $f$ are
respectively $\bar{\cal M}_g=\{ \sigma_{-1}=-\Omega; \sigma_0=0;
\sigma_{1}=\Omega \}$ and $\bar{\cal M}_f=\{ \sigma_{1}=\Omega;
\sigma_{2}=1 \}$ whereas the set of minima of the potential in
${\mathbb R}^2$ is, using the notation of \cite{Modeloa},  ${\cal
M}=\{ A_\pm =(\pm 1,0); B_\pm =(0,\pm \sqrt{1-\sigma^2}); O=(0,0)
\}$. The reticulum in ${\mathbb R}^2$ is formed by two type 2 cells
joined by the edge on the ordinate axis. There are three singular
kinks, denoted as $K_1^{AO}$, $K_1^{BO}$ and $K_1^{AB}$, and two
families of topological composite kinks, one of them stable,
$K_2^{AO}(\gamma_1)$, and the other one, $K_2^{BB}(\gamma_1)$,
unstable.

To close this article we shall study a two-parametric family of
generalized MSTB models with an eight-order polynomial as potential
energy density:
\begin{equation}
U(\phi_1,\phi_2) = (\phi_1^2+\phi_2^2-\sigma^2)^2 (\phi_1^2+
\phi_2^2 -\tau^2)^2+3 \Omega^2\phi_2^2\left( \phi_1^2+\phi_2^2
+\textstyle\frac{\Omega^2-2 \sigma^2-2\tau^2}{3}\right)^2 +\alpha
\phi_2^2+\Omega^2\tau^2\phi_2^4 \label{eq:meta1}
\end{equation}
where $\alpha=\frac{\Omega^2}{3} (2 \Omega^4-2 \sigma^2 \Omega^2- 2
\tau^2\Omega^2 +4 \Omega^2-\sigma^4-\tau^4)$ and
$\Omega^2=\sigma^2\tau^2$.
\begin{figure}[htb]
\centerline{
\includegraphics[height=4.5cm]{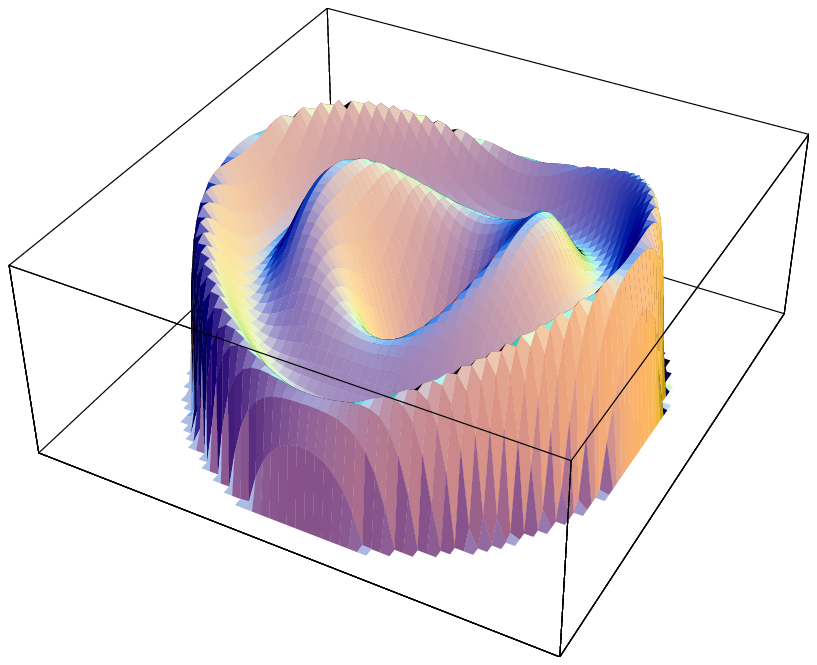}\hspace{1cm}
\includegraphics[height=4.5cm]{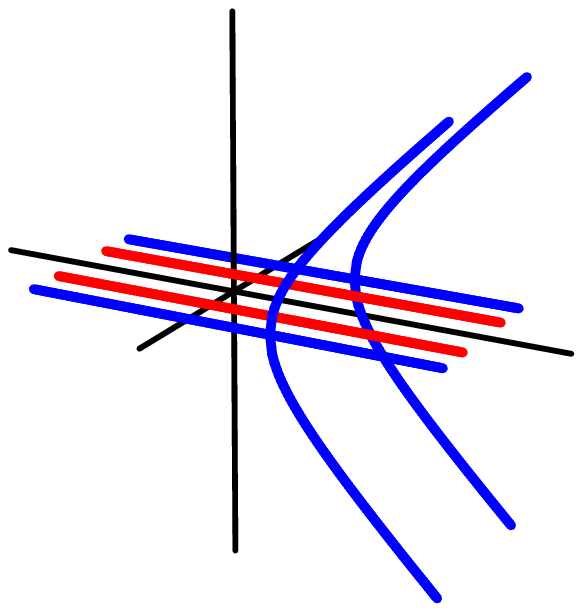}}
\caption{\small \textit{Potential density energy in the model ($\tau
< \sigma <1$) and bifurcation of the elements of ${\cal M}$.}}
\end{figure}

\noindent The parameters $\sigma$ and $\tau$ are the coupling
constants of these generalized MSTB models and the foci of the
elliptic coordinate system are determined from their product. The
four superpotentials are:
\begin{eqnarray}
W(\phi_1,\phi_2)&=&\pm\frac{\sqrt{2}}{12} \sqrt{(\phi_1 \pm
\Omega)^2 +\phi_2^2} \left[ 3(\phi_1^2+ \phi_2^2 \mp \Omega
\phi_1+\Omega^2 )
(\phi_1^2+ \phi_2^2 \pm \Omega \phi_1+\Omega^2 )- \right. \nonumber\\
& & \left. - 5(\sigma^2+\tau^2) (\phi_1^2 +\phi_2^2 \mp \Omega
\phi_1+\Omega^2 ) \mp 3 \Omega \phi_1 (\phi_1^2+
\phi_2^2+\Omega^2)+15 \Omega^2   \right] \label{eq:esup} \qquad .
\end{eqnarray}
The secret to finding a complete solution to the zero mechanical
energy Hamilton-Jacobi equation of the analogous mechanical system
is that the mechanical system is of Stackel type. There are two
invariants in involution -the mechanical energy, $I_1$, and a
generalized momentum, $I_2$- that are quadratic in the momenta:
{\small\begin{eqnarray} I_1 & =&  \frac{1}{2}\left(
\frac{d\phi_1}{dx}\right)^2+\frac{1}{2} \left( \frac{d\phi_2}{dx}
\right)^2-U(\phi_1,\phi_2)
\label{eq:meta2} \\
I_2 &=& \frac{1}{2} \left(
\phi_1\frac{d\phi_2}{dx}-\phi_2\frac{d\phi_1}{dx}\right)^2-\frac{\Omega^2}{2}
\left( \frac{d\phi_2}{dx}\right)^2-\Omega^2  \phi_2^2 (\phi_1^2+
\phi_2^2+\delta) \left[(\phi_1^2+
\phi_2^2-\kappa)^2+2\Omega^2\phi_2^2+\beta\right] \,
.\label{eq:meta3}
\end{eqnarray}}
The mechanical system is amenable to the separation of variables.
Here, $\beta=\Omega^2(\Omega^2-\sigma^2 -\tau^2 +\frac{3
}{2})-\frac{1}{4}(\sigma^4 +\tau^4)$,
$\delta=\Omega^2-\sigma^2-\tau^2$ and
$\kappa=\frac{\sigma^2+\tau^2}{2}$.

In the count and localization of the elements of ${\cal M}$ a
complicated process of bifurcation arises, depending on the values
of the parameters $\tau$ and $\sigma$. To analyze the most
interesting cases, we shall consider a fixed value of
$\sigma\in(0,1)$ and let allow $\tau$ to vary along the real line.
If both parameters are greater than or equal to one the model
presents only isolated singular kinks. There are three cases - 1)
$\tau<1$; 2) $\tau=1$, and 3) $\tau>1$- where degenerate kink
families arise in the model. In Figure 9b we show the elements of
${\cal M}$ when $\tau$ is varied from 0 to 3. If $\tau \in (0,1)$,
${\cal M}$ is formed by four elements living at the intersection
points of the two enveloping ellipses of the separatrix trajectories
with the $\phi_1$-axis. For $\tau> 1$ there are still four elements
of ${\cal M}$ in the $\phi_1$-axis but also another four elements of
${\cal M}$ at the intersection of the enveloping ellipse and
hyperbola of the separatrix trajectories arise. $\tau=1$ marks a
pitchfork bifurcation of the elements in the interior ellipse when
this curve becomes a hyperbola for $\tau>1$ . All these statements
become clear looking at the potential energy density in elliptic
variables:
\begin{equation}
\displaystyle \xi_{\sigma \tau}^{*}
U(\phi_1,\phi_2)=\frac{1}{u^2-v^2} \left[ (u^2-\sigma^2)^2
(u^2-\tau^2)^2 (u^2-\Omega^2)+(v^2-\sigma^2)^2 (v^2-\tau^2)^2
(\Omega^2 -v^2) \right] \label{eq:potmetaelip}
\end{equation}
Comparison between (\ref{eq:potmetaelip}) and (\ref{eq:deta}) shows
that we are dealing with a family of generalized MSTB models.

\subsection{The structure of the set ${\cal M}$}

We shall now describe the distribution of the zeroes of the
potential energy density both in ${\mathbb E}^2$ and ${\mathbb R}^2$
successively for $\tau<1$, $\tau>1$, and $\tau=1$.
\begin{itemize}
\item {\bf Regime I:} $0<\tau^2<\sigma^2<1$. The model belongs to type I-1a
generalized MSTB models. Because $\Omega<\tau<\sigma$ we have: $n=1$
, $m=2$. The set of roots of $g$ and $f$ are respectively $\bar{\cal
M}_g=\{ \sigma_{-1}=-\Omega; \sigma_1=\Omega \}$ and $\bar{\cal
M}_f=\{ \sigma_{1}=\Omega; \sigma_2=\tau; \sigma_3=\sigma \}$.
Therefore, the set of zeroes of $U(\phi_1,\phi_2)$ has four
elements, both in ${\mathbb E}^2$ and ${\mathbb R}^2$, in this range
of the parameters:
\[
\bar{\cal M}_{\rm I}= \{ \bar{A}_\pm=(\sigma,\pm\Omega);
\bar{B}_\pm=(\tau,\pm\Omega) \} \qquad  , \qquad  {\cal M}_{\rm I} =
\{ A_\pm=(\pm\sigma,0); B_\pm=(\pm \tau,0) \} \quad .
\]
The model exhibits symmetry with respect to the Klein four-group
${\mathbb Z}_2^2$ generated by the reflections of the fields:
$\phi_1 \rightarrow -\phi_1$ and $\phi_2 \rightarrow -\phi_2$. In
the quantum version of the model, the symmetry with respect to the
${\mathbb Z}^2$ sub-group generated by $\phi_1 \rightarrow -\phi_1$,
however, is broken by the choice of any $A$ or $B$ ground state. The
set ${\cal M}_{\rm I}$ is formed by the $A$- and $B$-orbits of this
sub-group, such that the moduli space of vacua has two elements:
${\cal M}_{\rm I}/{\mathbb Z}_2^2=\{A,B\}$.

In Figure 10 the reticulum corresponding to this regime  is shown.
It is formed by two cells: $C_{-11}^{12}$, a type 3 cell delimited
by the edges $B_+B_-$, $B_+F_+$, $F_+F_-$, $F_-B_-$, and
$C_{-11}^{23}$, a type 1 cell whose vertices are the points $A_+$,
$A_-$, $B_+$, and $B_-$. On the Cartesian plane, ${\mathbb R}^2$,
the cell $C_{-11}^{23}$ is the region confined between the two
ellipses
\begin{equation}
\phi_1^2+\frac{\phi_2^2}{\bar\sigma^2}=\tau^2 \qquad ({\rm left})
\qquad , \qquad \qquad \phi_1^2+\frac{\phi_2^2}{\bar\tau^2}=\sigma^2
\qquad ({\rm right}) \qquad , \label{eq:elipse2}
\end{equation}
whereas $C_{-11}^{12}$ is the region bounded by the ellipse
(\ref{eq:elipse2})(right). To simplify this and subsequent formulas
we use the following conventions: $\bar{\sigma}=\sqrt{1-\sigma^2}$,
$\bar{\tau}=\sqrt{|1-\tau^2|}$.

\begin{figure}[htb]
\centerline{
\includegraphics[height=3.5cm]{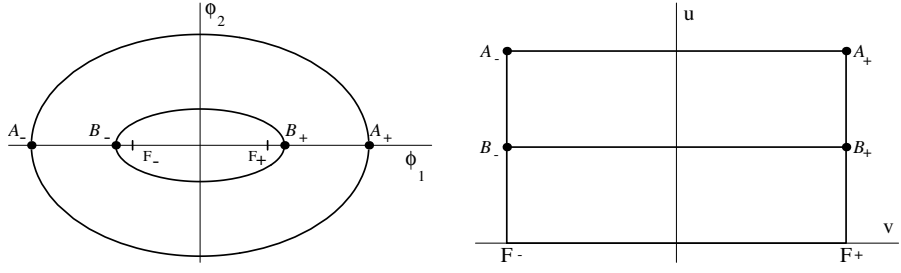}}
\caption{\small \textit{Reticulum in ${\mathbb E}^2$ and elements of
$\bar{\cal M}$ in the regime I. Image in ${\mathbb R}^2$.}}
\end{figure}

\item {\bf Regime II:} $0<\sigma^2<1<\tau^2$. Because
$\sigma<\Omega<\tau$, the model also belongs to Type I 1a
generalized MSTB models in this regime, but now: $n=2$, $m=1$. The
set of roots of $g$ and $f$ are respectively: $\bar{\cal
M}_g=\{\sigma_{-2}= -\Omega ; \sigma_{-1}=-\sigma; \sigma_1=\sigma ;
\sigma_2=\Omega \}$, $\bar{\cal M}_f=\{ \sigma_{2}=\Omega;
\sigma_3=\tau \}$. The cardinal of the set of zeroes of the
potential energy in the elliptic strip ${\mathbb E}^2$ differs from
the same number in the Cartesian plane ${\mathbb R}^2$. $\bar{\cal
M}_{\rm II}$ has six elements whereas ${\cal M}_{\rm II}$ contains
eight:
\[
\bar{\cal M}_{\rm II} =   \{ \bar{A}_\pm=(\tau,\pm\Omega) \, ;\,
\bar{B}_\pm=(\Omega,\pm\sigma) \, ; \, C_\pm^\pm=(\tau,\pm\sigma) \,
\}
\]
\[ {\cal M}_{\rm II} =   \{ A_\pm=(\pm\tau,0) \, ;\,
B_\pm=(\pm\sigma,0) \, ; \, C_\pm^+=(1,\pm\bar\Omega) \,
;\,C_\pm^-=(-1,\pm\bar\Omega)  \} \quad , \quad
\bar{\Omega}=\bar{\sigma}|\bar{\tau}| \quad .
\]
The key point in the bifurcation process is that the ellipse
(\ref{eq:elipse2})(right) arising in regime I becomes a hyperbola in
regime II. As in regime I, there are two orbits -one $A$, and one
$B$- in ${\cal M}_{\rm II}$ of the ${\mathbb Z}_2$ sub-group
generated by $\phi_1\rightarrow -\phi_1$. The four $C$ points,
however, form a single orbit of the full ${\mathbb Z}_2^2$ field
reflection group because the $C^+$ and $C^-$ orbits are connected by
the other reflection $\phi_2\rightarrow -\phi_2$, which would also
be spontaneously broken in the process of quantization if one of the
ground states, $C$, is chosen.

Figure 11 shows the reticulum in ${\mathbb E}^2$ in this regime. It
is formed by three cells: $C_{-11}^{23}$, $C_{-2-1}^{23}$, and
$C_{12}^{23}$. The images of $C_{-2-1}^{23}$ and $C_{12}^{23}$ in
${\mathbb R}^2$ are connected by the field reflection $\phi_1
\rightarrow - \phi_1$. $C_{\pm 2\pm 1}^{23}$ are type 2 cells with
vertices at the points $A_\pm$, $B_\pm$, $C_\pm^-\equiv C_\pm^+$,
and $F_\pm$. The vertices of the type 1 cell  $C_{-11}^{23}$ are the
points $B_-$, $B_+$, $C_-^\pm$, and $C_+^\pm$.

\begin{figure}[htb]
\centerline{
\includegraphics[height=3.5cm]{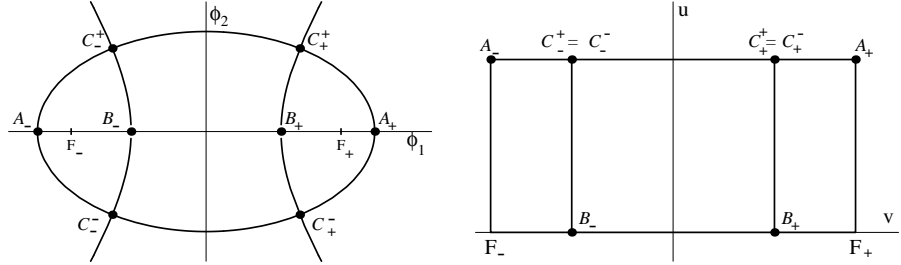}}
\caption{\small \textit{Reticulum in ${\mathbb E}^2$ and elements of
$\bar{\cal M}$ in the regime II.}}
\end{figure}

\item {\bf Regime III:} $0<\sigma^2<1=\tau^2$. This is the border between the
two previous regimes. The potential energy density
(\ref{eq:potmetaelip}) becomes
\begin{equation}
\displaystyle \xi^{*} U=\frac{1}{u^2-v^2} \left[ (u^2-\sigma^2)^3
(u^2-1)^2 -(v^2-\sigma^2)^3 (v^2-1)^2  \right] \label{eq:meta15}
\qquad .
\end{equation}
Because $\sigma=\Omega<1$, the model belongs to type I-2a
generalized MSTB models, with $n=1$, $m=1$. There are four elements
in both $\bar{\cal M}$ and ${\cal M}$:
\[
\bar{\cal M}_{\rm III} = \{ \bar{A}_\pm=(1,\pm\Omega) \, ;\,
\bar{B}_\pm= (\Omega,\pm\Omega)\, \}\qquad , \qquad {\cal M}_{\rm
III} = \{ A_\pm=(\pm 1,0) \, ;\, B_\pm= (\pm\Omega,0)\, \}
\]
The phase transition occurs when the points $B$ cross the foci,
which are zeroes of the potential $U$. In Figure 12, the reticulum
in ${\mathbb E}^2$ for this regime is shown. There is only the type
1 cell $C_{-11}^{12}$, with vertices at the points $A_+$, $A_-$,
$B_+$, and $B_-$
\begin{figure}[htb]
\centerline{
\includegraphics[height=3.5cm]{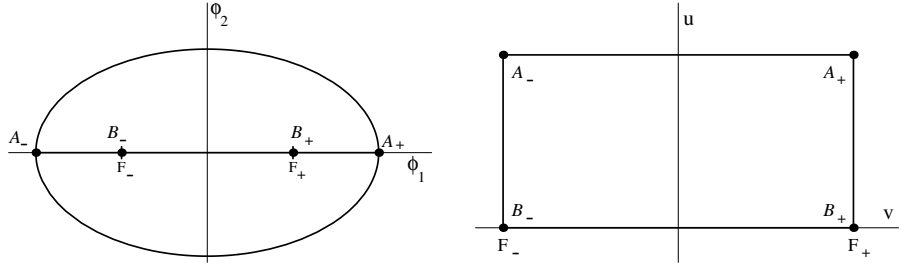}}
\caption{\small \textit{Reticulum in ${\mathbb E}^2$ and elements of
$\bar{\cal M}$ in the regime III.}}
\end{figure}
\end{itemize}

\subsection{Kink Variety in Regime I}

\subsubsection*{Singular kinks}

Let us try the orbit $\phi_2=0$ in the second-order static field
equations (\ref{eq:ordinary}). A simple integration plus some
technical manipulations provide several basic kink solutions
encompassed in the formula:
\begin{equation}
\left| \frac{\sigma+\phi_1}{\sigma-
\phi_1}\right|^{\frac{1}{\sigma}} \left| \frac{\tau-\phi_1}{\tau+
\phi_1}\right|^{\frac{1}{\tau}} =e^{\pm 2 \sqrt{2} (\sigma^2-\tau^2)
\bar x} \label{eq:meta3}   \qquad .
\end{equation}

1(a).  ${K}_1^{BF\bigvee FF\bigvee FB}$: If $|\phi_1(x)|<\tau$ for
all finite $x$ and $\lim_{x\rightarrow -\infty}\phi_1(x)=\pm\tau$,
$\lim_{x\rightarrow +\infty}\phi_1(x)=\mp\tau$, the solution in
(\ref{eq:meta3}) describes kinks ${K}_1^{BF\bigvee FF\bigvee FB}$
that join the points $B_+$ and $B_-$, crossing the foci $F_\pm$ such
that $\phi_1=0$, -where the field changes sign- occurs for
$\bar{x}=0$. These kinks are analogous to the TK1 kinks of the MSTB
model. Their energy is:
\begin{eqnarray*}
{E}[{K}_1^{BF\bigvee FF\bigvee FB}]&=& \left|W(-\tau
,0)-W(-\Omega,0)\right|+
\left|W(-\Omega ,0)-W(\Omega,0)\right|+\left|W(\Omega ,0)-W(\tau,0)\right|\\
&=& \displaystyle\frac{4 \sqrt{2}}{15} \tau [ 5
\Omega^2(1+3\sigma-\sigma^3)-\tau^4+\Omega^2\tau^2(3\sigma^3-5\sigma)]
\qquad \qquad .
\end{eqnarray*}
Because all of them cross a conjugate point (a focus), TK1 kinks are
unstable.

1(b). ${K}_1^{A_\pm B_\pm}$: if $\tau < |\phi_1(x)|<\sigma$ for all
finite $x$ and $\lim_{x\rightarrow -\infty}\phi_1(x)=\pm\tau$,
$\lim_{x\rightarrow +\infty}\phi_1(x)=\pm\sigma$ the solution in
(\ref{eq:meta3}) describes $K_1^{AB}$ kinks. The $K_1^{AB}$-kink
orbits join the points $A_\pm$  with $B_\pm$ along the $\phi_1$-axis
, see Figure 10. Their energy is:
\[
{E}[{K}_1^{A_\pm B_\pm}]=\left|W(\pm\sigma,0)-W(\pm\tau,0)\right|=
\displaystyle\frac{2 \sqrt{2}}{15} (\sigma-\tau)^3 (\sigma^2+3
\Omega +\tau^2) \qquad .
\]
These kinks are stable because their orbits are not crossed by any
other kink orbits.

\vspace{0.2cm}

2. ${K}_1^{B_\pm B_\mp}$: plugging the elliptic orbit
(\ref{eq:elipse2})(left) into the equations (\ref{eq:ordinary}),
solving in favor of $\phi_1$, and integrating the quadrature, we
obtain:
\begin{equation}
\left| \frac{\tau+\phi_1}{\tau-\phi_1} \right|^{\frac{1}{\tau}}
\left| \frac{1-\phi_1}{1+\phi_1} \right|= e^{2 \sqrt{2} \sigma^3
\bar\tau^2 \bar x } \qquad . \label{eq:meta5}
\end{equation}
The ${K}_1^{BB}$ solitary waves characterized by this equation have
kink orbits joining the $B_\pm$ with the $B_\mp$ points through the
upper and lower half-ellipses (\ref{eq:elipse2}) (left), see Figure
10. The second field component is given in terms of the first as:
$\phi_2(x)=\pm\bar\sigma\sqrt{\tau^2-\phi_1^2(x)}$. The energy of
the ${K}_1^{BB}$ kinks is:
\[
{E}[{K}_1^{B_\pm
B_\mp}]=\left|W(\pm\tau,0)-W(\mp\tau,0)\right|=\frac{2 \sqrt{2}}{15}
\, \Omega^3  (15-5 \sigma^2-5 \tau^2+3 \Omega^2) \qquad .
\]

\vspace{0.2cm}

3. ${K}_1^{A_\pm A_\mp}$: by trying the elliptic orbit
(\ref{eq:elipse2})(right) in the equations (\ref{eq:ordinary}),
solving in favor of $\phi_1$, and integrating the quadrature, we
find:
\begin{equation}
\left| \frac{\sigma+\phi_1}{\sigma-\phi_1}
\right|^{\frac{1}{\sigma}} \left| \frac{1-\phi_1}{1+\phi_1} \right|
= e^{2 \sqrt{2} \tau^3 \bar\sigma^2 \bar x } \qquad
.\label{eq:meta7}
\end{equation}
The ${K}_1^{AA}$ solitary waves determined by this equation have
kink orbits joining the $A_\pm$ with the $A_\mp$ points through the
upper and lower half-ellipses (\ref{eq:elipse2}) (right), see Figure
10. The second field component is given in terms of the first as:
$\phi_2(x)=\pm\bar\tau\sqrt{\sigma^2-\phi_1^2(x)}$. The energy of
the ${K}_1^{AA}$ kinks is:
\[
{E}[{K}_1^{A_\pm
A_\mp}]=\left|W(\pm\sigma,0)-W(\mp\sigma,0)\right|=\frac{2
\sqrt{2}}{15} \, \Omega^3  (15-5 \sigma^2-5 \tau^2+3 \Omega^2)
\qquad .
\]
Amazingly, these kinks have the same energy as the ${K}_1^{BB}$
kinks despite the complicated expression of the super-potential $W$
(\ref{eq:esup}). Two observations help to understand this point: (1)
The expression (\ref{eq:esup}) for $W$ is invariant under the
exchange $\sigma\leftrightarrow\tau$. (2) The kink orbits
${K}_1^{AA}$ and ${K}_1^{BB}$ have exactly the same length in
${\mathbb E}^2$, see Figure 10 (right).

Finally, the ${K}_1^{AA}$ and ${K}_1^{BB}$ kink orbits are not
crossed by other trajectories. Thus, these kinks are stable.

\subsubsection*{Generic kinks}

By applying the Hamilton-Jacobi procedure to find the kink orbits
and profiles respectively as the quadratures of (\ref{eq:diez}) and
(\ref{eq:once}), we identify the kinks confined inside the cells. In
this case, integration of the kink orbit equation (\ref{eq:diez})
corresponds to
\begin{equation}
\begin{array}{l}
{\displaystyle{  \left\{ \left( \frac{u-\Omega}{u+\Omega}
\right)^{\frac{1}{\bar\sigma^2 \bar\tau^2}} \left| \frac{u+
\tau}{u-\tau} \right|^{\frac{\sigma^3} { \bar\sigma^2
(\sigma^2-\tau^2)}} \left( \frac{\sigma-u}{\sigma+u}
\right)^{\frac{\tau^3} { \bar\tau^2 (\sigma^2-\tau^2)}}
\right\}^{ {\rm Sign} ({u}')}}} \times \\
{\displaystyle{ \left\{ \left( \frac{\Omega-v}{\Omega+v} \right)^{
\frac{1}{\bar\sigma^2 \bar\tau^2}} \left( \frac{ \tau+v}{\tau-v}
\right)^{\frac{\sigma^3} { \bar\sigma^2 (\sigma^2-\tau^2)}} \left(
\frac{\sigma-v}{\sigma+v} \right)^{\frac{\tau^3} { \bar\tau^2
(\sigma^2-\tau^2)}} \right\}^{ {\rm Sign} ({v}')}
 = e^{2 \sqrt{2} \Omega^3  \gamma_1}}}
\end{array}
\label{eq:meta8} \quad ,
\end{equation}
whereas the spatial distribution of these solitary waves, the result
of integrating (\ref{eq:once}), is given by:
\begin{equation}
\begin{array}{l}
{\displaystyle \left\{ \left( \frac{u-\Omega}{u+\Omega} \right)^{
\frac{1}{\bar\sigma^2 \bar\tau^2}} \left| \frac{u+ \tau}{\tau-u}
\right|^{\frac{\sigma} {\bar\sigma^2 (\sigma^2-\tau^2)}} \left(
\frac{\sigma-u}{\sigma+u} \right)^{\frac{\tau} {\bar\tau^2
(\sigma^2-\tau^2)}}
\right\}^{ {\rm Sign} ({u}')}} \times \\
{\displaystyle \left\{ \left( \frac{\Omega-v}{\Omega+v} \right)^{
\frac{1}{\bar\sigma^2 \bar\tau^2}} \left( \frac{\tau+v}{\tau-v}
\right)^{\frac{\sigma} {\bar\sigma^2 (\sigma^2-\tau^2)}} \left(
\frac{ \sigma-v}{\sigma+v} \right)^{\frac{\tau} {\bar\tau^2
(\sigma^2-\tau^2)}} \right\}^{ {\rm Sign} ({v}')}
 = e^{2 \sqrt{2} \Omega (x+\gamma_2)}}
\end{array}
\label{eq:meta9} \qquad .
\end{equation}

4. ${K}_2^{BF\bigvee FB}(\gamma_1)$: In the $C_{-11}^{12}$ cell
-$-\tau<\phi_1<\tau$, $-\tau\bar\sigma<\phi_2<\tau\bar\sigma$ in
${\mathbb R}^2$, or, $\Omega< u < \tau$, $-\Omega < v < \Omega$ in
${\mathbb E}^2$- there are no trajectories determined by
(\ref{eq:meta8}) and (\ref{eq:meta9}) with either ${\rm
Sign}\,\,u^\prime={\rm Sign}\,\,v^\prime$ or ${\rm
Sign}\,\,u^\prime\neq{\rm Sign}\,\,v^\prime$ that connect nodal $B$
points. Nevertheless, solutions of the second-order static field
equations are obtained by continuously gluing one trajectory from a
$B$ point to the farther focus with another trajectory coming back
from that focus to the $B$ point. Necessarily, one piece of the
joint two-step trajectory is determined by (\ref{eq:meta8}) and
(\ref{eq:meta9}) with a combination of signs, whereas the second
piece obeys the same equations with the other sign combination.

These solitary wave solutions form a family of non-topological
kinks. There are two real integration constants: $\gamma_1\in
(-\infty,\infty)$ and $\gamma_2\in (-\infty,\infty)$. The second one
sets the kink center, but $\gamma_1$ distinguishes between the
members of the kink family. All these kink orbits connect the points
$B_\pm$ with themselves, passing through one of the foci, see Figure
13. We shall refer to the members of this family as
$K_2^{BB}(\gamma_1)$ kinks. The behavior of these solutions is
completely analogous to that of the NTK kinks of the MSTB model: all
the kink orbits meet at a single focus and are thus unstable.
\begin{figure}[htb]
\centerline{
\includegraphics[height=3.cm]{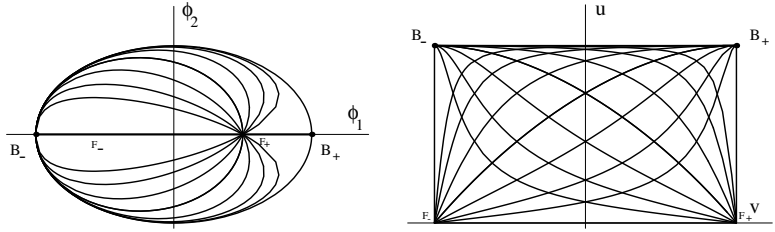}}
\caption{\small \textit{$K_2^{BB}(\gamma_1)$ Orbits: in ${\mathbb
R}^2$ (left) in ${\mathbb E}^2$ (right). The kink profiles and
energy densities in this type of cell are plotted in Figure 1.}}
\end{figure}

\noindent All the non-topological kinks are degenerated in energy,
${E}[{K}_2^{BF\bigvee FB}(\gamma_1)]= \left|W(-\tau
,0)-W(-\Omega,0)\right|+ \left|W(-\Omega
,0)-W(\Omega,0)\right|+\left|W(\Omega ,0)-W(\tau,0)\right| +
\left|W(\pm\tau,0)-W(\mp\tau,0)\right|={E}[{K}_1^{BF\bigvee FF
\bigvee FB }]+{E}[{K}_1^{B}]= \frac{2 \sqrt{2}}{15} \tau^3 (10
\sigma^2+45 \sigma^3- 15 \sigma^5-2 \tau^2-15 \sigma \Omega^2+9
\sigma^3 \Omega^2)$, and their energy is equal to the sum of the
energies of the topological kinks living at the boundary of the
cell. In keeping with this kink mass sum rule, one can check that
these topological kinks arise at the $\gamma_1\rightarrow\pm\infty$
limit of the non-topological kink moduli space ({\ref{eq:meta8}):
\[
\lim_{\gamma_1 \rightarrow \pm \infty} K_2^{BB}(\gamma_1) \equiv
K_1^{BF\bigvee FF \bigvee FB}+{K}_1^{BB} \qquad .
\]

5. ${K}_2^{A_\pm B_\mp}(\gamma_1)$: In the cell $C_{-11}^{23}$, the
domain in ${\mathbb R}^2$ bounded by the ellipses
(\ref{eq:elipse2})(left) and (\ref{eq:elipse2})(right), -
$\phi_1^2+\frac{\phi_2^2}{\bar\sigma^2}<\tau^2$,
$\phi_1^2+\frac{\phi_2^2}{\bar\tau^2}<\sigma^2$ in ${\mathbb R}^2$,
or, $\tau< u <\sigma$ , $-\Omega< v<\Omega$ in ${\mathbb E}^2$-
there are curves characterized by (\ref{eq:meta8}) and
(\ref{eq:meta9}) with the same sign combination joining the nodal
$A$ and $B$ points. These solitary wave solutions form another
one-parametric kink family. The kink orbits connect the points
$A_\pm$ with $B_\mp$ or viceversa -depending on the sign
combination- and are thus topological kinks, see Figure 14. The
orbits are not crossed by any other trajectory and these kinks are
stable. Their energy is: ${E}[{K}_2^{AB}(\gamma_1)] =
\left|W((\pm\sigma,0)-W(\mp\sigma,0)\right|+\left|W((\pm\sigma,0)-W(\pm\tau,0)\right|
=E[{K}_1^{AA}]+E[{K}_1^{AB}]=
\left|W((\pm\tau,0)-W(\mp\tau,0)\right|+
\left|W((\pm\sigma,0)-W(\pm\tau,0)\right|=E[{K}_1^{BB}]+E[{K}_1^{AB}]
=\frac{2 \sqrt{2}}{15} (\sigma^5-5 \sigma \Omega^2+5 \Omega^2 \tau+
15 \Omega^3 -\tau^5-5 \Omega^3 \tau^2+3 \Omega^5 )$; i.e., equal to
the sum of the energies of two of the topological kinks living in
the cell boundary.
\begin{figure}[htb]
\centerline{\includegraphics[height=3.cm]{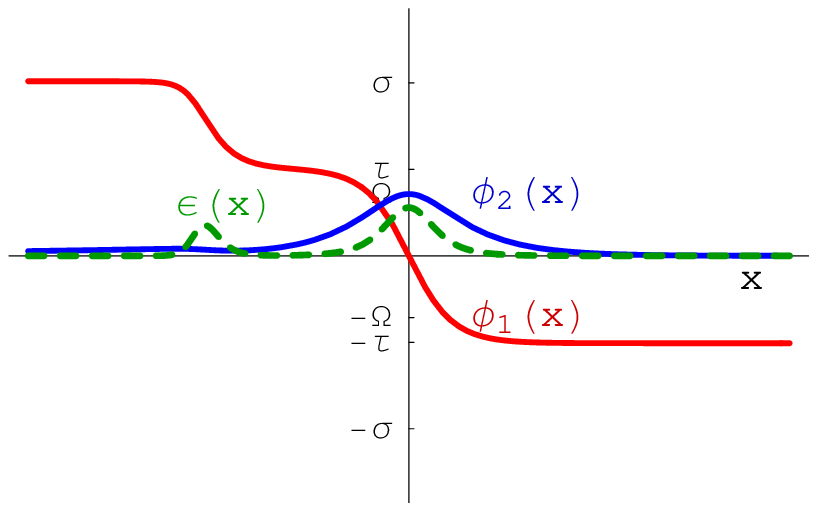}
\includegraphics[height=3.cm]{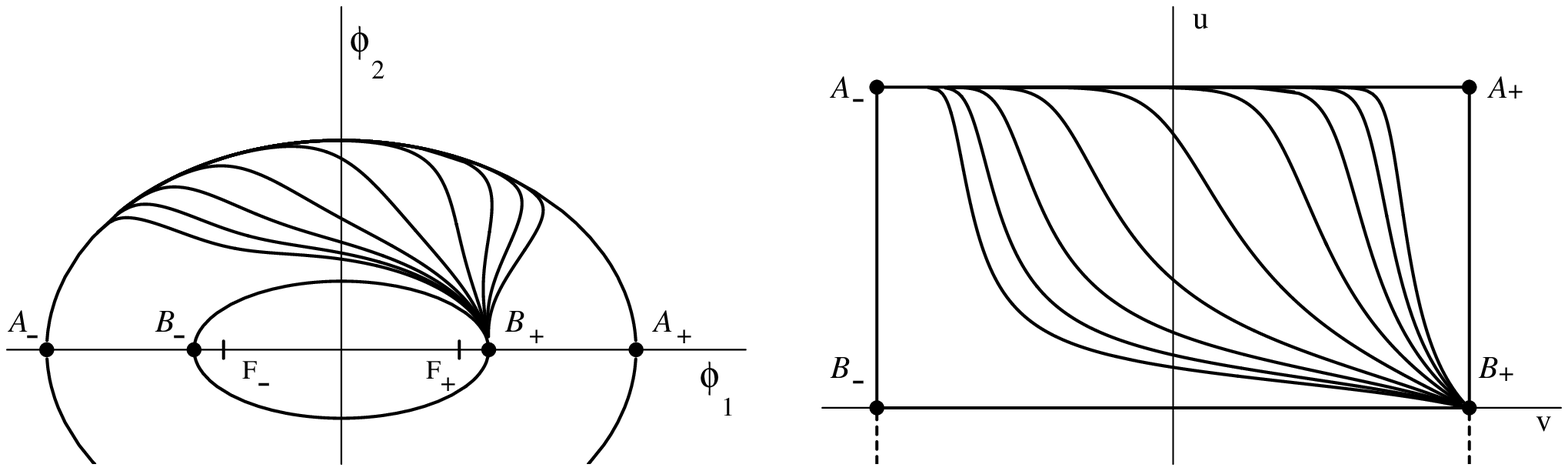}}
\caption{\small \textit{$K_2^{AB}(\gamma_1)$ kinks: Profiles and
energy density (left) Orbits in ${\mathbb R}^2$ (center) in
${\mathbb E}^2$ (right).}}
\end{figure}
Again, it is easy to check that these singular kink combinations
arise at the boundary of the $K_2^{AB}$ kink moduli space:
\[
\lim_{\gamma_1 \rightarrow \pm \infty} {K}_2^{AB}(\gamma_1) \equiv
\left\{
\begin{array}{l}
{K}_1^{AA}+{K}_1^{AB} \\
{K}_1^{BB}+{K}_1^{AB}
\end{array} \right. \qquad .
\]

In sum, there are six different types of kink solutions: four single
kinks, the ${K}_1^{BF\bigvee FF\bigvee FB}$, the ${K}_1^{AB}$, the
${K}_1^{AA}$ and the ${K}_1^{BB}$ kinks, and two kink families, the
${K}_2^{BB}(\gamma_1)$ and ${K}_2^{AB}(\gamma_1)$ kink moduli
spaces. The boundary of the ${K}_2^{BB}(\gamma_1)$ family is a
non-linear combination of the single kinks ${K}_1^{BF\bigvee
FF\bigvee FB}$ and ${K}_1^{BB}$ whereas the boundary of the
${K}_2^{AB}(\gamma_1)$ family is a non-linear combination of two
single kinks: either ${K}_1^{BB}$ and ${K}_1^{AB}$ or ${K}_1^{AA}$
and ${K}_1^{AB}$.

\subsection{Kink Variety in Regime II}

\subsubsection*{Singular kinks}

On the $\phi_1$ axis we again find the solutions (\ref{eq:meta3})
within the appropriate range of parameters.

1(a). ${K}_2^{A_\pm B_\pm}$: When $\sigma< |\phi(x) |< \tau$, the
kink orbits connect the $A$ with the $B$ points and viceversa. The
energy of these kinks is:
\[
{E}[{K}_2^{AB}]=\left|W(\pm\sigma,0)-W(\pm\tau,0)\right|=\frac{2
\sqrt{2}}{15} \,(\sigma+\tau)^3 (3 \Omega-\sigma^2 -\tau^2) \qquad .
\]
From the arguments of Section \S 3 one can identify these solitary
waves as composite kinks formed by two basic kinks. One subtle point
is the following: the kink orbits pass through one of the foci.
Nevertheless, on the $\phi_1$ axis the foci are regular points of
the superpotential. Although a pencil of kink orbits meet at the
foci, the ${K}_2^{AB}$ trajectories do not belong to this
congruence. We conclude that these composite kinks are stable.

\vspace{0.2cm}

1(b). ${K}_1^{B_\pm B_\mp}$: If $\sigma<|\phi_1(x)|$
(\ref{eq:meta3}), provides another type of solitary waves. The kink
orbits are also located on the $\phi_1$ axis but link the $B_+$ and
$B_-$ points and viceversa, see Figure 11. Their energy is:
\[
{E}[{K}_1^{B_\pm
B_\mp}]=\left|W(\sigma,0)-W(-\sigma,0)\right|=\frac{4 \sqrt{2}}{15}
\, \sigma^3 (5 \tau^2-\sigma^2) \qquad .
\]
These kink orbits do not cross the foci and are thus stable.
\vspace{0.2cm}

On the ellipse (\ref{eq:elipse2}) (left) two types of kinks live,
ruled by the formula (\ref{eq:meta5}).

2(a). ${K}_1^{AC}$: If $1<|\phi_1(x)|<\tau$ and
$\phi_2(x)=\pm\bar\sigma\sqrt{\tau^2-\phi_1^2(x)}$ the kink orbits
connect the $A_+$ and the $C^+_\pm$ points (and also the $A_-$ and
the $C^-_\pm$), see Figure 11. The energy of these kinks is:
\begin{eqnarray*}
{E}[{K}_1^{AC}]&=&\left|W(\tau,0)-W(1,\pm\bar\Omega)\right|=\left|W(-\tau,0)-W(-1,\pm\bar\Omega)\right|
\\&=&\displaystyle\frac{\sqrt{2}}{15}
\, \sigma^3 (\tau-1)^2 (2 \sigma^2+4 \sigma \Omega-10 \tau^2+6
\Omega^2-5 \tau^3+3 \Omega^2 \tau) \qquad .
\end{eqnarray*}
The kink orbits do not pass through the foci and these kinks are
stable.

\vspace{0.2cm}

2(b). ${K}_1^{C^-C^+}$:  If $1<|\phi_1(x)|$, formula
(\ref{eq:meta5}) provides the other type of kinks with orbits in the
quadric curve (\ref{eq:elipse2})(left). These kink orbits connect
the points $C^-_\pm$ with $C^+_\pm$, see Figure 11. The kink energy
is: $
{E}[{K}_1^{C^-C^+}]=\left|W(-1,\pm\bar\Omega)-W(1,\pm\bar\Omega)\right|=
\frac{4 \sqrt{2}}{15} \, \sigma^3 (5 \tau^2-\sigma^2)$ and these
kinks are also stable. In this regime the quadric
(\ref{eq:elipse2})(right) is a hyperbola:
\begin{equation}
\phi_1^2-\frac{\phi_2^2}{\tau^2-1}=\sigma^2 \label{eq:meta10} \qquad
.
\end{equation}

3. ${K}_1^{C^\pm B_\pm}$: When $1>|\phi_1(x)|>\sigma$ and
$\phi_2(x)=\pm|\bar\tau|\sqrt{\phi_1^2(x)-\sigma^2}$, formula
(\ref{eq:meta7}) gives these kinks. The kink orbits connect
$C^-_\pm$ with $B_-$ and $C^+_\pm$ with $B_+$, see Figure 11. The
energy of these stable kinks is: $E[{K}_1^{C
B}]=\left|W(-1,\pm\bar\Omega)-W(-\sigma,0)\right|=\left|W(1,\pm\bar\Omega)-W(\sigma,0)\right|
=E[{K}_2^{AB}]-E[{K}_1^{AC}]$.

\subsubsection*{Generic kinks}

In this regime of the model, the kink orbit and the kink profile
general equations (\ref{eq:diez}) and (\ref{eq:once}) provided by
the Hamilton-Jacobi method become respectively:
\begin{equation}
\begin{array}{l}
{\displaystyle  \left\{ \left( \frac{u-\sigma}{u+\sigma}
\right)^{\frac{\tau^3} { \bar\tau^2 (\tau^2-\sigma^2)}} \left(
\frac{u+\Omega}{u-\Omega} \right)^{ \frac{1}{\bar\sigma^2
\bar\tau^2}} \left( \frac{\tau-u}{\tau+u} \right)^{\frac{\sigma^3} {
\bar\sigma^2 (\tau^2-\sigma^2)}}
\right\}^{ {\rm Sign} ({u}')}} \times \\[0.2cm]
\cdot \hspace{0.2cm} {\displaystyle \left\{ \left|
\frac{v+\sigma}{v-\sigma} \right|^{\frac{\tau^3} { \bar\tau^2
(\tau^2-\sigma^2)}} \left( \frac{\Omega-v}{\Omega+v} \right)^{
\frac{1}{\bar\sigma^2 \bar\tau^2}} \left( \frac{ \tau+v}{\tau-v}
\right)^{\frac{\sigma^3} { \bar\sigma^2 (\tau^2-\sigma^2)}}
\right\}^{ {\rm Sign} ({v}')}
 = e^{2 \sqrt{2} \Omega^3  \gamma_1}}
\end{array}
\label{eq:meta11} \quad ,
\end{equation}
\begin{equation}
\begin{array}{l}
{\displaystyle \left\{ \left( \frac{u-\sigma}{u+\sigma}
\right)^{\frac{\tau} {\bar\tau^2 (\tau^2-\sigma^2)}} \left(
\frac{u+\Omega}{u-\Omega} \right)^{ \frac{1}{\bar\sigma^2
\bar\tau^2}} \left( \frac{\tau-u}{\tau+u} \right)^{\frac{\sigma}
{\bar\sigma^2 (\tau^2-\sigma^2)}}
\right\}^{ {\rm Sign} ({u}')}} \times \\[0.2cm]
\cdot \hspace{0.2cm} {\displaystyle \left\{ \left| \frac{
v-\sigma}{v+\sigma} \right|^{\frac{\tau} {\bar\tau^2
(\tau^2-\sigma^2)}} \left( \frac{\Omega+v}{\Omega-v} \right)^{
\frac{1}{\bar\sigma^2 \bar\tau^2}} \left( \frac{\tau-v}{\tau+v}
\right)^{\frac{\sigma} {\bar\sigma^2 (\tau^2-\sigma^2)}} \right\}^{
{\rm Sign} ({v}')}
 = e^{2 \sqrt{2} \Omega (x+\gamma_2)}}
\end{array}
\label{eq:meta12} \quad .
\end{equation}

In ${\mathbb R}^2$, the boundaries of the cells are formed by pieces
of the ellipse (\ref{eq:elipse2}) (left), the hyperbola
(\ref{eq:meta10}), and the $\phi_1$ axis, see Figure 15. Although
there are three cells in this regime, two of them are related by the
symmetries of the model. The analysis is thus restricted to the
$C_{-11}^{23}$ and $C_{12}^{23}$ cells. From the curves determined
by (\ref{eq:meta11}) and (\ref{eq:meta12}), we find the following
three types of kink families:

\vspace{0.2cm}

4(a). ${K}_2^{B_\pm C^+_\mp}(\gamma_1)$: In the cell $C_{-11}^{23}$
the trajectories determined by (\ref{eq:meta11}) and
(\ref{eq:meta12}) connect the points $B_+$ with $C^+_-$ -skipping
the saddle points $B_-$ and $C^+_+$- or $B_-$ with $C^+_+$ -skipping
the saddle points $B_+$ and $C^+_-$-, depending on the sign of the
combinations ${\rm Sign} ({u}') = {\rm Sign} ({v}')$ and ${\rm Sign}
({u}') \neq {\rm Sign} ({v}')$. The orbits of these ${K}_2^{B_\pm
C^+_\mp}(\gamma_1)$ one-parametric kink families do not cross the
foci, see Figure 15, and the kinks are stable.

\begin{figure}[htb]
\centerline{\includegraphics[height=3.cm]{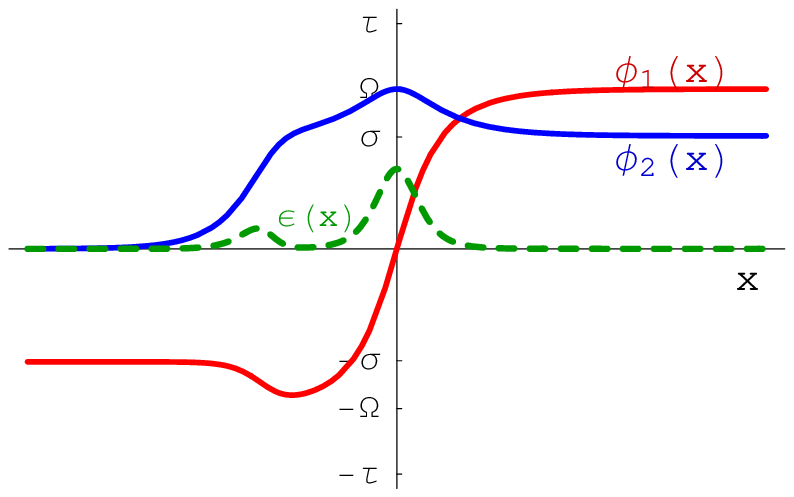}
\includegraphics[height=3cm]{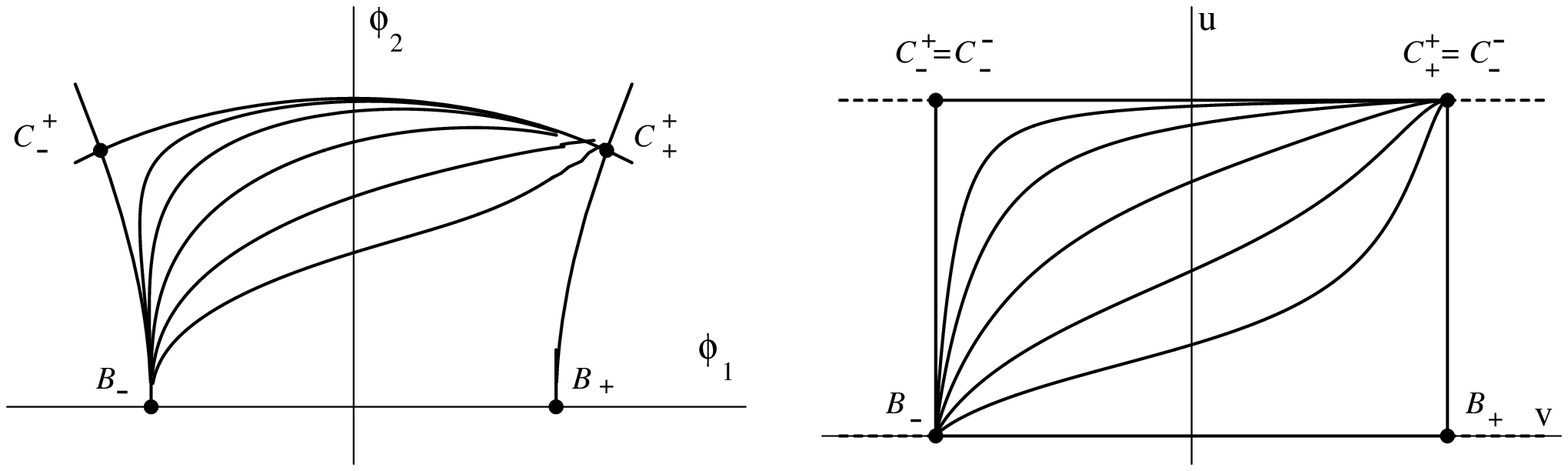}}
\caption{\small \textit{${K}_2^{B_\pm C^+_\mp}(\gamma_1)$ kinks:
Profiles and energy density (left) Orbits in ${\mathbb R}^2$
(center) in ${\mathbb E}^2$ (right).}}
\end{figure}

\noindent The energy of any member of these families of topological
kinks and the kink mass sum rules are:
\begin{eqnarray*}
{E}[{K}_2^{B_\pm C^+_\mp}(\gamma_1)]& = & \left|W(\pm\sigma,0)-W(1,\mp\bar\Omega)\right|
=E[{K}_1^{B_\pm C^+_\pm}]+E[{K}_1^{C_\mp^+ C^+_\pm}]=E[{K}_1^{B_\pm C^+_\pm}]+E[{K}_1^{B_\mp B_\pm}]\\
&=&\frac{\sqrt{2}}{15} (-4 \sigma^5+20 \sigma \Omega^2+10 \Omega^2
\tau+15 \Omega^3 -5 \sigma^2 \Omega^3-2 \tau^5-5 \Omega^3 \tau^2+3
\Omega^5 ) \qquad .
\end{eqnarray*}
This result perfectly fits the infinite $\gamma_1$ limit of the
family:
\[
\lim_{\gamma_1 \rightarrow \pm \infty} {K}_2^{B_\pm
C^+_\mp}(\gamma_1) \equiv \left\{
\begin{array}{l}
{K}_1^{B_\pm C^+_\pm}+{K}_1^{C^+_\mp C^+_\pm} \\
{K}_1^{B_\pm C^+_\pm}+{K}_1^{B_\mp B_\pm}
\end{array} \right. \qquad \qquad .
\]

\vspace{0.2cm}

4.(b) ${K}_2^{A_+B_+}(\gamma_1)$ and ${K}_2^{A_-B_-}(\gamma_1)$:
Inside the cells $C_{12}^{23}$ and $C_{-2-1}^{23}$, the trajectories
determined by the equations (\ref{eq:meta11}) and (\ref{eq:meta12})
for any sign combination respectively connect the $A_+$ ($A_-$) and
$B_+$ ($B_-$) nodal points without crossing any focal point, see
Figure 16. These heteroclinic orbits thus provide a family of stable
topological kinks (one family per cell).

\begin{figure}[htb]
\centerline{\includegraphics[height=3.cm]{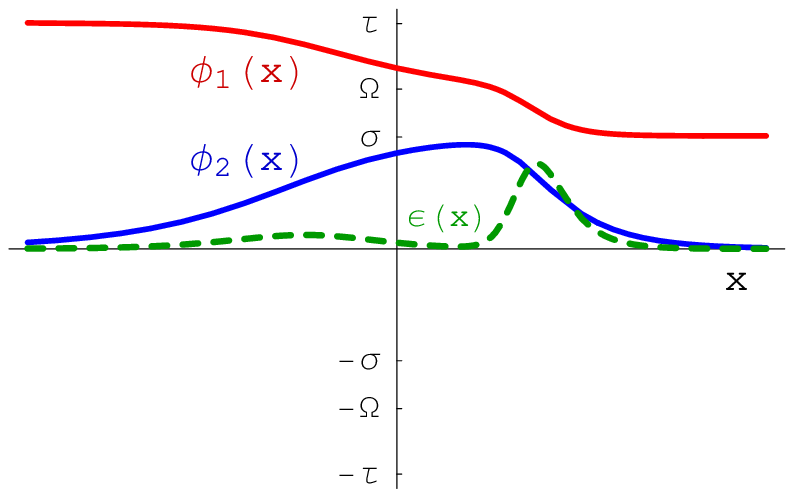}
\includegraphics[height=3cm]{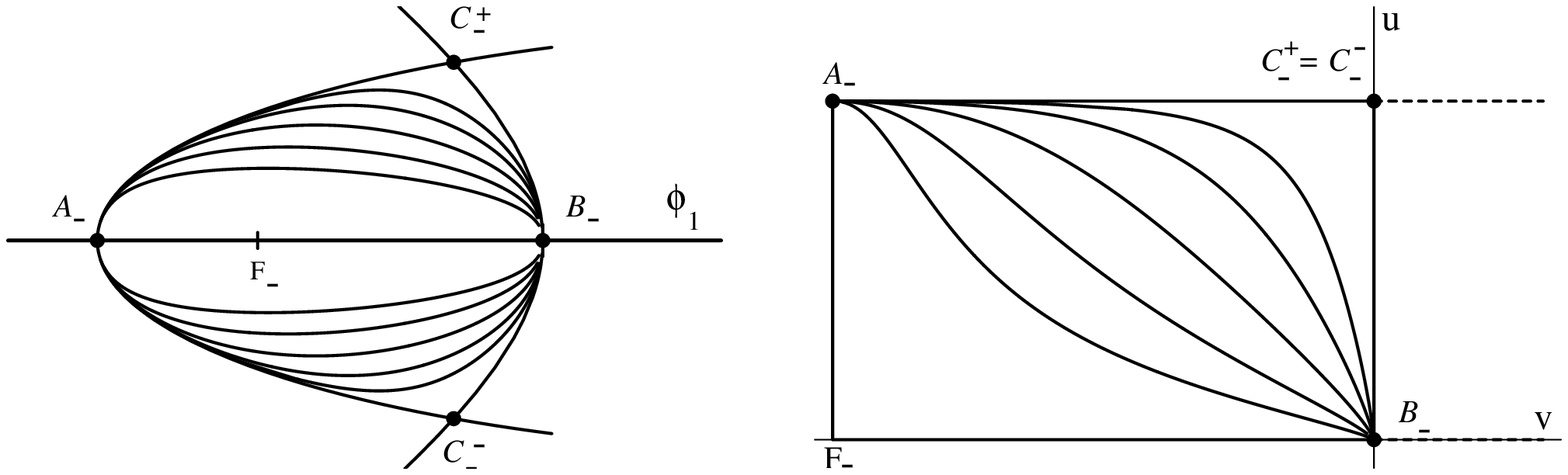}}
\caption{\small \textit{${K}_2^{BC}(\gamma_1)$ kinks: Profiles and
energy density  (left) Orbits in the Cartesian plane (center) and
elliptic plane (right).}}
\end{figure}

\noindent The energy of these topological kinks is:
\[
{E}[{K}_2^{A_\pm
B_\pm}(\gamma_1)]=\left|W(\pm\tau,0)-W(\pm\sigma,0)\right|= \frac{2
\sqrt{2}}{15} (\sigma+\tau)^3 (3 \Omega-\sigma^2-\tau^2) \qquad .
\]
The kink mass sum rule
\[
E[{K}_2^{A_\pm
B_\pm}(\gamma_1)]=E[{K}_1^{AC}]+E[{K}_1^{CB}]=E[{K}_2^{AB}]
\]
means that the infinite $\gamma_1$ limit of the family is:
\[
\lim_{\gamma_1 \rightarrow \pm \infty} {K}_2^{A_\pm B_\pm}(\gamma_1)
\equiv \left\{
\begin{array}{l}
{K}_1^{AC}+{K}_1^{CB} \\
{K}_2^{BC}
\end{array} \right.  \qquad \qquad .
\]

5. ${K}_4^{C^+_-F_-\bigvee F_-C^-_-}(\gamma_1)$ and
${K}_4^{C^+_+F_+\bigvee F_+C^-_+}(\gamma_1)$: In the $C_{12}^{23}$
and $C_{-2-1}^{23}$ cells equations (\ref{eq:meta11}) and
(\ref{eq:meta12}) include trajectories passing from one $C^+$ point
to a focus and from a focus to a $C^-$ point ( or viceversa) for any
combination of signs. Kink orbits (homoclinic in ${\mathbb E}^2$,
heteroclinic in ${\mathbb R}^2$) are formed by continuously gluing
at a focus two of these trajectories, one with ${\rm
Sign}(u)^\prime={\rm Sign}(v)^\prime$ and the other with ${\rm
Sign}(u)^\prime\neq{\rm Sign}(v)^\prime$. In this way, two families
of topological kinks are obtained with orbits starting from one of
the $C^+_\pm$ points, crossing the $F_\pm$ focus, and ending at the
correlative point $C^-_\pm$, see Figure 17. The energy of these
unstable topological kinks is:
\begin{eqnarray*}
{E}[{K}_4^{C^+F\bigvee FC^-}(\gamma_1)]&=&
\left|W(1,\pm\bar\Omega)-W(\pm\Omega,0)\right|+\left|W(\pm\Omega,0)-W(-1,\pm\bar\Omega)\right|\\&=&\frac{4
\sqrt{2}}{15} (\sigma+\tau)^3 (3 \Omega-\tau^2-\sigma^2)
\end{eqnarray*}
\begin{figure}[htb]
\centerline{\includegraphics[height=3.cm]{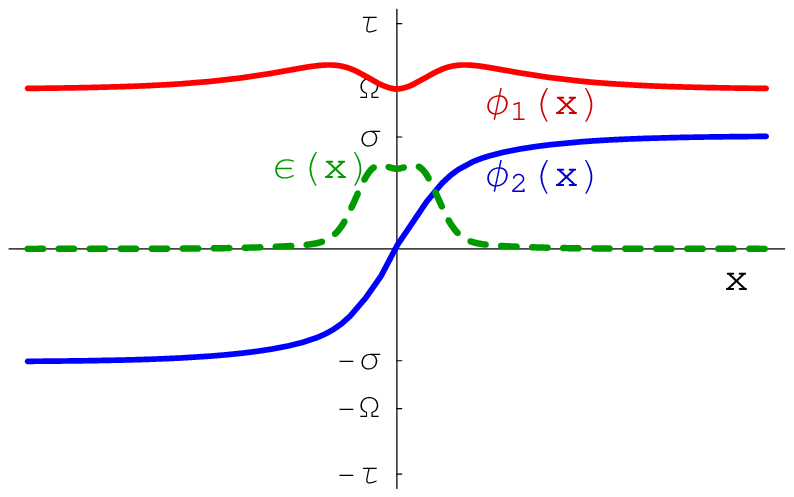}
\includegraphics[height=3cm]{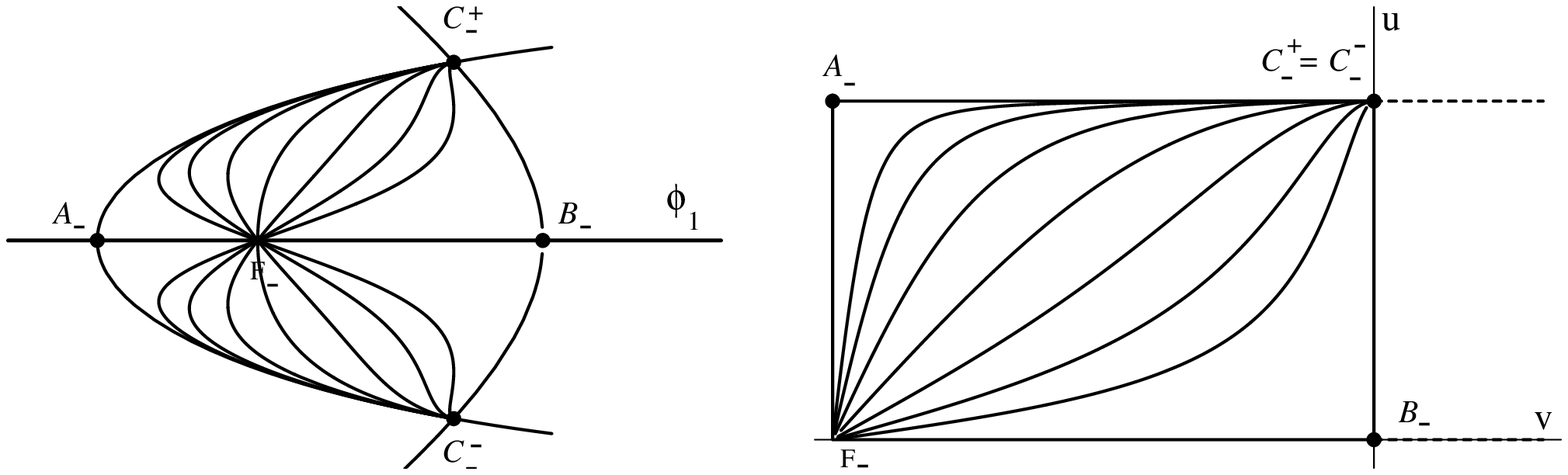}}
\caption{\small \textit{${K}_4^{C^+_+C^-_+}(\gamma_1)$ kinks:
Profiles and energy density (left) Orbits in ${\mathbb R}^2$
(center) in ${\mathbb E}^2$ (right).}}
\end{figure}

The kink mass sum rule is ${E}[{K}_4^{C^+F\bigvee
FC^-}(\gamma_1)]={E}[{K}_1^{CB}]+{E}[{K}_2^{AB}(\gamma_1)]+{E}[{K}_1^{AC}]$
in agreement with the $\gamma_1\rightarrow\infty$ limit of the
family:
\[
\lim_{\gamma_1 \rightarrow \pm \infty} {K}_4^{C^+F\bigvee
FC^-}(\gamma_1) \equiv {K}_1^{CB}+{K}_2^{AB}(\gamma_1)+{K}_1^{Ac}
\qquad .
\]

In sum, the kink variety in this regime is formed by seven kinds of
solitary waves. (1) Four singles kinks: ${K}_1^{AC}$, ${K}_1^{CB}$,
${K}_1^{BB}$ and ${K}_1^{CC}$. (2) Three kink families:
${K}_2^{BC}(\gamma_1)$, ${K}_2^{AB}(\gamma_1)$ and
${K}_4^{CC}(\gamma_1)$. Members of the ${K}_2^{BC}(\gamma_1)$ family
are composite kinks formed by a pair of either
${K}_1^{CC}$-${K}_1^{CB}$ or ${K}_1^{BC}$-${K}_1^{BB}$ kinks. The
basic kink inter-center distance in each pair is finite and is
determined by the $\gamma_1$ parameter. Basic kinks appear liberated
from each other (infinite inter-center distance) when
$|\gamma_1|=\infty$. The kinks of the ${K}_2^{AB}(\gamma_1)$ family
are similar non-linear combinations of two single kinks: the
${K}_1^{AC}$-${K}_1^{CB}$ kink pair arise at the $\gamma_1=\infty$
boundary of the ${K}_2^{AB}(\gamma_1)$ moduli space. There is,
however, a subtle point: at the $\gamma_1=-\infty$ boundary lives
the ${K}_2^{AB}$ kink, which is thus a composite but singular kink.
Finally, the structure of the ${K}_4^{CF\bigvee FC}(\gamma_1)$ kinks
is even more complex. These kinks can be thought of as non-linear
combinations of the ${K}_1^{AC}$, ${K}_2^{BA}$ and ${K}_1^{BC}$
basic kinks  when the distances between the three centers are
finite.

\subsection{Kink Variety in Regime III}

\subsubsection*{Singular kinks}

In this regime the ansatz $\phi_2=0$ plugged into (\ref{eq:meta3})
gives rise to two types of kink orbits.

1(a). ${K}_1^{A_\pm B_\pm}$:  These orbits connect the points
$A_+$ with $B_+$ or $A_-$ with $B_-$, see Figure 12. The energy of
the corresponding kinks is: ${E}[{K}_1^{AB}]= \left|W(\pm
1,0)-W(\pm\Omega,0)\right|=\frac{2 \sqrt{2}}{15}
(1-\Omega)^3(1+\Omega(3+\Omega))$.

\vspace{0.2cm}

1(b). ${K}_1^{B_+B_-}$: There is also a kink orbit on the
$\phi_1$-axis connecting $B_+$ with $B_-$. The kink energy is:
${E}[{K}_1^{B_+B_-}]=\left|W(
\Omega,0)-W(-\Omega,0)\right|=\frac{4 \sqrt{2}}{15} \Omega^3
(5-\Omega^2)$.

\vspace{0.2cm}

The third type of singular kink in regime III lives on the ellipse
(\ref{eq:elipse2})(left).

\vspace{0.2cm}

2. ${K}_1^{A_+A_-}$: These elliptic kink orbits are given by
(\ref{eq:meta5} and) connect the $A_+$ and $A_-$ points, see
Figure 12. The kink energy is:
${E}[{K}_1^{A_+A_-}]=\left|W(1,0)-W(-1,0)\right|=\frac{4
\sqrt{2}}{15} \Omega^3 (5-\Omega^2)={E}[{K}_1^{B_+B_-}]$.

\subsubsection*{Generic kinks}

The Hamilton-Jacobi procedure provides the kink orbits
\begin{equation}
\begin{array}{l}
{\displaystyle \left\{ \left( \frac{u+\sigma}{u-\sigma}
\right)^{\frac{1-3 \sigma^2}{2}} \left( \frac{1+u}{1-u}
\right)^{\sigma^3} {\rm exp} \left[ {\frac{- \sigma \bar\sigma^2
u}{u^2-\sigma^2}} \right]
\right\}^{{\rm Sign} ({u}')} \times} \\
{\displaystyle \left\{ \left( \frac{\sigma-v}{\sigma+v}
\right)^{\frac{1-3 \sigma^2}{2}} \left( \frac{1-v}{1+v}
\right)^{\sigma^3} {\rm exp} \left[ {\frac{ \sigma \bar\sigma^2
v}{v^2-\sigma^2}} \right] \right\}^{{\rm Sign} ({v}')}=e^{2
\sqrt{2}\sigma^3 \gamma_1} }
\end{array}
\label{eq:meta13}\qquad ,
\end{equation}
and the kink profiles
\begin{equation}
\begin{array}{l}
{\displaystyle \left\{ \left( \frac{u-\sigma}{u+\sigma}
\right)^{\frac{1-3 \sigma^2}{2 \sigma^3}} \left( \frac{1-u}{1+u}
\right) {\rm exp} \left[ {\frac{\bar\sigma^2 u}{\sigma^2
(u^2-\sigma^2)}} \right]
\right\}^{{\rm Sign} ({u}')} \times} \\
{\displaystyle \left\{ \left( \frac{\sigma+v}{\sigma-v}
\right)^{\frac{1-3 \sigma^2}{2 \sigma^3}} \left( \frac{1+v}{1-v}
\right) {\rm exp} \left[ {\frac{ -\bar\sigma^2 v}{\sigma^2
(v^2-\sigma^2)}} \right] \right\}^{{\rm Sign} ({v}')}=e^{2 \sqrt{2}
(x+\gamma_2)} }
\end{array}
\label{eq:meta14}
\end{equation}
of Regime III in the $C_{-11}^{12}$ cell.

\noindent 3. ${K}_2^{A_-B_+}(\gamma_1)$ and
${K}_2^{A_+B_-}(\gamma_1)$: Equations ({\ref{eq:meta13}) and
({\ref{eq:meta14}) describe two families of topological kinks
joining the $A_\pm$ with the $B_\mp$ points, see Figure 16..
\begin{figure}[htb]
\centerline{\includegraphics[height=3.cm]{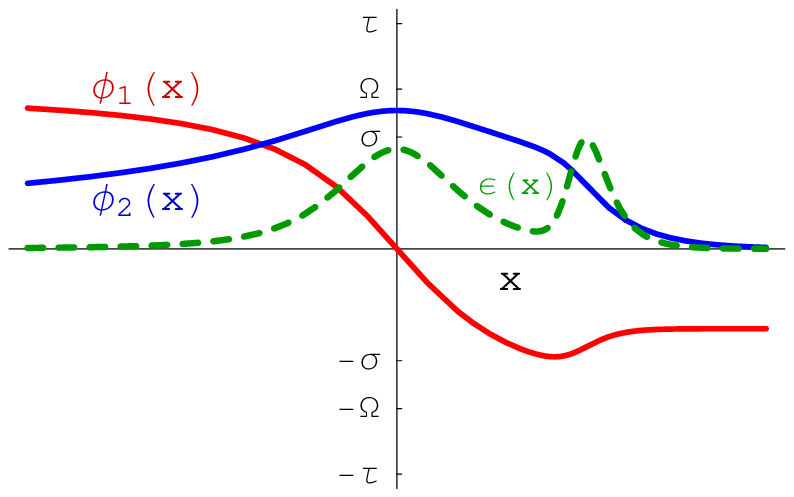}
\includegraphics[height=3cm]{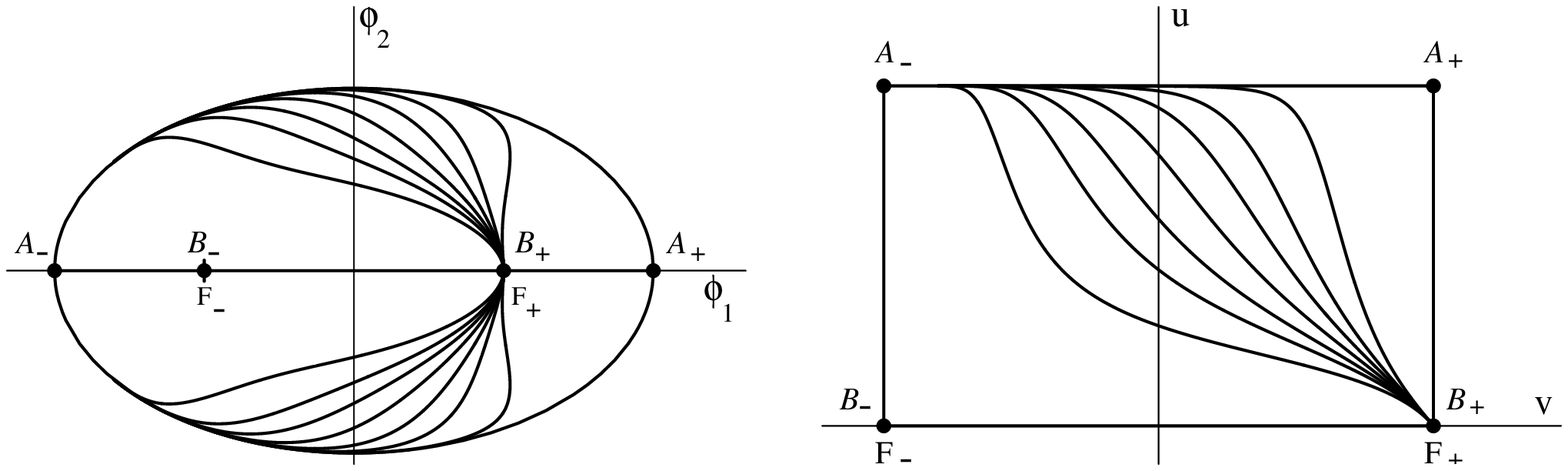}}
\caption{\small \textit{${K}_2^{A_-B_+}(\gamma_1)$ kinks: Profiles
and energy density (left) Orbits in ${\mathbb R}^2$ (center) in
${\mathbb E}^2$ (right).}}
\end{figure}

The kink energy is:
\[
{E}[{K}_2^{A_\pm B_\mp}(\gamma_1)]=\left|W(\pm
1,0)-W(\mp\Omega,0)\right|=\frac{2 \sqrt{2}}{15} (1-5 \Omega^2+15
\Omega^3 -3 \Omega^5) \qquad ,
\]
such that the kink mass sum rules hold:
\[
{E}[{K}_2^{A_\pm B_\mp}(\gamma_1)]=
E[{K}_1^{A_-A_+}]+E[{K}_1^{A_+B_+}]=E[{K}_1^{B_-B_+}]+E[{K}_1^{A_-B_-}]
\qquad \qquad ,
\]
in agreement with the $|\gamma_1|=\infty$ limit of the family
\[
\lim_{\gamma_1 \rightarrow \pm \infty} {K}_2^{A_\pm B_
\mp}(\gamma_1) \equiv \left\{
\begin{array}{l}
{K}_1^{A_-A_+}+{K}_1^{A_+B_+} \\
{K}_1^{B_-B_+}+{K}_1^{A_-B_-}
\end{array} \right. \qquad \qquad .
\]

\subsection{Kink mass sum rules}

A summary of the different kink mass sum rules is as follows:

\hspace{1cm} \begin{tabular}{l}
- Kink mass sum rules in regime I: \\[0.2cm]
\hspace{2cm} $E[K_1^{AA}]=E[K_1^{BB}]$   \\
\hspace{2cm} $E[K_2^{AB}(\gamma_1)]=E[K_1^{AB}]+E[K_1^{BB}]$    \\
\hspace{2cm} $E[K_2^{BB}(\gamma_1)]= E[{K}_1^{BF\bigvee FF\bigvee
FB}]+E[K_1^{BB}]$
\end{tabular}

\vspace{0.2cm}

\hspace{1cm} \begin{tabular}{l}
- Kink mass sum rules in regime II: \\[0.2cm]
\hspace{2cm} $E[K_1^{BB}]=E[K_1^{C^+C^+}]$   \\
\hspace{2cm} $E[K_2^{B_\mp C_\pm^+}(\gamma_1)]=E[K_1^{BC^+}]+E[K_1^{C^+_\mp C^+_\pm}]$    \\
\hspace{2cm} $E[K_2^{AB}]= E[{K}_1^{AC^+}]+E[K_1^{C^+B}]$ \\
\hspace{2cm} $E[K_2^{AB}(\gamma_1)]=E[K_2^{AB}]$    \\
\hspace{2cm} $E[K_4^{C^+F\bigvee FC^+}(\gamma_1)]=2
E[K_2^{AB}(\gamma_1)]$
\end{tabular}

\vspace{0.2cm}

\hspace{1cm} \begin{tabular}{l}
- Kink mass sum rules in regime III: \\[0.2cm]
\hspace{2cm} $E[K_1^{AA}]=E[K_1^{BB}]$   \\
\hspace{2cm} $E[K_2^{AB}(\gamma_1)]=E[K_1^{AB}]+E[K_1^{BB}]$
\end{tabular}

\section{Summary and Outlook}

The developments discussed in this paper unveil the structure of a
broad class of two-component scalar field theory models, which
generalize the well-studied MSTB model. It is well known that the
search for static solutions in (1+1)D scalar field theory is
tantamount to the solving of an analogous mechanical system with
flipped potential with respect to the field theoretical model. The
generalized MSTB models are characterized by: (1) Having a Type I
Liouville analogous mechanical system. These mechanical systems are
Hamilton-Jacobi separable by using elliptic coordinates. (2) Having
a discrete set of unstable critical points of the mechanical
potential, homogeneous solutions or \lq\lq vacua" in the field
theory. This feature supports the existence of very rich varieties
of heteroclinic and homoclinic separatrix trajectories in the
mechanical problem, and topological and non-topological solitary
waves or kinks in the field theory setting.

We have shown the following structural properties:
\begin{itemize}

\item The \lq\lq vacua" of the field theoretical system and the foci of the elliptic
and hyperbolic coordinating curves are placed at the junctions of a
rectangular mesh displayed on the elliptic strip, which is divided
into different cells where the dynamics runs separately.

\item The distribution of the vacua on this reticulum depends on
the details of the potential energy density of the model.

\item There exist only three types of cells in the mesh. In Type I
cells, the vacua of the model sit on the four vertices. Only three
vertices are vacua in Type II cells, whereas one of the two foci
sits at the fourth vertex. Finally, in Type III cells two vacua and
two foci occupy  the four vertices.

\item The cell of the reticulum is the cornerstone that determines the
structure of the kink variety. There exist two families of kink
solutions confined in each cell and single kinks living on the edges
of the cell. The characteristics of these kinks, in particular their
stability, do not depend on the details of the potential density but
only on the type of the cell in which they are confined.

\item The different kink families are
described thoroughly. Because the separability of the mechanical
model, kink mass sum rules arise between at least two types of
kinks.

\item For this reason, each
member of a kink family (confined inside the cell) is indeed a
composite kink that can be understood as a non-linear combination of
several single kinks (settled at the edges of the cell).

\item In sum, the potential energy density of
the field theoretical model determines the reticulum bounding the
cells and the cells determine the behavior and structure of the kink
families. Therefore, a qualitative description of the kink variety
of generalized MSTB models is do-able by merely observing the
expression of the potential energy density in elliptic coordinates.

\end{itemize}

We finish by noticing that the scheme developed in this paper can be
extended to other two-component scalar field theoretical models. In
particular, models with Type III and II Liouville analogous
mechanical systems, respectively Hamilton-Jacobi separable using
parabolic and polar coordinates, have very similar structures. We
believe, however, that the Type I models are primordial because
parabolic and polar coordinates are special limits of the system of
elliptic coordinates.

\section{Acknowledgements}

We are very grateful to Hernan Augusto Piragua Ariza for reporting us some typos in the introduction included in the first version of the manuscript.

\end{document}